\documentclass[twocolumn,10pt,a4paper,romanappendices]{IEEEtran-v17} 
\let\labelindent\relax 
\usepackage{enumitem}
\usepackage{hhline}

\usepackage{color}

\ifCLASSINFOpdf
  \usepackage[pdftex]{graphicx}
\else
\fi
\usepackage[cmex10]{amsmath}
\usepackage[tight,footnotesize]{subfigure}

\usepackage{hyperref}

\usepackage[utf8]{inputenc}
\usepackage{multirow}
\usepackage{relsize}
\usepackage{amssymb}
\usepackage{mathrsfs}
\usepackage{dsfont}
\usepackage{bbold}
\usepackage{xspace}
\usepackage[final,nomargin,footnote]{fixme}
\usepackage{microtype}
\usepackage{tikz}
\usepackage{pgfplots}
\pgfplotsset{compat=newest}
\pgfplotsset{plot coordinates/math parser=false}
\usepgfplotslibrary{fillbetween}
\usetikzlibrary{patterns}
\newlength\figureheight
\newlength\figurewidth
\newlength\smallfigureheight
\newlength\smallfigurewidth
\usetikzlibrary{decorations.pathreplacing}

\usetikzlibrary{shapes,arrows, positioning, fit}

\def\cdf(#1)(#2)(#3){0.5*(1+(erf((#1-#2)/(#3*sqrt(2)))))}%

\makeatletter
\def\@IEEEinterspaceratioM{0.265}
\def\@IEEEinterspaceMINratioM{0.1651}
\def\@IEEEinterspaceMAXratioM{0.38}

\def\@IEEEinterspaceratioB{0.31}
\def\@IEEEinterspaceMINratioB{0.19}
\def\@IEEEinterspaceMAXratioB{0.38}
\@IEEEtunefonts
\makeatother
\newtheorem{definition}{Definition}

\newtheorem{theorem}{Theorem}
\newtheorem{lemma}[theorem]{Lemma}
\newtheorem{corollary}[theorem]{Corollary}
\newtheorem{proposition}[theorem]{Proposition}

\providecommand{\ee}[1]{\exp\mathopen{}\left(#1\right)}

\providecommand{\eeBig}[1]{\exp\mathopen{}\Big(#1\Big)}
\providecommand{\eebigg}[1]{\exp\mathopen{}\bigg(#1\bigg)}
\providecommand{\eeBigg}[1]{\exp\mathopen{}\Bigg(#1\Bigg)}
\providecommand{\e}[1]{e^{#1}}

\newcommand{\sch}{r}
\newcommand{\Sch}{R}

\providecommand{\vectornorm}[1]{\left\lVert#1\right\rVert}

\providecommand{\indi}[1]{\mathds{1}\mathopen{}\left\{#1\right\}}

\providecommand{\X}{\mathbf{X}}

\providecommand{\x}{\mathbf{x}}
\providecommand{\y}{\mathbf{y}}

\providecommand{\len}[1]{\text{len}\farg{#1}}

\newcommand{\const}{\mathbb{c}}

\newcommand{\E}[1]{\mathbb{E}\mathopen{}\left[#1\right]}
\newcommand{\Ebig}[1]{\mathbb{E}\mathopen{}\big[#1\big]}
\newcommand{\EBig}[1]{\mathbb{E}\mathopen{}\Big[#1\Big]}
\newcommand{\Ebigg}[1]{\mathbb{E}\mathopen{}\bigg[#1\bigg]}
\newcommand{\EBigg}[1]{\mathbb{E}\mathopen{}\Bigg[#1\Bigg]}
\newcommand{\inff}[1]{\inf\mathopen{}\left\{#1\right\}}

\newcommand{\supp}[1]{\sup\mathopen{}\left\{#1\right\}}
\newcommand{\minn}[1]{\min\mathopen{}\left\{#1\right\}}

\newcommand{\minnbigg}[1]{\min\mathopen{}\bigg\{#1\bigg\}}
\newcommand{\maxxBig}[1]{\max\mathopen{}\Big\{#1\Big\}}
\newcommand{\maxxBigg}[1]{\max\mathopen{}\Bigg\{#1\Bigg\}}

\newcommand{\maxxbigg}[1]{\max\mathopen{}\bigg\{#1\bigg\}}
\newcommand{\maxx}[1]{\max\mathopen{}\left\{#1\right\}}
\newcommand{\EE}[2]{\mathbb{E}_{#1}\mathopen{}\left[#2\right]}

\newcommand{\Vaa}[2]{\text{Var}_{#1}\mathopen{}\left[#2\right]}

\newcommand{\farg}[1]{\mathopen{}\left( #1 \right)}

\newcommand{\fargBigg}[1]{\mathopen{}\Bigg( #1 \Bigg)}

\newcommand{\pr}[1]{\mathbb{P}\mathopen{}\left[#1\right]}
\newcommand{\prbig}[1]{\mathbb{P}\mathopen{}\big[#1\big]}
\newcommand{\prBig}[1]{\mathbb{P}\mathopen{}\Big[#1\Big]}
\newcommand{\prbigg}[1]{\mathbb{P}\mathopen{}\bigg[#1\bigg]}
\newcommand{\prBigg}[1]{\mathbb{P}\mathopen{}\Bigg[#1\Bigg]}

\newcommand{\Msf}{M^*_{\text{sf}}}

\newcommand{\lefto}{\mathopen{}\left}
\newcommand{\vect}[1]{\boldsymbol{\mathbf{#1}}}
\newcommand{\tildeinf}{\tilde \imath_k}
\newcommand{\diff}{\nabla}
\newcommand{\dd}{\mathop{}\!\mathrm{d}}

\newcommand{\intdiff}{\mathrm{d}}

\newcommand{\sprob}{q}
\newcommand{\tvRV}{H}
\newcommand{\tvRVbar}{\bar H}
\newcommand{\transpose}{^\mathrm{T}} 
\newcommand{\convnorm}[1]{w\farg{#1}}
\newcommand{\bscprob}{\delta}
\newcommand{\optx}{\vect{\hat v}}

\newcommand{\optbar}{\vect{\bar v}} 
\newcommand{\optbarnbold}{\bar v}
\DeclareMathOperator*{\argmin}{arg\,min}
\DeclareMathOperator*{\argmax}{arg\,max}

\newcommand{\preffree}[1]{\mathscr{#1}}

\newcommand*{\PathTITVAR}{.}%

\newcounter{MYtempeqncnt}

\newboolean{journal}
\setboolean{journal}{TRUE}
\newboolean{arxiv}
\setboolean{arxiv}{FALSE}
\newboolean{thesis}
\setboolean{thesis}{false}
\newboolean{finalmanuscript}
\setboolean{finalmanuscript}{false}

\begin{document}
%
\title{Common-Message Broadcast Channels with Feedback in the Nonasymptotic Regime: Stop Feedback}

\author{\thanks{The work of K. F. Trillingsgaard and P. Popovski  was supported by the European Research Council (ERC Consolidator Grant Nr. 648382 WILLOW) within the Horizon 2020 Program. The work of G. Durisi was supported by the Swedish Research Council under the grant 2016-032931. 
The material of this paper was presented in part at the 2016 IEEE International Symposium on Information Theory \cite{Trillingsgaard2016}.}Kasper Fløe Trillingsgaard\thanks{K. F. Trillingsgaard and P. Popovski are with the Department of Eletronic Systems, Aalborg University, 9220, Aalborg Øst, Denmark (e-mail: \{kft,petarp\}@es.aau.dk).}, \emph{Member, IEEE}, Wei Yang\thanks{W. Yang is with Qualcomm Technologies, Inc.,  San Diego, 92121, USA (e-mail: weiyang@qti.qualcomm.com).}, \emph{Member, IEEE}
,
Giuseppe Durisi\thanks{G. Durisi is with the Department of Electrical Engineering, Chalmers University of Technology, 41296, Gothenburg, Sweden (e-mail: durisi@chalmers.se).}, \emph{Senior Member, IEEE}, and
Petar Popovski, \emph{Fellow, IEEE}\ifthenelse{\boolean{finalmanuscript}}{\thanks{Copyright (c) 2018 IEEE. Personal use of this material is permitted.  However, permission to use this material for any other purposes must be obtained from the IEEE by sending a request to pubs-permissions@ieee.org.}}{}} 


%


\maketitle
\thispagestyle{empty}

\begin{abstract}
We investigate the maximum coding rate for a given average blocklength and error probability over a $K$-user discrete memoryless broadcast channel for the scenario where a common message is transmitted using variable-length stop-feedback codes. 
For the point-to-point case, Polyanskiy \emph{et al.} (2011) demonstrated that variable-length coding combined with stop-feedback significantly increases the speed of convergence of the maximum coding rate to capacity. This speed-up manifests itself in the absence of a square-root penalty in the asymptotic expansion of the maximum coding rate for large blocklengths, i.e., \emph{zero dispersion}. 
In this paper, we present nonasymptotic achievability and converse bounds on the maximum coding rate of the common-message $K$-user discrete memoryless broadcast channel, which strengthen and generalize the ones reported in Trillingsgaard \emph{et al.} (2015) for the two-user case. An asymptotic analysis of these bounds reveals that zero dispersion cannot be achieved for certain common-message broadcast channels (e.g., the binary symmetric broadcast channel). Furthermore, we identify conditions under which our converse and achievability bounds are tight up to the second order. Through numerical evaluations, we illustrate that our second-order  expansions approximate accurately the maximum coding rate and that the speed of convergence to capacity is indeed slower than for the point-to-point case.
\end{abstract}


\IEEEpeerreviewmaketitle 

\section{Introduction}
\IEEEPARstart{W}{e} consider the setup in which an encoder wishes to convey a common message over a discrete memoryless broadcast channel with feedback from $K$ decoders. Similarly to the single-decoder case,~\emph{full feedback} (i.e., instantaneous feedback of the received symbols) combined with \emph{fixed-blocklength codes} does not improve capacity, which is given by \cite[p. 126]{Gamal2011}
\begin{align}
  C = \sup_{P} \min_{k\in\{1,\cdots,K\}} I(P, W_k).\label{eq:capacity_def}
\end{align}
Here, $W_1,\cdots,W_K$ denote the channels to the decoders $1,\dots,K$, respectively, and the supremum is over all input distributions $P$. 
For the case of no feedback, the common-message broadcast channel is equivalent to a compound channel, and the speed at which $C$ is approached as the blocklength $n$ increases is of the order ${1}/{\sqrt{n}}$ (see \cite{Polyanskiy}),  which is the same as in the single-decoder (point-to-point) case \cite{Polyanskiy2010b}. Specifically, the logarithm of the maximum number of codewords $M^*(n,\epsilon)$ that can be transmitted with  blocklength $n$ and maximum error probability $\epsilon$ can be expanded as~\cite{Polyanskiy}
\begin{IEEEeqnarray}{rCl}
  \frac{1}{n}\log M^*(n,\epsilon) &=& C - \sqrt{\frac{V_{\text{no-fb}}}{n}}Q^{-1}(\epsilon) + o\lefto(\frac{1}{\sqrt{n}}\right)\label{eq:compound_asymp}
\end{IEEEeqnarray}
where
\begin{IEEEeqnarray}{rCl}
  \sqrt{V_{\text{no-fb}}} = \min_{\vect{v}: \sum_{x} v_x = 0} \max_{k\in\{1,\ldots,K\}} \Big\{\diff I_k(\vect{v}) + \sqrt{V_k}\Big\}.
\end{IEEEeqnarray}
Here, $V_k$ denotes the conditional information variance of component channel $k$ evaluated at the unique capacity-achieving distribution $P^*$ (see \eqref{eq:cond_inf_var}) and $\diff I_k(\vect{v})$ denotes the directional derivative of the mutual information of decoder $k$ at $P^*$ (see \eqref{eq:differential}).

For point-to-point channels, although feedback does not increase capacity, it improves dramatically the error exponent, provided that \emph{variable-length codes} are used. 
This was first demonstrated by Burnashev  who found that the error exponent for this setting is given by \cite{Burnashev1976}
\begin{IEEEeqnarray}{rCl}
  E(R) = \frac{\widetilde C_1}{\widetilde C}(\widetilde C-R)\label{eq:err_exp_feedback}
\end{IEEEeqnarray}
for all rates $0 <R < \widetilde C$. Here, $\widetilde C$ denotes the channel capacity for the point-to-point case and $\widetilde C_1$ denotes the maximum relative entropy between two conditional output distributions. Yamamoto and Itoh \cite{Yamamoto1979} proposed a two-phase scheme that attains the error exponent in \eqref{eq:err_exp_feedback}. Furthermore, Berlin \emph{et al.} \cite{Berlin2009} provided an alternative and simpler converse proof to \eqref{eq:err_exp_feedback}, which parallels the two-phase scheme proposed in \cite{Yamamoto1979}. 

In the fixed-error regime, Polyanskiy \emph{et al}. \cite{Polyanskiy2011} found that the speed at which the maximum coding rate converges to capacity is significantly improved in the presence of full feedback and variable-length codes. Specifically, they showed that
\begin{align}
  \frac{1}{\ell}\log \widetilde M_{\text{f}}^*(\ell, \epsilon) = \frac{\widetilde C}{1-\epsilon} -\mathcal{O}\farg{\frac{\log \ell}{\ell}}\label{eq:SD_vlf_rate}
\end{align} 
where $\ell$ denotes the average blocklength (average transmission time) and $\widetilde M_{\text{f}}^*(\ell,\epsilon)$ is the maximum number of codewords that can be transmitted with average transmission time $\ell$ and average error probability $\epsilon$ in the point-to-point case. 
One sees from~\eqref{eq:SD_vlf_rate} that no square-root penalty occurs (\emph{zero dispersion}), which implies a fast convergence to the asymptotic limit. This fast convergence is demonstrated numerically in \cite{Polyanskiy2011} by means of nonasymptotic bounds.  

When fixed-length codes are used, it has been shown in \cite{Polyanskiy2011} and \cite{feedback_improve} that feedback does not improve the second-order term in the large-blocklength expansion of the maximum number of codewords  for a large class of channels with certain symmetry properties.
However, for some channels with nonunique capacity-achieving input distributions~\cite{feedback_improve} feedback results in a larger second-order term.

In this paper, we shall be concerned with the scenario in which the feedback channel is only used to stop transmissions (stop/decision feedback).
Following \cite{Polyanskiy2011}, we shall refer to variable-length coding schemes relying on stop feedback as variable-length stop-feedback (VLSF) codes. 
It was shown in \cite{Polyanskiy2011,Aslan2002,Aslan2006} that the error exponent
\begin{IEEEeqnarray}{rCl}
E(R) = \widetilde C - R\label{eq:vlsf_err_exp}
\end{IEEEeqnarray}
is achievable using VLSF codes. 
However, the tightest converse bound known is the full-feedback error exponent~\eqref{eq:err_exp_feedback}. 
Stop feedback is sufficient to achieve the zero-dispersion result in~\eqref{eq:SD_vlf_rate}.
However, also in this case,  the tightest nonasymptotic converse bound available for VLSF codes is the full-feedback converse reported in~\cite{Polyanskiy2011}.\footnote{An exception is the binary erasure channel, for which a nonasymptotic converse bound for the case of stop feedback that is tighter than the ones for full feedback is reported in~\cite{Devassy2016}.}

When only stop feedback is available, the zero-dispersion result~\eqref{eq:SD_vlf_rate} does not necessarily hold for the common-message discrete memoryless broadcast channels (CM-DMBC) considered in this paper. 
Specifically, we showed in~\cite{Trillingsgaard2015} that there exist CM-DMBCs for which the second term in the asymptotic expansion of the maximum coding rate achievable with VLSF codes is of order $1/\sqrt{\ell}$ (cf. \eqref{eq:SD_vlf_rate}). 
Our analysis in \cite{Trillingsgaard2015} is limited to the two-user case and relies on the restrictive assumption that there exists a unique input distribution $P^*$ that simultaneously maximizes $I(P,W_1)$ and $I(P,W_2)$. Furthermore, the upper and lower bounds on the maximum coding rate provided in \cite{Trillingsgaard2015} do not match up to the second order.
In this paper, we refine the results obtained in \cite{Trillingsgaard2015} and extend them to a broader class of common-message broadcast channels.

\subsubsection*{Contribution}
Focusing on VLSF codes, we obtain nonasymptotic achievability and converse bounds on the maximum number of codewords $\Msf(\ell,\epsilon)$ with average blocklength $\ell$ that can be transmitted on a CM-DMBC with reliability $1-\epsilon$. Here, the subscript ``sf'' stands for stop feedback.
By analyzing these bounds in the large-$\ell$ regime, we prove that when the $K$ component channels are independent (in the sense made precise in \eqref{eq:factorization}) and when the mutual information evaluated at the capacity-achieving input distribution equals $C$ for two or more component channels, then the asymptotic expansion of $\log\Msf(\ell,\epsilon)$ contains a square-root penalty, provided that some mild technical conditions are satisfied. 
Thus, we cannot expect the same fast convergence to capacity  as in the point-to-point case.
The intuition behind this result is as follows: in the point-to-point case, the stochastic overshoots of the information density that result in the square-root penalty can be virtually eliminated by using variable-length coding with stop-feedback.
Indeed, decoding is stopped after the information density exceeds a certain threshold, which yields only negligible stochastic variations. In the multiuser setup, however, the stochastic variations in the difference between the stopping times at the decoders make the square-root penalty reappear.
Note that our result does not necessarily imply that feedback is useless. It only shows that VLSF codes cannot be used to speed-up convergence to the same level as in the point-to-point case.
We also obtain upper and lower bounds on the second-order term in the asymptotic expansion of $\log \Msf(\ell,\epsilon)$ that generalize and tighten the ones reported in~\cite{Trillingsgaard2015}. The bounds turn out to match in certain special cases, e.g., when, in a two-user case, $P^*$ simultaneously maximizes  $I(P,W_1)$ and $I(P,W_2)$ (the case treated in \cite{Trillingsgaard2015}). Numerical evaluations of our nonasymptotic achievability and converse bounds reveal that the asymptotic expansion of $\log\Msf(\ell,\epsilon)$ obtained in this paper yields an accurate approximation for the maximum coding rate. 

We remark that many of the results in this paper first appeared in the conference paper \cite{Trillingsgaard2016}. The present paper includes proofs that were omitted in \cite{Trillingsgaard2016} as well as several intuitive remarks, discussion, and additional numerical results.

\subsubsection*{Notation}
We denote the $n$-dimensional all-zero vector and the $n$-dimensional all-one vector by $\vect{0}_n$ and $\vect{1}_n$, respectively. Vectors are denoted by boldface letters (e.g., $\vect{x}$), while their entries are denoted by roman letters (e.g., $x_i$).
The length of a vector is denoted by $\len{\cdot}$ and the Euclidean norm by $\vectornorm{\cdot}$. For a differentiable function $f(\cdot)$, we let $f'(\cdot)$ denote its derivative. Upper case, lower case, and calligraphic letters indicate random variables (RV), deterministic quantities, and sets, respectively. The cardinality of a set is denoted by $\lvert\cdot\rvert$ (e.g., $|\mathcal{A}|$). We let $x^n_m$ denote the tuple $(x_m,\cdots,x_n)$. For the channel outputs at decoder $k$, we let $y_{k,m}^n$ denote the tuple $(y_{k,m},\ldots,y_{k,n})$. When $m=1$, the subscript is sometimes omitted. We denote the set of probability distributions on $\mathcal{A}$ by $\mathcal{P}(\mathcal{A})$ and the support of a probability mass function $P$ by $\text{supp}(P)$. For a RV $X$ with probability distribution $P$, we let $P^n$ denote the joint probability distribution of the vector $[X_1,\cdots,X_n]$, where $\{X_i\}$ are independently and identically distributed (i.i.d.) according to $P$. The probability density function of a standard Gaussian RV  is denoted by  $\phi(\cdot)$. Furthermore, $\Phi(x)\triangleq 1- Q(x)$ is its cumulative distribution function, with $Q(\cdot)$ being the $Q$ function. We let $x^+$ denote $\max(0,x)$. Throughout the paper, $\log(\cdot)$ is the base $e$ logarithm and the index $k$ always belongs to the set $\mathcal{K}\triangleq \{1,\cdots,K\}$, although this is sometimes not explicitly mentioned. We use ``$\const$'' to denote a finite nonnegative constant. Its value may change at each occurrence. For two functions $f(\cdot)$ and $g(\cdot)$, the notation $f(x) = \mathcal{O}(g(x))$, as $x\rightarrow \infty$, means that  $\limsup_{x\rightarrow \infty} |f(x)/g(x)| < \infty$,  $f(x)=o(g(x))$, as $x\rightarrow \infty$, means that $\lim_{x\rightarrow \infty} |f(x)/g(x)| = 0$, and $f(x)  = \Theta(g(x))$, as $x\rightarrow \infty$, means that $c \leq  f(x)/g(x) \leq C$ for two positive constants $c$ and $C$, $c < C$, and for all sufficiently large $x$. Finally, $\mathbb{N}$ denotes the set of positive integers, $\mathbb{Z}_+ \triangleq \mathbb{N}\cup \{0\}$, the symbol $\mathbb{R}$ indicate the set of real numbers, and $\mathbb{R}_0^n$ denotes the set $\Big\{\vect{x} \in\mathbb{R}^n: \sum_{i=1}^n x_i = 0\Big\}$.

\section{System Model}
A CM-DMBC with $K$ decoders consists of a finite-cardinality input alphabet $\mathcal{X}$, and finite-cardinality output alphabets $\{\mathcal{Y}_k\}$, along with $K$ stochastic matrices $\{W_k\}$, where $W_k(y_k|x)$ denotes the probability that $y_k\in\mathcal{Y}_k$ is observed at decoder $k$ given the channel input $x\in\mathcal{X}$. We assume, without loss of generality, that $\mathcal{X} = \{1,\cdots,|\mathcal{X}|\}$. The outputs at time~$t$ are assumed to be conditionally independent given the input, i.e.,
\begin{equation}\label{eq:factorization}
  P_{Y_{1,t},\cdots,Y_{K,t}|X_t}(y_{1,t},\cdots,y_{K,t}|x_t) \triangleq \prod_k W_k(y_{k,t}|x_t).
\end{equation}
Let $P \times W_k: (x,y_k) \mapsto P(x)W_k(y_k|x)$ denote the joint probability distribution of input and output at decoder $k$. Finally, let $PW_k: y_k \mapsto \sum_{x\in\mathcal{X}} P(x)W_k(y_k | x)$ denote the induced marginal distribution on $\mathcal{Y}_k$.
For every $P\in\mathcal{P}(\mathcal{X})$ and $n\in\mathbb{N}$, the information density is defined as
\begin{align}
  i_{P, W_k}(x^n; y_k^n) \triangleq\sum_{i=1}^n \log \frac{W_k(y_{k,i}|x_i)}{PW_k(y_{k,i})}.\label{eq:inf_dens_def}
\end{align}
We let $I_k(P)\triangleq \EE{P\times W_k}{i_{P, W_k}(X; Y_k)}$ be the mutual information, 
\begin{IEEEeqnarray}{rCl}
V_k(P)\triangleq \EE{P}{\Vaa{P\times W_k}{i_{P, W_k}(X; Y_k)|X}}\label{eq:cond_inf_var}
\end{IEEEeqnarray}
be the conditional information variance, and 
\begin{IEEEeqnarray}{rCl}
T_k(P)\triangleq \EE{P\times W_k}{|i_{P, W_k}(X; Y_k) - I_k(P)|^3}
\end{IEEEeqnarray}
be the third absolute moment of the information density. Here, $\EE{P\times W_k}{\cdot}$ and $\Vaa{P\times W_k}{\cdot}$ denote the expectation and the variance, respectively, when the joint probability distribution of $(X,Y_k)$ is $P\times W_k$, and $\EE{P}{\cdot}$ denotes the expectation when the probability distribution on $X$ is $P$. 
The capacity of the CM-DMBC is given by \eqref{eq:capacity_def}, where the supremum is over all probability distributions $P\in \mathcal{P}(\mathcal{X})$.
We restrict ourselves to the case where the supremum in \eqref{eq:capacity_def} is attained by a unique probability distribution $P^*$.
 The corresponding (unique) capacity-achieving output distribution for decoder $k$ is denoted by $P_{Y_k}^*$. 
Furthermore, the individual capacities of each of the discrete memoryless component channels $\{W_k\}$ are denoted by
\begin{IEEEeqnarray}{rCl}
C_k \triangleq \sup_{P\in\mathcal{P}(\mathcal{X})} I_k(P).\label{eq:indi_capacity_def}
\end{IEEEeqnarray}
Finally, we let $V_{k} \triangleq V_k(P^*)$ and let $\diff I_k(\vect{v})$ denote the directional derivative of the mutual information $I_k(P)$ along the direction $\vect{v}\in\mathbb{R}_0^{|\mathcal{X}|}$ at the point $P^*$
\begin{IEEEeqnarray}{rCl} 
  \diff I_k(\vect{v}) \triangleq \sum_{x\in\mathcal{X}} v_x D(W_k(\cdot | x) || P_{Y_k}^* ).\label{eq:differential}
\end{IEEEeqnarray}
Here, $D(\cdot ||\cdot)$ denotes the Kullback-Leibler divergence.

In addition to \eqref{eq:factorization} and to the uniqueness of $P^*$, we shall also assume that the channel laws $\{W_k\}$ satisfy the following conditions:
\begin{enumerate}
  \item $I_k(P^*) = C$ for every $k\in\mathcal{K}$.
  \item $V_k(P^*) > 0$ for every $k\in\mathcal{K}$.
  \item $P^*(x)>0$ for all $x\in\mathcal{X}$.
\end{enumerate}
The first condition is not critical, and it is added only to simplify the statement of our results. Indeed, the nonasymptotic bounds we shall present in Theorem~\ref{thm:simple_achiev} and \ref{thm:converse_bound} also hold  for the case when $I_k(P^*)>C$ for some $k$. Furthermore, our positive dispersion result (Theorem~\ref{thm:asymp}) also holds  when the first condition is violated, provided that there exist at least two component channels $k_1$ and $k_2$ such that $I_{k_1}(P^*) = I_{k_2}(P^*) = C$. This is because the decoders whose component channels satisfy $I_k(P^*)>C$ feed back their stop signals much earlier than the remaining decoders and therefore do not contribute to the asymptotic expansion of $\log \Msf(\ell,\epsilon)$. If $I_k(P^*) = C$ for a single component channel, then zero dispersion can be attained. This may happen in certain practical scenarios, e.g., when all receivers are at different distances from the transmitter.

We are now ready to formally define a VLSF code for the CM-DMBC.
\begin{definition}
\label{def:VLSFcode}
An $(\ell, M,\epsilon)$-VLSF code for the CM-DMBC consists of:
\begin{enumerate}[leftmargin=*]
\item A RV $U\in\mathcal{U}$, with $|\mathcal{U}|\leq K+1$, which is known at both the encoder and the decoders.\footnote{Some remarks on the role of $U$ can be found after this definition.}
\item A sequence of encoders   $f_n: \mathcal{U}\times \mathcal{M} \mapsto \mathcal{X}$, each one mapping the message $J$, drawn uniformly at random from the set $\mathcal{M}\triangleq\{1,\ldots,M\}$, to the channel input $X_n = f_n(U,J)$.
\item Nonnegative integer-valued RVs $\tau_1,\cdots,\tau_K$ that are stopping times with respect to the filtrations (see \cite[p.~488]{billingsley}) $\mathcal{F}_{k,n} \triangleq \sigma\{U, Y_{k}^n\}$ and satisfy
\begin{IEEEeqnarray}{rCl}
  \Ebig{\max_k \tau_k}\leq \ell.\label{eq:avg_blocklength_const}
\end{IEEEeqnarray}
\item A sequence of decoders $g_{k,n}: \mathcal{U}\times \mathcal{Y}_k^n \mapsto \mathcal{M}$ satisfying
\begin{align}
  \prbig{J \not= g_{k,\tau_k}(U, Y_k^{\tau_k}) } \leq \epsilon, \qquad k\in\mathcal{K}.\label{eq:def_prob_error}
\end{align}
\end{enumerate}
\end{definition}
The maximum number of codewords with average length $\ell$ and error probability not exceeding $\epsilon$ is denoted by
 \begin{IEEEeqnarray}{rCl}
   \Msf(\ell, \epsilon) &\triangleq \maxx{M : \exists (\ell,M,\epsilon)\text{-VLSF code}}.
 \end{IEEEeqnarray}

 Some remarks on Definition~\ref{def:VLSFcode}  are in order.
 %
VLSF codes require a feedback link from each decoder to the encoder.
This feedback consists of a $1$-bit ``stop signal'' per decoder which is sent by decoder $k$ at time~$\tau_k$. The encoder continuously transmits until all decoders have fed back a stop signal. 
Hence, the blocklength is  $\max_k \tau_k$.  Note also that, differently from the full-feedback case, the encoder output at time $n$ depends on the message and on the common randomness $U$, but does not depend on the past output signals $\{Y_k^{n-1}\}$ and it is also independent of the stop signals received before time $n$. 
The stop signals are also only available at the encoder but not at the other decoders. 
This implies that $\tau_{k}$ and $g_{k,\tau_{k}}$ depend only on the common randomness $U$ and on the output sequence  $Y_{k}^{\tau_{k}}$, but does not depend on  $\{\tau_{  k'}\}_{  k' \not=  k}$.
Allowing the encoder output at time $n$ to depend on the previously received stop signal may yield to a faster convergence to capacity than what reported in this paper. 
From a practical perspective, however, this dependency complicates the design of the encoder and the decoders. 
Specifically, the encoder may need to use multiple codebooks depending on which stop signals are received and the decoders need to detect when the encoder switches between codebooks.

Note also that our definition of average blocklength~\eqref{eq:avg_blocklength_const} is inherently ``encoder-centric''.
An alternative, decoder-centric approach would be to require that  $\max_k\E{\tau_k} \leq \ell$. 
Under such an alternative definition, the zero-dispersion result from~\cite{Polyanskiy2011} continues to hold.

The RV $U$ serves as common randomness between the transmitter and all receivers, and enables the use of randomized codes~\cite{Lapidoth1998}. 
As for the proof of~\cite[Th.~3]{Polyanskiy2011}, randomized codes are necessary to prove our achievability bound. This is because we need to prove the existence of a code simultaneously satisfying \eqref{eq:avg_blocklength_const} and \eqref{eq:def_prob_error}. 
Note that the classic random-coding argument would allow us to  establish the existence of a deterministic VLSF-code (a VLSF-code with $|\mathcal{U}|=1$) satisfying only one of the constraints.
To establish the bound on the cardinality of $U$ provided in Definition~\ref{def:VLSFcode}, one can proceed as in \cite[Th.~19]{Polyanskiy2011} and use Caratheodory theorem to show that $|\mathcal{U}|\leq K+2$.
This bound can be further improved to $|\mathcal{U}|\leq K+1$ by using the Fenchel-Eggleston theorem \cite[p.~35]{Eggleston} in place of Caratheodory theorem. 


\section{Main Results}
\subsection{Nonasymptotic Achievability Bound}
We provide below a $K$-user generalization of the nonasymptotic achievability bound reported in \cite[Th.~1]{Trillingsgaard2015}.\footnote{Note that there is a typo in~\cite[Eq. (12)]{Trillingsgaard2015}: a maximization over $k$ is missing.}
\begin{theorem}
\label{thm:simple_achiev}
  Fix a probability distribution $P_{X^\infty}$ on $\mathcal{X}^\infty$. Let $\gamma\geq 0$ and $0 \leq \sprob \leq 1$  be arbitrary scalars.  Let the joint probability distribution of $(X^n, \bar X^n, Y_1^n,\cdots, Y_K^n)$ be
  \begin{multline}
    P_{X^n, \bar X^n,Y_1^n, \cdots,Y^n_K}(x^n, \bar x^n, y_1^n, \cdots,y_K^n)\\= P_{Y_1^n,\cdots, Y^n_K|X^n}(y_1^n, \cdots,y_K^n|x^n) P_{X^n}(x^n)P_{X^n}(\bar x^n)\label{eq:joint_probability_dist}
  \end{multline} 
  for all $n\in\mathbb{Z}_+$ and define the stopping times $\tau_k^{(0)}$ and $\bar \tau_k^{(0)}$, $k\in\mathcal{K}$, as follows:
  \begin{align}
  \tau_k^{(0)} &\triangleq \inff{n \geq 0: i_{P_{X^n}, W_k^n}(X^n ; Y_k^n)\geq \gamma}\label{eq:tau_k_def}\\
  \bar \tau_k^{(0)} &\triangleq \inff{n \geq 0: i_{P_{X^n}, W_k^n}(\bar X^n ; Y_k^n)\geq \gamma}.
  \end{align}
For every $M$, there exists an $(\ell, M, \epsilon)$-VLSF code such that
\begin{align}
    \ell &\leq (1-\sprob) \EBig{\max_k \tau_k^{0)}}\label{eq:achiev_Emax}\\
    \epsilon &\leq  \max_k\mathopen\Big\{\sprob + (1-\sprob)(M-1)\pr{\tau_k^{(0)}\geq \bar\tau_k^{(0)}}\Big\}\label{eq:achiev_Proberror1}\\
    &\leq \sprob + (1-\sprob)(M-1)\exp\left\{-\gamma\right\}.\label{eq:achiev_Proberror2}
\end{align}
\end{theorem}

\begin{IEEEproof}
The proof of Theorem~\ref{thm:simple_achiev} follows closely the proof of~\cite[Th. 3]{Polyanskiy2011}. See Appendix~\ref{app:achievability_proof} for details.
\end{IEEEproof}
 If the constant $q$ is set to $0$, then we obtain a straightforward generalization of \cite[Th. 3]{Polyanskiy2011}. The constant $q$ in Theorem~\ref{thm:simple_achiev} is used to enable time-sharing. With probability $q$, the decoders simultaneously send stop signals to the encoder at time $0$. The common randomness $U$ can be used to enable this weak form of cooperation among the decoders.

\subsection{Nonasymptotic Converse Bound}
\label{sec:main_converse}
Let $\preffree{Y}_k$ denote all possible sequences (of arbitrary length) of symbols from $\mathcal{Y}_k$, i.e., $\preffree{Y}_k\triangleq \{[\ ] \}\cup \bigcup_{n=1}^\infty \mathcal{Y}_k^n$, where $[\ ]$ denotes the vector of length $0$.
A subset $\preffree{\overline Y}_k$ of $\preffree{Y}_k$ is called complete prefix-free if and only if, for every $\mathbf{y} \in\mathcal{Y}_k^\infty$, there exists a unique $\mathbf{\bar y}\in\preffree{\overline Y}_k$ such that $\mathbf{\bar y}$ is a prefix to $\mathbf{y}$, i.e., $\mathbf{\bar y} = [y_1,\cdots,y_{\len{\vect{\bar y}}}]$. The role of the complete prefix-free subsets of $\preffree{Y}_k$ is to provide an equivalent representation of the stopping time $\tau_k$. Indeed, given a stopping time $\tau_k$, there exists a complete prefix-free subset $\preffree{\overline Y}_k^{(u)}$ of $\preffree{Y}_k$ for each $u\in\mathcal{U}$ such that $Y_k^{\tau_k} \in\preffree{\overline Y}_k^{(U)}$. Conversely, every set of complete prefix-free subsets $\{\preffree{\overline Y}_k^{(u)}\}_{u\in\mathcal{U}}$ also defines a stopping time $\tau_k = \min\{n\in\mathbb{Z}_+: Y_k^n \in \preffree{\overline Y}_k^{(U)}\}$, i.e., $\tau_k$ is a RV that depends only on the realizations of $U$ and of $Y_k^\infty$.
Let $Q_k^{(\infty)}$ be an arbitrary probability measure on $\preffree{Y}_k$ and define the mapping $Q_k: \preffree{Y}_k \mapsto [0,1]$ as follows:
\begin{IEEEeqnarray}{rCl}
  Q_{k}(\mathbf{\bar y}) \triangleq \sum_{\substack{\mathbf{y}\in\mathcal{Y}_k^\infty:\\ [y_1, \cdots,y_{\len{\bar y}}] = \mathbf{\bar y}}}Q_k^{(\infty)}(\y), \qquad \vect{\bar y}\in\preffree{Y}_k. \label{eq:Qk_def1}
\end{IEEEeqnarray}
We shall use the convention that $[y_1, \cdots,y_{\len{\vect{\bar y}}}]=[\ ]$ when $\len{\vect{\bar y}} = 0$.
For every complete prefix-free subset $\preffree{\overline Y}_k\subset\preffree{Y}_k$, we observe that $Q_k(\cdot)$ defines a probability measure on $\preffree{\overline Y}_k$. Indeed,
\begin{IEEEeqnarray}{rCl}
1 &=& \sum_{\mathbf{y}\in\mathcal{Y}_k^\infty} Q_k^{(\infty)}(\mathbf{y}) \\
&=& \sum_{\mathbf{\bar y} \in\preffree{\overline Y}_k} \sum_{\substack{\mathbf{y}\in\mathcal{Y}_k^\infty:\\ [y_1, \cdots,y_{\len{\vect{\bar y}}}]= \mathbf{\bar y }}} Q_k^{(\infty)}(\mathbf{y}) \\
&=&  \sum_{\mathbf{\bar y} \in\preffree{\overline Y}_k} Q_{k}(\mathbf{\bar y}).
\end{IEEEeqnarray}
 Based on $Q_k(\cdot)$, we define the \emph{log-likelihood ratio}
\begin{IEEEeqnarray}{rCl}
  i_k(x^n; y_k^n) \triangleq  \log \frac{P_{Y_k^n | X^n}(y_k^n|x^n)}{Q_k(y_k^n)}\label{eq:mismatched_inf_def}
\end{IEEEeqnarray}
for $x^n\in\mathcal{X}^n$, $y_k^n \in\mathcal{Y}^n$, and $n\in\mathbb{N}$ with the convention that $i_k([\ ]; [\ ]) = 0$.

To prove our nonasymptotic converse bound, we shall make use of the following lemma, which provides an information spectrum-type converse for VLSF codes. 
The proof of this lemma relies on a non-standard application of the meta-converse theorem \cite[Th.~26]{Polyanskiy2010b}. 
Specifically, the meta-converse is applied to a general channel whose channel inputs are infinite-dimensional vectors and whose channel outputs are variable-length vectors belonging to a complete prefix-free subset of $\preffree{Y}_k$. 
\begin{lemma}\label{lem:vlsf_metaconverse}
Fix arbitrary probability measures $Q_k^{(\infty)}$ on $\preffree{Y}_k$, an arbitrary constant $\eta>0$, and an $(\ell,M,\epsilon)$-VLSF code whose encoders  induce a conditional probability distribution $P_{\vect{X}}^{(u)}$ on $\mathcal{X}^\infty$ given $U=u$ and whose stopping times are equivalently defined by the complete prefix-free subsets $\{\preffree{\overline Y}^{(u)}_k\}_{u\in\mathcal{U}}$ of the set $\preffree{Y}_k$. 
There exist positive constants $\varepsilon_{k}^{(u)}$, 
defined for all  $u\in\mathcal{U}$, and 
satisfying $\EE{U}{\varepsilon_{k}^{(U)}} \leq \epsilon+\eta$, such that
\begin{IEEEeqnarray}{rCl}
\mathds{P}^{(u)}\mathopen{}\Big[i_k(\X ; \vect{\overline Y}_k) < \log(\eta M) \Big] &\leq& \varepsilon_{k}^{(u)}  \label{eq:conve_const_before_tildeinf_eval2}
\end{IEEEeqnarray}
for every $\bar \x\in \text{supp}(P^{(u)}_\mathbf{X})$ and $k\in\mathcal{K}$. 
The probability measure on $\mathcal{X}^\infty \times \preffree{\overline Y}_1^{(u)}\times \cdots\times \preffree{\overline Y}_K^{(u)}$ required to evaluate~\eqref{eq:conve_const_before_tildeinf_eval2} is
\begin{multline}
\mathds{P}^{(u)}_{\X, \vect{\overline Y}_1,\cdots, \vect{\overline Y}_K}(\x, \vect{\bar y}_1,\cdots,\vect{\bar y}_K) \\\triangleq  P_{\X}^{(u)}(\x) \prod_{k=1}^K \prod_{i=1}^{\len{\vect{\bar y}_k}}W_k(\bar y_{k,i} |x_i).\label{eq:Pk_def_eval}
\end{multline}
Here, we use the convention that $\prod_{i=1}^0 W_k(\bar y_{k,i}|x_i) = 1$. 
\end{lemma}
\begin{IEEEproof}
See Appendix~\ref{proof:vlsf_metaconverse}.
\end{IEEEproof}
We are now ready to state and prove our converse bound, which provides us with a lower bound on the average blocklength given $M$, an arbitrary probability measure $Q_k^{(\infty)}$, and an arbitrary positive constant $\eta$.
\begin{theorem}
\label{thm:converse_bound}
For arbitrary probability measures $Q_k^{(\infty)}$ on $\preffree{Y}_k$, and arbitrary $M\in\mathbb{N}$, $t\in \mathbb{Z}_+$, $\eta>0$, and $\varepsilon_k\in(0,1)$, $k\in\mathcal{K}$, define the following function:\footnote{As clarified in the proof of the theorem, for a fixed $U=u$ and a set of error probabilities $\varepsilon_k^{(u)}$ for the decoders, the function $L_t\farg{\varepsilon_1^{(u)},\cdots,\varepsilon_K^{(u)}}$ is an upper bound on the conditional probability that $\max_k \tau_k\leq t$ given $U=u$.}
\begin{IEEEeqnarray}{rCl}
  \IEEEeqnarraymulticol{3}{l}{L_t(\varepsilon_1,\cdots,\varepsilon_K)}\nonumber\\
  \quad&\triangleq& \max_{x^t\in\mathcal{X}^t} \prod_{k} \minnbigg{1,\nonumber\\
  &&\quad\pr{\max_{0\leq n \leq t}i_k(x^n; Y_k^n)\geq \log M + \log \eta}+  \varepsilon_k }.\label{eq:Lt_def}\IEEEeqnarraynumspace
\end{IEEEeqnarray}
Here, the vector $x^n$ contains the first $n$ entries of $x^t$, and $Y_k^t\sim P_{Y_k^t|X^t=x^t}$.
Then, every $(\ell,M,\epsilon)$-VLSF code must satisfy
\begin{equation}
  \ell \geq \!\! \min_{ \substack{P_U \in \mathcal{P}(\mathcal{U)}, \varepsilon_k^{(u)}\in [0,1]: \\  \EE{U}{\varepsilon_k^{(U)}} \leq \epsilon+ \eta\\ }  } \!\!\EE{U}{\sum_{t=0}^{\infty}\!\left(1- L_t\farg{\varepsilon_1^{(U)},\cdots,\varepsilon_K^{(U)}}  \right)}.\label{eq:conve_l_lower_bound}
\end{equation}
\end{theorem}
\begin{IEEEproof}
To establish Theorem~\ref{thm:converse_bound}, we derive a lower bound on the average blocklength $\ell$ that holds for all VLSF codes having $M$ codewords and error probability no larger than $\epsilon$.
Fix an arbitrary $(\ell,M,\epsilon)$-VLSF code.  
 It follows from~\eqref{eq:avg_blocklength_const} and from the conditional independence of the stopping times $\{\tau_k\}$ given $U$ and $\X$ that
\begin{IEEEeqnarray}{rCl}
  \ell &\geq& \EBig{\EBig{\max_k \tau_k|U, \X}}\\
  &=& \E{ \sum_{t=0}^\infty \Big(1- \prBig{\max_k\tau_k \leq t | U,\vect{X}} \Big)}\\
  &=&  \E{\sum_{t=0}^\infty \left(1 - \prod_k \mathds{P}^{(U)}\mathopen{}\Big[\len{\mathbf{\overline Y}_k}\leq t|  \X\Big]\right)}.\label{eq:conve_Etau1_tau2_in_terms_of_pr_tau1_tau_2_eval}
\end{IEEEeqnarray}
Here, we have used that $\tau_k = \len{\mathbf{\overline Y}_k}$.
Hence, we can lower-bound $\ell$ by upper-bounding $\mathds{P}^{(u)}\mathopen{}\Big[\len{\mathbf{\overline Y}_k}\leq t|\X=\x\Big]$ for every $t\in\mathbb{Z}_+$. 

Now, set 
\begin{IEEEeqnarray}{rCl}
  \varepsilon_{k}^{(u)}(\bar \x) &\triangleq& \mathds{P}^{(u)}\mathopen{}\Big[i_k(\X ; \vect{\overline Y}_k) < \lambda \Big| \X=\bar \x\Big].\label{eq:i_ineq10}
\end{IEEEeqnarray}
Then, it follows by Lemma~\ref{lem:vlsf_metaconverse} that we must have
\begin{IEEEeqnarray}{rCl}
  \EE{\vect{X},U}{\varepsilon_{k}^{(U)}(\bar \X)} \leq \epsilon + \eta.\label{eq:i_ineq}
\end{IEEEeqnarray}
From an intuitive perspective, \eqref{eq:i_ineq} serves as a constraint on the stopping times. Namely, given $M$, $k$, and $\{\varepsilon_k^{(u)}(\vect{\bar x})\}$, the  information density must exceed a threshold  with probability larger than or equal $1-\varepsilon_k^{(u)}(\vect{\bar x})$ when the stop signals are sent. Note also that \eqref{eq:i_ineq} depends on the choice of $Q^{(\infty)}_k$ through the information density and on $\{\tau_k\}$ through $\{\vect{\overline Y}_k\}$.

Next, we upper-bound $\mathds{P}^{(U)}\mathopen{}\Big[\len{\mathbf{\overline Y}_k}\leq t|  \X=\vect{\bar x}\Big]$ for every $u\in\mathcal{U}$ and $\bar \x \in \text{supp}(P_{\mathbf{X}}^{(u)})$.
Since the stopping times $\{\tau_k\}$ are conditionally independent given $U=u$ and $\X=\bar \x$,  we have the steps \eqref{eq:conve_maxtau1_tau2_u0}--\eqref{eq:conve_maxtau1_tau2_u_Lt}, shown in the top of the next page.\begin{figure*}[!t]
\normalsize
\setcounter{MYtempeqncnt}{\value{equation}}
\setcounter{equation}{35}
\begin{IEEEeqnarray}{rCl}
\IEEEeqnarraymulticol{3}{l}{\prod_{k} \mathds{P}^{(u)}\mathopen{}\left[\len{\vect{\bar Y}_k}\leq t \Big| \X=\bar \x\right] } \nonumber\\
  \quad &=& \prod_{k} \bigg(\mathds{P}^{(u)}\mathopen{}\bigg[\max_{0\leq n \leq \len{\vect{\bar Y}_k}} i_{k}(\X;\bar Y_k^{n})\geq \lambda,\len{\vect{\bar Y}_k} \leq t\bigg|\X=\bar \x\bigg] \nonumber\\
 && \qquad\qquad\qquad\qquad\qquad\qquad\qquad\qquad\qquad{}+ \mathds{P}^{(u)}\mathopen{}\bigg[ \max_{0\leq n \leq\len{\vect{\bar Y}_k}} i_{k}(\X;\bar Y_k^{n})< \lambda,\len{\vect{\bar Y}_k} \leq t\bigg|\X=\bar \x\bigg] \bigg)\label{eq:conve_maxtau1_tau2_u0}\\
   &\leq& \prod_{k} \min\mathopen{}\bigg\{1,\mathds{P}^{(u)}\mathopen{}\bigg[\max_{0\leq n \leq \minn{t, \len{\vect{\bar Y}_k} }} i_{k}(\X;\bar Y_k^{n})\geq \lambda\bigg|\X=\bar \x\bigg]  + \mathds{P}^{(u)}\mathopen{}\bigg[ \max_{0\leq n \leq \len{\vect{\bar Y}_k}} i_{k}(\X;\bar Y_k^{n})< \lambda\bigg|\X=\bar \x\bigg] \bigg\}\label{eq:PrMax_tau_k_t}\\
    &\leq& \prod_{k}  \min\mathopen{}\bigg\{1,\mathds{P}^{(u)}\mathopen{}\bigg[\max_{0\leq n \leq \minn{t, \len{\vect{\bar Y}_k} }} i_{k}(\X;\bar Y_k^{n})\geq \lambda\bigg|\X=\bar \x\bigg] +\mathds{P}^{(u)}\mathopen{}\bigg[i_k(\X ; \vect{\bar Y}_k) < \lambda \bigg| \X=\bar \x\bigg] \bigg\}\label{eq:conve_maxtau1_tau2_u1}\\
     &=& \prod_{k} \min\mathopen{}\bigg\{1,\mathds{P}^{(u)}\mathopen{}\bigg[\max_{0\leq n \leq \minn{t, \len{\vect{\bar Y}_k}} } i_{k}(\X;\bar Y_k^n)\geq \lambda\bigg|\X=\bar \x\bigg]+\varepsilon_k^{(u)}(\bar \x) \bigg\}\label{eq:PrMax_tau_k_t2}\\
    &\leq& \max_{x^t \in \mathcal{X}^t}\prod_{k} \minn{1,\pr{\max_{0\leq n \leq t}i_{k}( x^n;Y_k^n) \geq \lambda} +\varepsilon_{k}^{(u)}(\bar \x)}\label{eq:conve_maxtau1_tau2_u}\\
    &=& L_t(\varepsilon_{1}^{(u)}(\bar \x),\cdots, \varepsilon_{K}^{(u)}(\bar \x)).\label{eq:conve_maxtau1_tau2_u_Lt}
\end{IEEEeqnarray}

\setcounter{equation}{\value{MYtempeqncnt}}
\hrulefill
\vspace*{4pt}
\end{figure*}\addtocounter{equation}{6} 
Here, \eqref{eq:PrMax_tau_k_t2} follows from \eqref{eq:i_ineq}; in \eqref{eq:conve_maxtau1_tau2_u}, we let $Y^n_k$ be distributed according to $P_{Y_k^n|X^n=x^n}$; finally, \eqref{eq:conve_maxtau1_tau2_u_Lt} follows from \eqref{eq:Lt_def}. Note that the probability term in \eqref{eq:conve_maxtau1_tau2_u} does not depend on the code.

 Roughly speaking, the steps \eqref{eq:conve_maxtau1_tau2_u0}--\eqref{eq:conve_maxtau1_tau2_u}  dispose of the dependency on $\{\tau_k\}$. 
 The intuition behind these steps is as follows: First, define the auxiliary stopping times
\begin{IEEEeqnarray}{rCl}
  \IEEEeqnarraymulticol{3}{l}{\hat \tau_k\triangleq}\nonumber\\
   && \left\{\begin{array}{ll} 
  \min\{n: \imath_k(\X; Y^n_k) >\lambda\} & \text{if } \max_{0\leq n\leq t}\imath_k(\X; Y^n_k) >\lambda\\
  t &  \text{if } \max_{0\leq n\leq t} \imath_k(\X; Y^n_k) \leq \lambda\\
  &  \quad\text{or } A_k^{(U)}(\vect{\overline x}) = 1 \\
  \infty & \text{if } \max_{0\leq n\leq t} \imath_k(\X; Y^n_k) \leq \lambda\\
  &  \quad\text{and } A_k^{(U)}(\vect{\overline x}) = 0
  \end{array}
    \right.\label{eq:opt_stopping_time_cond}
\end{IEEEeqnarray}
where $\{A_k^{(u)}(\vect{\overline x})\}$ are independent Bernoulli distributed RVs with parameters $\max_{0\leq n\leq t}\mathds{P}^{(u)}\mathopen{}\Big[\imath_k(\X; Y^n_k) < \lambda|\X=\vect{\overline x}\Big]/\varepsilon_k^{(u)}(\vect{\bar x})$. The stopping time in \eqref{eq:opt_stopping_time_cond} roughly states that if the information density of decoder $k$ exceeds $\lambda$ before time $t$, decoder $k$ should send a stop signal when this happens. If this does not happen, the decoder should choose randomly between sending a stop signal at time $t$ or letting $\hat \tau_k = \infty$. The key observation is that by replacing the stopping times $\tau_k$ by $\hat \tau_k$ in  \eqref{eq:conve_maxtau1_tau2_u0}, the probability in~\eqref{eq:conve_maxtau1_tau2_u0} equals \eqref{eq:conve_maxtau1_tau2_u}.  We note that $\tau_k$ cannot be chosen equal to $\hat \tau_k$ in an achievability scheme because $\tau_k$ is only defined with respect to the filtration $\{\sigma(U,Y^n_k)\}_n$, whereas $\hat \tau_k$ is defined with respect to the larger  filtration $\{\sigma(U,X^n,Y^n_k)\}_n$. 
This is, however, not issue in the converse argument. Note also that the auxiliary stopping times $\{\hat \tau_k\}$ are different for each $t$.

 Next, by substituting \eqref{eq:conve_maxtau1_tau2_u_Lt} in \eqref{eq:conve_Etau1_tau2_in_terms_of_pr_tau1_tau_2_eval}, we conclude that
\begin{multline}
  \EBig{\max_k \tau_k | U=u, \X=\bar\x}\\ \geq \sum_{t=0}^\infty \left(1-L_t(\varepsilon_{1}^{(u)}(\bar \x),\cdots,\varepsilon_{K}^{(u)}(\bar \x))\right).
\end{multline}
Hence, $\E{\max_k \tau_k}$ can be lower-bounded as follows:
\begin{IEEEeqnarray}{rCl}
 \EBig{\max_k \tau_k}&\geq & \min_{\substack{P_{U,\X}\in \mathcal{P}( \mathcal{U}\times \mathcal{X}^\infty ), \varepsilon_{k}^{(u)}(\x)\in[0,1]:\\ \E{\varepsilon_{k}^{(U)}(\X)} \leq \epsilon +\eta}}\nonumber\\
 &&{}\E{\sum_{t=0}^\infty \left(1-L_t(\varepsilon_{1}^{(U)}(\X),\cdots,\varepsilon_{K}^{(U)}(\X))\right)}.\IEEEeqnarraynumspace\label{eq:Emaxtau_exp}
 \end{IEEEeqnarray}
 The right-hand side of \eqref{eq:Emaxtau_exp} depends on the code only through the random quantities $\varepsilon_k^{(U)}(\vect{X})$. Now, by defining $\overline U = (U,\vect{X})$ and $\mathcal{\overline U}=\mathcal{U}\times \mathcal{X}^\infty$, we obtain
 \begin{multline}
 \EBig{\max_k \tau_k}  \\\quad\geq  \min_{\substack{P_{\overline U}\in \mathcal{P}( \mathcal{\overline U}), \varepsilon_{k}^{(\overline u)}\in[0,1]:\\ \E{\varepsilon_{k}^{(\overline U)}} \leq \epsilon +\eta}} \E{\sum_{t=0}^\infty \left(1-L_t(\varepsilon_{1}^{(\overline U)},\cdots,\varepsilon_{K}^{(\overline U)})\right)}.\label{eq:conve_final_bound1}
 \end{multline}
This concludes the proof provided that one shows that the cardinality of $\mathcal{\overline U}$ in \eqref{eq:conve_final_bound1} can be upper-bounded by $K+1$. 
This cardinality bound, which can be established by an  application of Caratheodory's theorem, is provided in Appendix~\ref{sec:converse_bound}
\end{IEEEproof}
We remark that the converse bound in Theorem~\ref{thm:converse_bound} also provides a new converse for the single-decoder setup when $K=1$. However, when evaluated numerically, it turns out that this bound is less tight than the converse bounds for full feedback provided in \cite{Polyanskiy2011}.

As we shall see next, choosing $Q^{(\infty)}_k$ as a simple product distribution  yields a computable and tight nonasymptotic bound for symmetric channels.\footnote{A channel is symmetric if the rows and columns of the stochastic channel matrix are permutations of each other \cite[p.~189]{Cover2012}.}
For general CM-DMBCs, choosing $Q^{(\infty)}_k$ as a convex combination of product distributions (cf., \eqref{eq:Qinf_def}) appears necessary to obtain tight large-$\ell$ asymptotic expansions.

If $\{W_k\}$ are identical and symmetric, we have the following particularization of Theorem~\ref{thm:converse_bound}.
\begin{corollary}
\label{thm:converse_bound2}
For arbitrary $M\in\mathbb{N}$, $t\in \mathbb{Z}_+$, $\eta>0$, and an arbitrary sequence $\x \in\mathcal{X}^\infty$, let
\begin{IEEEeqnarray}{rCl}
  v_t = \prBig{\max_{0\leq n \leq t}i_{P^*, W_1}(x^n; Y_1^n)\geq \log M + \log \eta}
\end{IEEEeqnarray}
where $Y_1^t\sim P_{Y_1^t|X^t=x^t}$ and $i_{P^*, W_1}(\cdot; \cdot)$ is defined in \eqref{eq:inf_dens_def}. When $W_1=\cdots =W_K$ and $W_1$ is symmetric,  every $(\ell,M,\epsilon)$-VLSF code must satisfy
\begin{align}
  \ell \geq \min_{ \substack{P_U \in\mathcal{P}(\mathcal{U}), \varepsilon^{(u)}\in [0,1]: \\  \EE{U}{\varepsilon^{(U)}} \leq \epsilon+ \eta\\ }  } \EE{U}{\sum_{t=0}^{\infty}\left(1- \minn{1,v_t + \varepsilon^{(U)} }^K \right)}.\label{eq:conve_l_lower_bound2}
\end{align}
\end{corollary}
\begin{IEEEproof}
  We apply the converse bound in Theorem~\ref{thm:converse_bound} with $Q_1^{(\infty)} =\cdots = Q_K^{(\infty)}$. Furthermore, we choose $Q_1^{(\infty)}$ as a product distribution with marginal $Q_1(y) = 1/|\mathcal{Y}_1|$ for all $y\in\mathcal{Y}_1$. By \eqref{eq:Qk_def1}, we have that
$Q_k(\vect{\overline y}_k) = |\mathcal{Y}_k|^{-\len{\vect{\overline y}_k}} $.
Since the capacity-achieving output distribution of a symmetric channel is uniform \cite[Eq.~(7.22)]{Cover2012}, we conclude that
\begin{IEEEeqnarray}{rCl}
i_k(x^t;Y_k^t) \sim i_{P^*,W_1}(x^t,Y_1^t)
\end{IEEEeqnarray}
for all $k\in\mathcal{K}$ given that $X^t=x^t$. One can verify that, when the channel is symmetric, the conditional probability distribution of the information density $i_{P^*,W_1}(x^t, Y_1^t)$ given $X^t=x^t$ does not depend on $x^t$. This allows us to drop the maximization over $x^t$ in \eqref{eq:Lt_def}. Finally, to express the minimization problem \eqref{eq:conve_l_lower_bound} in the form given in \eqref{eq:conve_l_lower_bound2}, we note that, for every $[\varepsilon_1,\cdots,\varepsilon_K]\in[0,1]^K$, we have
\begin{IEEEeqnarray}{rCl}
\IEEEeqnarraymulticol{3}{l}{  \left(\prod_{k=1}^K \minn{1,v_t + \varepsilon_k} \right)^{1/K} }\nonumber\\
\qquad &\leq& \frac{1}{K} \sum_{k=1}^K \minn{1,v_t + \varepsilon_k}\label{eq:geometric_aritmetic_mean}\\
  &\leq& \minn{1,v_t + \frac{1}{K} \sum_{k=1}^K \varepsilon_k}.\label{eq:jensen}
\end{IEEEeqnarray}
Here, \eqref{eq:geometric_aritmetic_mean} follows because the geometric mean is no larger than the arithmetic mean and \eqref{eq:jensen} follows from Jensen's inequality.
\end{IEEEproof}

\subsection{Asymptotic Expansion}


Analyzing~\eqref{eq:achiev_Emax}, \eqref{eq:achiev_Proberror2} and~\eqref{eq:conve_l_lower_bound} in the  limit $\ell \to\infty$, we obtain the following asymptotic characterization of $\log\Msf(\ell,\epsilon)$.

\begin{theorem}
\label{thm:asymp}
Let $V \triangleq (\prod_k V_k)^{1/K}$ and $\varrho_k \triangleq \sqrt{V_k/V}$ and assume, without loss of generality, that $C_1  \geq \cdots \geq C_K$.
For every CM-DMBC satisfying
\begin{IEEEeqnarray}{rCl}
  \frac{1}{C_i}+\frac{i}{C_K}  &> & \frac{i}{C}, \qquad i\in\{1,\cdots,K-1\}\label{eq:thm_asymp_cond}
\end{IEEEeqnarray}
and for every $\epsilon\in(0,1)$, we have\footnote{The subscripts ``a'' and ``c'' in $\Xi_{\text{a}}$ and $\Xi_{\text{c}}$ stand for achievability and converse, respectively.}
\begin{multline}
   \frac{C \ell}{1-\epsilon}-   \sqrt{\frac{V\ell}{1-\epsilon}}\Xi_{\text{a}}+o(\sqrt{\ell}) \\\leq \log \Msf(\ell,\epsilon)   \leq \frac{C \ell}{1-\epsilon}-\sqrt{\frac{V \ell}{1-\epsilon} } \Xi_{\text{c}} +o(\sqrt{\ell}). \label{eq:asymp_expansion}\IEEEeqnarraynumspace
\end{multline}
Here,
\begin{align}\label{eq:constant_LHS_as}
  \Xi_{\text{a}} \triangleq \min_{\substack{\vect{v} \in\mathbb{R}^{|\mathcal{X}|}_0}} \EBig{ \max_k \diff I_k(\vect{v})  +\varrho_k Z_k }
\end{align}
and
\begin{IEEEeqnarray}{rCl}\label{eq:constant_RHS_as}
  \Xi_{\text{c}} \triangleq \EBig{ \max_k   \tvRV_k  }\label{eq:Xi_c_def}
\end{IEEEeqnarray}
where $Z_k\stackrel{\text{i.i.d.}}{\sim}\mathcal{N}(0,1)$ and $\{\tvRV_k\}$ are independent RVs with cumulative distribution functions
\begin{align}
  F_{\tvRV_k}(w) \triangleq \Phi\farg{ \frac{w+ \diff I_k(\optx(w))}{\varrho_{ k}}}.\label{eq:Wk_def}
\end{align}
The function $\optx(\cdot)$ is defined as follows:\footnote{If the maximizer in \eqref{eq:hatx_def} is not unique, $\optx(w)$ is chosen arbitrarily from the set of maximizers.}
\begin{IEEEeqnarray}{rCl}
\optx(w) & \triangleq& \argmax_{\vect{v}\in\mathbb{R}_0^{|\mathcal{X}|}  } \prod_k \Phi\farg{\frac{w +  \diff I_k(\vect{v}) }{\varrho_{ k}} }.\label{eq:hatx_def}
\end{IEEEeqnarray}
\end{theorem}
\begin{IEEEproof}
  The converse bound in \eqref{eq:asymp_expansion} is proved in Appendix~\ref{sec:converse_asymptotics} and the achievability bound in \eqref{eq:asymp_expansion} is proved in Appendix~\ref{sec:achievability_asymptotics}.
  We next provide a heuristic argument that sheds light on the achievability part. 
  The key step  is to obtain a tight upper bound  on $\E{\max_k \tau_k^{(0)}}$ in \eqref{eq:achiev_Emax}. 
  The desired asymptotic expansion then follows by properly choosing $\gamma$ and $q$ in Theorem~\ref{thm:simple_achiev} (see Appendix~\ref{sec:achievability_asymptotics} for details). 
  For the purpose of this heuristic argument, consider the special case $C_1 = \ldots = C_K = C$, and  $V_1 = \ldots = V_K$. Also let us assume that, when $P_{X^\infty} = (P^*)^\infty$, the information densities can be well-approximated by the  Brownian motions with drift
  \begin{IEEEeqnarray}{rCl}
    \imath_k(X^n; Y^n) &\approx& n C + \sqrt{V_1} B(n)
  \end{IEEEeqnarray}
  where $B(n)$ is a standard Brownian motion. 
  It now follows from the Bachelier-Levy formula (see \cite{Lerche} or \cite{BL_formula}) that the probability density function of the first passage time $\inf\{t\in\mathbb{R}_+: t C + \sqrt{V_1}B(t) \geq \gamma \}$ is given by
  \begin{IEEEeqnarray}{rCl}
    \frac{\gamma}{\sqrt{V_1} t^{3/2}} \phi\farg{\frac{\gamma - t C }{\sqrt{t V_1}}}
  \end{IEEEeqnarray}
  where $\phi(\cdot)$ is the probability density function for the standard Gaussian RV.
  This shows that 
  \begin{IEEEeqnarray}{rCl}
    \frac{\tau_k^{(0)} - \gamma/C}{\sqrt{\gamma V_1/C^3}} \stackrel{\text{d}}{\rightarrow} \mathcal{N}(0,1)\label{eq:tau_asymp_norm}
  \end{IEEEeqnarray}
  as $\gamma\rightarrow \infty$ and, as a consequence, we have that
  \begin{IEEEeqnarray}{rCl}
    \E{\max_k\frac{\tau_k^{(0)} - \gamma/C}{\sqrt{\gamma V_1/C^3}}} \rightarrow \EBig{\max_k Z_k}.\label{eq:heuristic_Emax1}
  \end{IEEEeqnarray}
  Rewriting \eqref{eq:heuristic_Emax1}, we obtain
  \begin{IEEEeqnarray}{rCl}
    \EBig{\max_k\tau_k^{(0)}} = \frac{\gamma}{C} + \sqrt{\frac{\gamma V_1}{C^3}} \EBig{\max_k Z_k} + o(\sqrt{\gamma}).\label{eq:Emaxtau0_intuition}
  \end{IEEEeqnarray}
  This result is a particularization of Lemma~\ref{lem:time_varying} in Appendix~\ref{sec:achievability_asymptotics} for the case where $C_1=\ldots=C_K = C$ and where $V_1 = \ldots = V_K$. 
  Next, let $\delta$ be an arbitrary positive constant. By choosing $\gamma = \frac{\ell C}{1-\epsilon} -(1+\delta) \sqrt{\frac{V_1\ell}{1-\epsilon}} \EBig{\max_k Z_k}$ and by setting $q = \epsilon - \Theta(1/\ell)$, we observe from \eqref{eq:Emaxtau0_intuition} that $(1-q)\EBig{\max_k \tau_k^{(0)}} \leq \ell$ for all sufficiently large $\ell$. As a result, Theorem~\ref{thm:simple_achiev} implies the following asymptotic expansion of $\log \Msf(\ell,\epsilon)$
  \begin{IEEEeqnarray}{rCl}
    \IEEEeqnarraymulticol{3}{l}{\log \Msf(\ell,\epsilon)}\nonumber\\
    \qquad &\geq& \gamma+ \log \frac{\epsilon - q}{1-q}\\
    &\geq& \frac{\ell C}{1-\epsilon} -(1+\delta)\sqrt{\frac{V_1\ell}{1-\epsilon}} \EBig{\max_k Z_k} + o(\sqrt{\ell})\IEEEeqnarraynumspace
  \end{IEEEeqnarray}
  This argument is made rigorous and further generalized in Appendix~\ref{sec:achievability_asymptotics}.
\end{IEEEproof}

Some remarks are in order.
The condition \eqref{eq:thm_asymp_cond} is needed only for the converse part.
Furthermore, when $K=2$, the condition \eqref{eq:thm_asymp_cond} reduces to $1/C_1 + 1/C_2 > 1/C$. Note that for every CM-DMBC, we have that $1/C_1+1/C_2\geq 1/C$. 
Indeed, suppose on the contrary that $1/C_1 + 1/C_2 < 1/C$. 
Then one can achieve a rate larger than $C$ by sequential transmission to the two decoders:
\begin{IEEEeqnarray}{rCl}
  \IEEEeqnarraymulticol{3}{l}{\max_{\alpha\in[0,1]}\min\mathopen{}\left\{ \alpha C_1, (1-\alpha)C_2 \right\}}\nonumber\\
  \qquad &=& \min\mathopen{}\left\{ \frac{C_2}{C_1+C_2} C_1, \left(1-\frac{C_2}{C_1+C_2}\right)C_2 \right\}\IEEEeqnarraynumspace\\
  & =& \frac{1}{1/C_1+1/C_2} \\
  &>& C.
\end{IEEEeqnarray}
But this contradicts the fact that $C$ is the capacity. 
Theorem~\ref{thm:asymp} does not hold for the special case $1/C_1 + 1/C_2 = 1/C$. 

As we shall show next, the constants $\Xi_{\text{a}}$ and $\Xi_{\text{c}}$ defined in \eqref{eq:constant_LHS_as} and \eqref{eq:constant_RHS_as}, respectively, satisfy $\Xi_{\text{a}} \geq \Xi_{\text{c}} > 0$. 
This implies that the second-order term in the asymptotic expansion of $\log \Msf(\ell,\epsilon)$ is positive for every $\epsilon\in(0,1)$ (see \eqref{eq:asymp_expansion})---a result that strengthens \cite[Th.~3]{Trillingsgaard2015}. 
Proving that $\Xi_{\text{a}} \geq \Xi_{\text{c}}$ will also allow us to shed light on the reason behind the gap between the achievability and converse bound and the role of the RVs $\{H_k\}$ 
in~\eqref{eq:constant_RHS_as}. 
\begin{proposition}
 Under the conditions described in Theorem~\ref{thm:asymp}, the constants in \eqref{eq:constant_LHS_as} and \eqref{eq:Xi_c_def} satisfy
  \begin{IEEEeqnarray}{rCl}
    0 < \Xi_{\text{a}} &\leq& \Xi_{\text{c}}.
  \end{IEEEeqnarray}
\end{proposition}
\begin{IEEEproof}
 It is convenient to rewrite $\Xi_{\text{a}}$ as follows:
  \begin{IEEEeqnarray}{rCl}
     \Xi_{\text{a}}
     &=& \min_{\vect{v} \in\mathbb{R}_0^{|\mathcal{X}|}} \lim_{n\rightarrow \infty } \E{ \max_k (\diff I_k(\vect{v})+n  +\varrho_k Z_k)^+-n }.\label{eq:convergence_in_dist}\IEEEeqnarraynumspace
  \end{IEEEeqnarray}
To obtain \eqref{eq:convergence_in_dist}, we used that for every set $\{X_k\}$ of integrable RVs, $\max_k (X_k+n)^+ - n$ converges in distribution to $\max_k X_k$ as $n\rightarrow \infty$. Hence, by using that $|\max_k (X_k+n)^+ - n| \leq \max_k |\maxx{-n,X_k}| \leq \max |X_k|$ almost surely for all $n\in\mathbb{N}$, we invoke Lebesgue's dominated convergence theorem 
  \cite[Th.~16.4]{billingsley} to conclude that
\begin{IEEEeqnarray}{rCl}
\lim_{n\rightarrow \infty}\EBig{\max_k (X_k+n)^+-n} = \EBig{\max_k X_k}.\label{eq:lebes_result}
\end{IEEEeqnarray}
This implies \eqref{eq:convergence_in_dist}.
Next, we bound \eqref{eq:convergence_in_dist} as follows
\begin{IEEEeqnarray}{rCl}
 \IEEEeqnarraymulticol{3}{l}{\Xi_{\text{a}}}\nonumber\\
  &=& \min_{\vect{v} \in\mathbb{R}_0^{|\mathcal{X}|}} \lim_{n\rightarrow {\infty}} \int_{0}^\infty \bigg(1-\prod_k \Phi\farg{ \frac{w-n - \diff I_k(\vect{v})}{\varrho_k}}\bigg)\intdiff  w \nonumber\\
  &&\qquad\qquad\qquad{}- n\label{eq:min_lim_max_k_max0}\\
       &\geq&  \lim_{n\rightarrow {\infty}} \int_{0}^\infty \bigg(1-\prod_k \Phi\farg{ \frac{w-n + \diff I_k(\optx(w-n))}{\varrho_k} }\bigg)\intdiff w \nonumber\\
       &&\qquad{}- n\label{eq:remark_ineq_ach_conv}\\
       &=& \lim_{n\rightarrow \infty} \E{\max_k (H_k + n)^+ - n}\label{eq:remark_convert_back_Xic}\\
       &=& \E{\max_k H_k}\label{eq:remark_convert_back2}\\
       &=& \Xi_{\text{c}}.\label{eq:remark_Xic}
\end{IEEEeqnarray}
To obtain \eqref{eq:min_lim_max_k_max0}, we used that for every nonnegative RV $X$, we have $\int_{0}^\infty (1-\pr{X<x})\intdiff x=\E{X}$; we also used that $\{Z_k\}$ are i.i.d. Gaussian and that for every real-valued RV $T$ and every $w\geq 0$, we have that
$\pr{(T)^+ < w} = \pr{T < w}$. The inequality \eqref{eq:remark_ineq_ach_conv} follows from \eqref{eq:hatx_def}. Finally, \eqref{eq:remark_convert_back_Xic} and \eqref{eq:remark_convert_back2} follow from steps similar to the ones leading to \eqref{eq:min_lim_max_k_max0} and \eqref{eq:convergence_in_dist}, respectively.

The inequality \eqref{eq:remark_ineq_ach_conv} reveals the origin of the gap between $\Xi_{\text{a}}$ and $\Xi_{\text{c}}$. The constant $\Xi_{\text{a}}$ is obtained by evaluating the achievability bound in Theorem~\ref{thm:simple_achiev} for an i.i.d. process $X^\infty$. Instead, in the computation of $\Xi_{\text{c}}$, we find the input distribution that maximizes $\pr{\max_k \tau_k \leq t}$ for each $t$. The resulting process is not i.i.d.

To prove that $\Xi_{\text{c}}>0$, we proceed as follows:
\begin{IEEEeqnarray}{rCl}
 \IEEEeqnarraymulticol{3}{l}{\Xi_{\text{c}}}\nonumber\\
  &=& \lim_{n\rightarrow {\infty}} \int_{0}^\infty \bigg(1-\prod_k \Phi\farg{ \frac{w-n + \diff I_k(\optx(w-n))}{\varrho_k} }\bigg)\intdiff w \nonumber\\
  &&\qquad{}- n\label{eq:Xic_positive_strict1}\\
   &>&  \lim_{n\rightarrow {\infty}} \int_{0}^\infty \bigg(1-\min_k \Phi\farg{ \frac{w-n + \diff I_k(\optx(w-n))}{\varrho_k}}\bigg)\intdiff w \nonumber\\
   &&\qquad {}-n\label{eq:Xic_positive_strict2}\\
       &\geq&  \lim_{n\rightarrow {\infty}} \int_{0}^\infty \bigg(1- \min_k \Phi\farg{ \frac{w-n}{\varrho_k}}\bigg)\intdiff w - n\label{eq:Xic_positive_strict3}\\
              &\geq&  \lim_{n\rightarrow {\infty}} \min_k\int_{0}^\infty \bigg(1-  \Phi\farg{ \frac{w-n}{\varrho_k}}\bigg)\intdiff w - n\label{eq:Xic_positive_strict4}\\
         &=& \lim_{n\rightarrow \infty} \min_k \E{(\varrho_k Z_k + n)^+ - n} = \min_k\E{\varrho_k Z_k}= 0.\label{eq:Xic_positive_strict5}
\end{IEEEeqnarray}
Here, \eqref{eq:Xic_positive_strict1} follows from \eqref{eq:remark_ineq_ach_conv}; the inequality in \eqref{eq:Xic_positive_strict2} holds because $\prod_k a_k < \min_k a_k$ for all $a_k\in(0,1)$; finally, \eqref{eq:Xic_positive_strict3} follows because $\min_k \diff I_k(\optx(w)) \leq 0$ for all $w\in\mathbb{R}$. Indeed, $\diff I_k(\cdot)$ is the directional derivative of $I_k(P)$ computed at the unique capacity-achieving input distribution $P^*$. The last equation \eqref{eq:Xic_positive_strict5} follows  from an argument similar to the one leading to \eqref{eq:lebes_result}.
\end{IEEEproof}

There are cases where $\Xi_{\text{a}}=\Xi_{\text{c}}$, and hence \eqref{eq:asymp_expansion} provides a complete second-order characterization of $\log \Msf(\ell,\epsilon)$. This happens when $\optx(\cdot)$ in \eqref{eq:hatx_def} equals $\vect{0}_{|\mathcal{X}|}$, which occurs for example when $P^*$ simultaneously maximizes $I_k(P)$ for all $k\in\mathcal{K}$. In the following corollary, we provide sufficient conditions for Theorem~\ref{thm:asymp} to  yield an asymptotic expansion that is tight up to the second order.
\begin{corollary}
\label{thm:asymp_sym}
We have that $\Xi_{\text{a}} = \Xi_{\text{c}}$ and, hence,
\begin{IEEEeqnarray}{rCl}
  \log M^*_{\text{sf}}(\ell, \epsilon) = \frac{C \ell}{1-\epsilon} - \sqrt{\frac{V \ell}{1-\epsilon}} \EBig{\max_k   Z_k} + o(\sqrt{\ell})\IEEEeqnarraynumspace\label{eq:sym_normal_approx}
\end{IEEEeqnarray}
if either of the following conditions hold
\begin{enumerate}
\item The capacity-achieving input distribution $P^*$  simultaneously maximizes $I_k(P)$ for all $k\in\mathcal{K}$, or
\item $V_1=\cdots=V_K$ and
\begin{IEEEeqnarray}{rCl}
  \sum_{k} \diff I_k(\vect{e}_{|\mathcal{X}|}(x)) = 0, \qquad x\in \{1,\cdots,|\mathcal{X}|\}. \label{eq:tight_condition}
\end{IEEEeqnarray}
Here, $\vect{e}_{|\mathcal{X}|}(i)$ denotes the $|\mathcal{X}|$-dimensional vector whose $i$th entry is equal to one and whose remaining entries are equal to zero.
\end{enumerate}
\end{corollary}
\begin{IEEEproof}
  We shall prove that $\Xi_{\text{a}} = \Xi_{\text{c}}$ under the stated conditions by characterizing $\optx(\cdot)$ in \eqref{eq:hatx_def}. When the component channels $W_1, \cdots, W_K$ have the same capacity-achieving input distribution, we have $\diff I_k(\vect{v}) = 0$ for all $\vect{v} \in\mathbb{R}_0^{|\mathcal{X}|}$ and all $k\in\mathcal{K}$. Hence, $\vect{0}_{|\mathcal{X}|}$ is a maximizer of \eqref{eq:hatx_def} which implies that \eqref{eq:remark_ineq_ach_conv} holds with equality.

  Consider now the case that $\diff I_k(\vect{v}) \not= 0$ for some $\vect{v} \in\mathbb{R}_0^{|\mathcal{X}|}$ and some $k\in\mathcal{K}$. Let $\text{ker}(\diff I_k)$ denote the kernel of $\diff I_k$, and let $\mathcal{D}^\dagger = \mathbb{R}_0^{|\mathcal{X}|}\cap \bigcap_{k=1}^K \text{ker}(\diff I_k)$ and $\mathcal{D} = \mathcal{R}_0^{|\mathcal{X}|} \setminus \mathcal{D}^\dagger$. We note that $\mathcal{D}\not= \emptyset$ since, by assumption, there exists a $\vect{v}\in\mathbb{R}_0^{|\mathcal{X}|}$ and a $k\in\mathcal{K}$ such that $\diff I_k(\vect{v}) \not =0$. Let the dimension of the linear subspace $\mathcal{D}$ be $m$
  and let $\mathds{D}\in\mathbb{R}^{|\mathcal{X}|\times m}$ be an $|\mathcal{X}|$-by-$m$ matrix with columns spanning the linear subspace $\mathcal{D}$.
  Now, define
  \begin{IEEEeqnarray}{rCl}
    \optx_{\mathcal{D}}(w) \triangleq \argmax_{\vect{v}_{\mathcal{D}}\in \mathbb{R}^m} \prod_k\Phi\farg{\frac{w + \diff I_k(\mathds{D} \vect{v}_{\mathcal{D}})}{\varrho_k}}\label{eq:hatx_m_def}
  \end{IEEEeqnarray}
  and
  \begin{IEEEeqnarray}{rCl}
&&\vect{\tilde v}_{\mathcal{D}} \triangleq \argmin_{\vect{v}_{\mathcal{D}}\in \mathbb{R}^m} \lim_{n\rightarrow {\infty}} \nonumber\\
&& \left(\int_{0}^\infty \bigg(1-\prod_k \Phi\farg{\frac{w-n - \diff I_k(\mathds{D} \vect{v}_{\mathcal{D}})}{\varrho_k} }\bigg)\intdiff w - n\right).\IEEEeqnarraynumspace
  \end{IEEEeqnarray}
  We note that $\optx_{\mathcal{D}}(w)$ is continuous in $w$ and that log-concavity of the objective function in \eqref{eq:hatx_m_def} implies that $\optx_{\mathcal{D}}(w)$ is the unique maximizer of the optimization problem.
  Moreover, $\mathds{D}\optx_{\mathcal{D}}(w)$ is a maximizer of \eqref{eq:hatx_def} and $\mathds{D}\vect{\tilde v}_{\mathcal{D}}$ is a minimizer in \eqref{eq:min_lim_max_k_max0}. This implies that the steps \eqref{eq:min_lim_max_k_max0}--\eqref{eq:remark_Xic} can be equivalently written as
  \begin{IEEEeqnarray}{rCl}
  \IEEEeqnarraymulticol{3}{l}{\Xi_{\text{a}}}\nonumber\\
  &=& \min_{\vect{v} \in\mathbb{R}_0^{|\mathcal{X}|}} \lim_{n\rightarrow {\infty}}\nonumber\\
  &&{} \int_{0}^\infty \bigg(1-\prod_k \Phi\farg{ \frac{w-n - \diff I_k(\vect{v})}{\varrho_k} }\bigg)\intdiff w - n\\
   &=& \lim_{n\rightarrow {\infty}} \nonumber\\
  &&{}\int_{0}^\infty \bigg(1-\prod_k \Phi\farg{ \frac{w-n - \diff I_k(\mathds{D}\vect{\tilde v}_{\mathcal{D}})}{\varrho_k} }\bigg)\intdiff w - n \\
    &\geq&  \lim_{n\rightarrow {\infty}} \int_{0}^\infty \bigg(1\nonumber\\
  &&\quad{}-\prod_k \Phi\farg{ \frac{w-n + \diff I_k(\mathds{D}\optx_{\mathcal{D}}(w-n))}{\varrho_k} }\bigg)\intdiff w - n\label{eq:corr_inequality}\IEEEeqnarraynumspace\\
   &=&  \lim_{n\rightarrow {\infty}} \int_{0}^\infty \bigg(1\nonumber\\
  &&\quad{}-\prod_k \Phi\farg{ \frac{w-n + \diff I_k(\optx(w-n))}{\varrho_k} }\bigg)\intdiff w - n \\
   &=& \Xi_{\text{c}}.
  \end{IEEEeqnarray}
Since $\optx_{\mathcal{D}}(w)$ is continuous and is the unique maximizer of \eqref{eq:hatx_m_def}, the inequality in \eqref{eq:corr_inequality} holds with equality if and only if $\optx_{\mathcal{D}}(w) = \mathbf{a}$ almost everywhere for some vector $\mathbf{a} \in\mathbb{R}^m$ that does not depend on $w$.

Suppose that $\optx_{\mathcal{D}}(w) = \mathbf{a}$. The objective function in \eqref{eq:hatx_m_def} is positive, strictly log-concave in $\vect{v}_{\mathcal{D}} \in\mathbb{R}^m$, and tends to zero as $\vectornorm{\vect{v}_{\mathcal{D}}}\rightarrow \infty$. Thus, the unique maximum is at the unique stationary point, which can be found by differentiating the logarithm of the objective function in \eqref{eq:hatx_m_def} in each of the $m$ dimensions and by equating it to zero. This yields
  \begin{multline}
    \sum_k \psi\farg{\frac{w+\diff I_k(\mathds{D}\mathbf{a})}{\varrho_k}}  \frac{\diff I_k(\mathds{D}\vect{e}_m(i))}{\varrho_k}  = 0,\\ \qquad i\in\{1,\cdots,m\}, w\in\mathbb{R}\label{eq:hatx_kkt}
  \end{multline}
  where $\psi(w) \triangleq \phi(w)/\Phi(w)$. It follows that \eqref{eq:hatx_kkt} cannot be satisfied for every $w\in\mathbb{R}$ unless $\vect{a} = \vect{0}_{m}$ and $\varrho_1=\cdots=\varrho_K$. In this case \eqref{eq:hatx_kkt} reduces to \eqref{eq:tight_condition}.
\end{IEEEproof}

For broadcast channels that do not satisfy the conditions of Corollary~\ref{thm:asymp_sym}, we can tighten the left-hand side of \eqref{eq:asymp_expansion} by using an input distribution that is not stationary memoryless. This yields the following theorem.
\begin{theorem}
Let $V \triangleq (\prod_k V_k)^{1/K}$ and $\varrho_k \triangleq \sqrt{V_k/V}$. Fix a differentiable function $\optbar: \mathbb{R} \mapsto \mathbb{R}^{|\mathcal{X}|}_0$ such that 
\begin{IEEEeqnarray}{rCl}
P^* + C \optbar'(w) \in\mathcal{P}(\mathcal{X})\label{eq:tv_input_dist_cond}
\end{IEEEeqnarray}
for all $w\in\mathbb{R}$. Additionally, define
\begin{IEEEeqnarray}{rCl}
E_k(s) \triangleq C - I_k(P^* + C \optbar'(s)) + C\diff I_k(\optbar'(s))\label{eq:Eks_cond}
\end{IEEEeqnarray}
and assume that 
\begin{IEEEeqnarray}{rCl}
\int_{-\infty}^{\infty} E_k(s)\intdiff s < \infty\label{eq:int_Ek_bound}
\end{IEEEeqnarray}
and that 
\begin{IEEEeqnarray}{rCl}
\sup_s |E'_k(s)|<\infty.\label{eq:bounded_Eks}
\end{IEEEeqnarray}
Then, for every CM-DMBC, we have
\begin{IEEEeqnarray}{rCl}
  \log M_{\text{sf}}^*(\ell,\epsilon)  \geq \frac{C \ell}{1-\epsilon} -\sqrt{\frac{V \ell}{1-\epsilon}} \bar \Xi_{\text{a}} - o(\sqrt{\ell}).\label{eq:TheoremTV_asymp}
\end{IEEEeqnarray}
Here,
\begin{IEEEeqnarray}{rCl}
  \bar \Xi_{\text{a}} \triangleq \E{\max_k \tvRVbar_k}
\end{IEEEeqnarray}
where the independent RVs $\{\tvRVbar_k\}$ are defined by the cumulative distribution functions
\begin{multline}
  F_{\tvRVbar_k}\farg{w} \\\triangleq \Phi\farg{\frac{1}{\varrho_k} \left(w +\diff I_k(\optbar(w)) - \int_{-\infty}^w \frac{E_k(s)}{C} \intdiff s  \right)}.\label{eq:FtvRVk_def}
\end{multline}
\label{thm:time_varying}
\end{theorem}
\begin{IEEEproof}
See Appendix~\ref{sec:achievability_asymptotics}.
\end{IEEEproof}
If one sets $\optbar(\cdot)$ in Theorem~\ref{thm:time_varying} equal to $\vect{\hat v}(\cdot)$ in \eqref{eq:hatx_def}, the resulting gap between $\Xi_{\text{c}}$ and $\bar \Xi_a$ is  caused only by the ``error'' term $E_k(s)$. 
Interestingly, there are channels beyond the ones for which Corollary~\ref{thm:asymp_sym} applies where $E_k(s)=0$ and, hence, a complete second-order characterization of $\log \Msf(\ell,\epsilon)$ is available. The next corollary describes a class of channels for which this is the case.
\begin{corollary}\label{cor:tv_asymp_time_sharing}
  Let $\mathcal{X}_1,\cdots,\mathcal{X}_\Sch$ be disjoint sets and let $\mathcal{X}=\cup_{\sch=1}^\Sch \mathcal{X}_\sch$. Moreover, for $k\in\mathcal{K}$ and $\sch\in\{1,\cdots,\Sch\}$, let $W_{k,\sch}$ be a channel from $\mathcal{X}_\sch$ to $\mathcal{Y}_k$ with capacity-achieving input distribution $P_\sch^*$ (independent of $k$), capacity-achieving output distribution $P_{Y_k}^*$ (independent of $\sch$), and capacity $C_{k,\sch}$. Define for all $x\in\mathcal{X}$ and $y\in\mathcal{Y}_k$ the channel $W_k(y|x) = W_{k ,\sch(x)}(y|x)$, where the function $\sch: \mathcal{X}\mapsto \{1,\cdots,\Sch\}$ is such that $x\in\mathcal{X}_{\sch(x)}$. Assume that $C_1  \geq \cdots \geq C_K$ and that 
\begin{IEEEeqnarray}{rCl}
  \frac{1}{C_i}+\frac{i}{C_K}  &> & \frac{i}{C}, \qquad  i\in\{1,\cdots,K-1\}.
\end{IEEEeqnarray}
Define
\begin{IEEEeqnarray}{rCl}
\vect{\beta}(w) & \triangleq& \argmax_{\vect{\beta}\in\mathbb{R}_0^{\Sch}} \prod_k \Phi\farg{\frac{1}{\varrho_{k}} \left(w + \sum_{\sch=1}^\Sch \beta_\sch C_{k,\sch}\right)}\label{eq:betaw_def}
\end{IEEEeqnarray}
and assume that 
\begin{IEEEeqnarray}{rCl}
P^*(x) + C P_{\sch(x)}^*(x)\beta'_{\sch(x)}(w) \in[0,1]\label{eq:cor_probability_assumption}
\end{IEEEeqnarray}
for every $x\in\mathcal{X}$ and $w\in\mathbb{R}$.
   Then, for every $\epsilon\in(0,1)$,
  \begin{IEEEeqnarray}{rCl}
    \log M^*_{\text{sf}}(\ell,\epsilon) = \frac{C \ell}{1-\epsilon} - \sqrt{\frac{V \ell}{1-\epsilon}}\Xi_{\text{c}} + o(\sqrt{\ell})\label{eq:thm_tv_tight}
  \end{IEEEeqnarray}
  where $\Xi_{\text{c}}$ is defined in \eqref{eq:Xi_c_def}.
\end{corollary}
\begin{IEEEproof}
  We shall first evaluate the mutual information and the directional derivative of the mutual information. Define the input distribution
  \begin{IEEEeqnarray}{rCl}
    P_{\vect{\alpha}} \triangleq [\alpha_1 P_1^*,\alpha_2 P_2^*, \cdots,\alpha_\Sch P_\Sch^*]\label{eq:tv_input_distribution}
  \end{IEEEeqnarray}
  for all nonnegative vectors $\vect{\alpha}$ with $\sum_{\sch=1}^\Sch \alpha_\sch = 1$.  The mutual information $I_k(P_{\vect{\alpha}})$ is given by
  \begin{IEEEeqnarray}{rCl}
    \IEEEeqnarraymulticol{3}{l}{I_k(P_{\vect{\alpha}})}\nonumber\\
    \quad &=& \sum_{x\in\mathcal{X}} \sum_{y\in\mathcal{Y}_k} P_{\vect{\alpha}}(x) W_k(y|x) \log \frac{W_k(y|x)}{P_{Y_k}^*(y) }\\
    &=& \sum_{\sch=1}^\Sch \alpha_\sch \sum_{x\in\mathcal{X}_r} \sum_{y\in\mathcal{Y}_k} P^*_{r}(x) W_{k,\sch}(y|x) \log \frac{W_{k,\sch}(y|x)}{P_{Y_k}^*(y) }\label{eq:same_output_dist}\IEEEeqnarraynumspace\\
    &=& \sum_{\sch=1}^\Sch \alpha_\sch C_{k,\sch}.\label{eq:ts_mut_inf}
  \end{IEEEeqnarray}
  In \eqref{eq:ts_mut_inf}, we used that the channels $W_{k,\sch}$ have the same capacity-achieving output distribution for $\sch\in\{1,\cdots,\Sch\}$.
  
  Next, we let $\vect{\alpha}^*$ be the maximizer of $\vect{\alpha}\mapsto \min_k I_k(P_{\vect{\alpha}})$ and compute the directional derivative of the mutual information at $P_{\vect{\alpha}^*}$ along the direction $\vect{v}\in\mathbb{R}_0^{|\mathcal{X}|}$:
  \begin{IEEEeqnarray}{rCl}
    \diff_{P_{\vect{\alpha}^*}} I_k(\vect{v}) &=& \sum_{r=1}^\Sch \sum_{x\in\mathcal{X}_\sch} v_x D(W_{k,\sch}(\cdot|x)||P_{Y_k}^*)\label{eq:tv_differentials0}\\
    &=& \sum_{\sch=1}^\Sch \left(\sum_{x\in\mathcal{X}_\sch} v_x\right) C_{k,\sch}.\label{eq:tv_differentials}
  \end{IEEEeqnarray}
   \begin{sloppypar}\noindent Here, \eqref{eq:tv_differentials0} follows because the output distribution at decoder $k$ given an input distribution of the form \eqref{eq:tv_input_distribution} is $P_{Y_k}^*$ and \eqref{eq:tv_differentials} follows from the assumption $P^*(x)>0$, which implies that $P_\sch^*(x)>0$ for $x\in\mathcal{X}_\sch$ and $\sch\in\{1,\cdots,\Sch\}$. In turn, this implies that $D(W_{k,\sch}(\cdot|x)||P_{Y_k}^*)=C_{k,\sch}$ (see, e.g., \cite[Th.~4.5.1]{Gallager68}). It follows from \eqref{eq:ts_mut_inf} and \eqref{eq:tv_differentials} that the capacity $C$ is achieved using time-sharing and that the capacity-achieving input distribution $P^*$ must have the form given by \eqref{eq:tv_input_distribution}. Indeed, by the concavity of mutual information and by the definition of $\vect{\alpha}^*$, we have  that, for all $P\in \mathcal{P}(\mathcal{X})$,
   \begin{IEEEeqnarray}{rCl}
    \IEEEeqnarraymulticol{3}{l}{\min_k I_k(P)}\nonumber\\
      &\leq& \min_k\Big\{ I_k(P_{\vect{\alpha}^*}) +  \diff_{P_{\vect{\alpha}^*}} I_k(P - P_{\vect{\alpha}^*})\Big\}\\
    &=& \min_k\bigg\{  \sum_{\sch=1}^\Sch \alpha^*_\sch C_{k,\sch} \nonumber\\
    &&\qquad\qquad{}+ \sum_{\sch=1}^\Sch \left(\sum_{x\in\mathcal{X}_\sch} (P(x)-P_{\vect{\alpha}^*}(x))\right) C_{k,\sch}\bigg\}\IEEEeqnarraynumspace\\
    &=& \min_k\bigg\{  \sum_{\sch=1}^\Sch \left(\alpha^*_\sch + \left(\sum_{x\in\mathcal{X}_\sch} (P(x)-P_{\vect{\alpha}^*}(x))\right)\right) C_{k,\sch} \bigg\}\IEEEeqnarraynumspace\\
    &\leq& \min_k I_k(P_{\vect{\alpha}^*}).
   \end{IEEEeqnarray}
   Thus, we must have that $P_{\vect{\alpha}^*} = P^*$.

  By substituting \eqref{eq:tv_differentials} in \eqref{eq:Wk_def}, we obtain\end{sloppypar}
\begin{align}
  F_{\tvRV_k}(w) \triangleq \Phi\farg{ \frac{1}{\varrho_{ k}} \left( w+\sum_{\sch=1}^\Sch \beta_\sch(w) C_{k,\sch} \right)}\label{eq:tv_Wk_def}
\end{align}
where the function $\vect{\beta}:  \mathbb{R}\mapsto \mathbb{R}^\Sch_0$ is given by \eqref{eq:betaw_def}.

Next, we shall prove that we can achieve \eqref{eq:thm_tv_tight} using Theorem~\ref{thm:time_varying}. Define the function $\optbar: \mathbb{R}\mapsto \mathbb{R}^{|\mathcal{X}|}$ as follows:
\begin{IEEEeqnarray}{rCl}
  \optbarnbold_x(w)= P^*_{\sch(x)}(x)\beta_{\sch(x)}(w), \qquad x\in\mathcal{X}.\label{eq:hatx_simplified}
\end{IEEEeqnarray}
Note that $\optbar(w)$ maximizes \eqref{eq:hatx_def}. Indeed,
\begin{IEEEeqnarray}{rCl}
\IEEEeqnarraymulticol{3}{l}{\max_{\vect{v}\in\mathbb{R}_0^{|\mathcal{X}|}} \prod_k \Phi\farg{\frac{w + \diff I_k(\vect{v})}{\varrho_k}}}\nonumber\\
 &=& \max_{\vect{v}\in\mathbb{R}_0^{|\mathcal{X}|}} \prod_k \Phi\farg{\frac{1}{\varrho_k}\left(w + \sum_{\sch=1}^\Sch\left(\sum_{x\in\mathcal{X}_\sch} v_x\right) C_{k,\sch} \right)}\label{eq:beta_max_hatx}\\
&=& \max_{\vect{\beta}\in\mathbb{R}_0^{\Sch}} \prod_k \Phi\farg{\frac{1}{\varrho_{ k}} \left(w + \sum_{\sch=1}^\Sch \sum_{x\in\mathcal{X}_\sch} P_\sch^*(x)\beta_\sch C_{k,\sch}\right)}\IEEEeqnarraynumspace\\
 &=& \prod_k \Phi\farg{\frac{1}{\varrho_{ k}} \left(w +  \sum_{x\in\mathcal{X}} P_{\sch(x)}^*(x) \beta_{\sch(x)}(w) C_{k,l}\right)}\label{eq:beta_from_def} \\
 &=& \prod_k \Phi\farg{\frac{1}{\varrho_{ k}} \left(w +  \diff I_k(\optbar(w)) \right)}.
\end{IEEEeqnarray}
Here, \eqref{eq:beta_max_hatx} follows from \eqref{eq:tv_differentials} and \eqref{eq:beta_from_def} follows from the definition of $\vect{\beta}(w)$ in \eqref{eq:betaw_def}.
Note that the definition of $\optbar(w)$ implies that $P^* + C \optbar'(w) = [(\alpha_1^*+ C\beta_1'(w) )P_1^*,\cdots,(\alpha_\Sch^*+C\beta_\Sch'(w)) P_\Sch^*]$, which is a probability distribution by the condition in \eqref{eq:cor_probability_assumption} and has the form \eqref{eq:tv_input_distribution}.
Next, we have that
\begin{IEEEeqnarray}{rCl}
  \IEEEeqnarraymulticol{3}{l}{E_k(s)}\nonumber\\
  \quad &=& C - I_k(P^* + C \optbar'(s)) + C \diff I_k(\optbar'(s))\\
&=& C - \sum_{\sch=1}^\Sch (\alpha_\sch^*+C\beta_\sch'(w))C_{k,\sch} + C \sum_{\sch=1}^\Sch \beta_\sch'(s) C_{k,\sch}\label{eq:ts_Ek}\IEEEeqnarraynumspace\\
&=&0.
\end{IEEEeqnarray}
Here, \eqref{eq:ts_Ek} follows from \eqref{eq:ts_mut_inf} and \eqref{eq:tv_differentials}. Since $E_k(s)=0$, we have that $\tvRV_k$ has the same distribution as $\tvRVbar_k$ for $k\in\mathcal{K}$. Furthermore, $\int_{-\infty}^\infty E_k(s)\intdiff s = 0$ and $|E_k'(s)|<\infty$ for every $s\in\mathbb{R}$. The conditions in Theorem~\ref{thm:time_varying} are thus satisfied, which implies that \eqref{eq:thm_tv_tight} is indeed achievable.
\end{IEEEproof} 
The following lemma shows that there exist nontrivial channels that satisfy the conditions of Corollary~\ref{cor:tv_asymp_time_sharing}.
\begin{lemma}
\label{lem:tv_opt_cond}
Let $\Sch=2$, $K=2$, and $\Delta_1 \triangleq C_{11}-C_{12} > 0 >  C_{21}-C_{22} \triangleq \Delta_2$. Let also
  \begin{IEEEeqnarray}{rCl}
D \triangleq
-\frac{\Delta_1 \varrho_2^2 + \Delta_2 \varrho_1^2 }{\Delta_1^2 \varrho_2^2 +\Delta_2^2 \varrho_1^2}.\label{eq:chi}
  \end{IEEEeqnarray}
  Then, the condition $P^*(x)+C P^*_{\sch(x)}(x)\beta_{\sch(x)}'(w)\in[0,1]$ holds for every $x\in\mathcal{X}$ and every $w\in\mathbb{R}$ provided that
  \begin{IEEEeqnarray}{rCl}
P^*(x) + (-1)^{\sch(x)+1} C P_{\sch(x)}^*(x) D &\in&[0,1]\label{eq:tv_opt_cond}
\end{IEEEeqnarray}
for every $x\in\mathcal{X}$ and
\begin{IEEEeqnarray}{rCl}
\left(\frac{\Delta_1}{\varrho_1} + \frac{\Delta_2}{\varrho_2}\right)(\varrho_2 - \varrho_1) \geq 0.\label{eq:lem_cond2}
\end{IEEEeqnarray}
\end{lemma}
\begin{IEEEproof}
See Appendix~\ref{app:tv_opt_cond}.
\end{IEEEproof}

\begin{figure*}[!t]
\begin{center}
\subfigure[$K=2$.]{ 
\setlength\smallfigureheight{7.5cm}
\setlength\smallfigurewidth{8.5cm}
\includegraphics{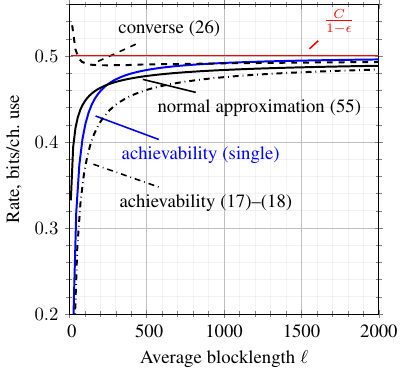}
}
\subfigure[$K=3$.]{
\setlength\smallfigureheight{7.5cm}
\setlength\smallfigurewidth{8.5cm}
\includegraphics{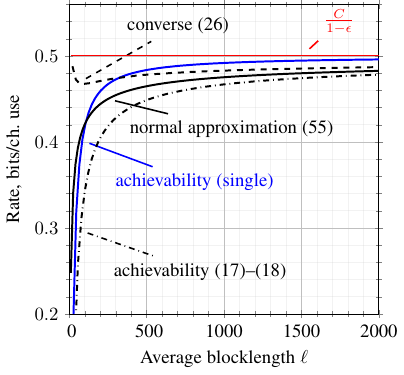}
}
\subfigure[$K=4$.]{
\setlength\smallfigureheight{7.5cm}
\setlength\smallfigurewidth{8.5cm}
\includegraphics{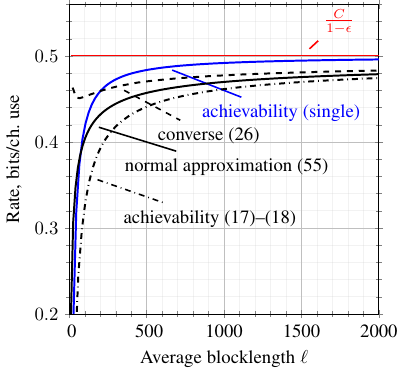}
}
\subfigure[Normal approximation for $K\in\{2,\cdots,8\}$.]{
\setlength\smallfigureheight{7.5cm}
\setlength\smallfigurewidth{8.5cm}
\includegraphics{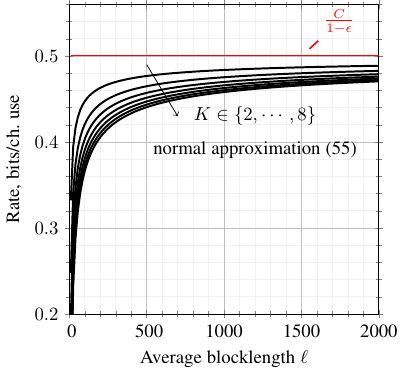}
}
\end{center}
\caption{Comparison between the achievability bound in Theorem~\ref{thm:simple_achiev}, the converse bound in Corollary~\ref{thm:converse_bound2}, and the normal approximation~\eqref{eq:sym_normal_approx} for $\epsilon=10^{-3}$.
The component channels in the CM-DMBC are BSCs with crossover probability $0.11$. The normal approximation corresponds to the asymptotic expansion in  \eqref{eq:sym_normal_approx} with the $o(\cdot)$ term neglected. The blue curve labeled ``achievability, $K=1$'' corresponds to the single-user achievability bound in \cite[Th.~3]{Polyanskiy2011} evaluated for a BSC with crossover probability $0.11$ and $\epsilon=10^{-3}$.}
\label{fig:num_bsc}
\end{figure*}
\begin{figure}[!t]
  \begin{center}
    \setlength\smallfigureheight{7.1cm}
    \setlength\smallfigurewidth{7.1cm}
    \includegraphics{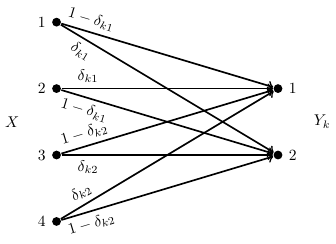}
  \end{center}
  \caption{Asymmetric channels $\{W_k\}$ that obey the conditions in Corollary~\ref{cor:tv_asymp_time_sharing} for the channel parameters $\bscprob_{11}=0.01, \bscprob_{12}=0.40, \bscprob_{21} = 0.15$, and $\bscprob_{22}=0.10$. The channels consist of two BSCs with common outputs.}
  \label{fig:W1W2}
\end{figure}

\section{Numerical Examples}

\subsection{Binary Symmetric Channels}
Let $W_1$ and $W_2$ be two BSCs, each with crossover probability $\bscprob$. Note that $W_1$ and $W_2$ are symmetric \cite[p.~185]{Cover2012} and have the same capacity-achieving input distribution. We evaluate the bounds presented in Theorem~\ref{thm:simple_achiev}, Corollary~\ref{thm:converse_bound2}, and Corollary~\ref{thm:asymp_sym} for the CM-DMBC having $W_1$ and $W_2$ as its component channels. The bounds are depicted in Fig.~\ref{fig:num_bsc} for the case $\bscprob=0.11$ and $\epsilon=10^{-3}$.
The capacity-achieving input distribution $P^*$ of the individual BSCs  is $\text{Bern}(1/2)$, their capacity is given by $1-H_b(\bscprob)$, where $H_b(\cdot)$ denotes the binary entropy function, and the directional derivatives $\diff I_k(\cdot)$ of the mutual information at $P^*$ are zero.
Furthermore, for $Y_k^n \sim P_{Y_k^n|X^n =x^n}$, the information densities $i_{P^*,W_k}(x^n; Y_k^n)$ satisfy
\begin{multline}
i_{P^*,W_k}(x^n; Y_k^n) \sim  n \log(2-2 \bscprob) + \log\farg{\frac{\bscprob}{1-\bscprob}}\sum_{j=1}^n Z_{k,j}\label{eq:num_inf_dens}
\end{multline}
where $Z_{k,j} \stackrel{\text{i.i.d.}}{\sim} \text{Bern}(\bscprob)$. We observe that the distribution of the information density in \eqref{eq:num_inf_dens} is independent of $x^n$. The converse bound in the figure is obtained from Corollary~\ref{thm:converse_bound2}, where the value of $\eta$ is optimized numerically. The achievability bound is obtained from Theorem~\ref{thm:simple_achiev} for the choice $P_{X^\infty} = (P^*)^\infty$. To evaluate the bound, we use that $\tau_1$ and $\tau_2$ are i.i.d. RVs. This allows us to compute $\E{\maxx{\tau_1,\tau_2}}$ by evaluating $\sum_{t=0}^\infty (1-F_{\tau_1}(t)^2)$ where $F_{\tau_1}\farg{\cdot}$ is the cumulative distribution function of $\tau_1$.
To estimate \eqref{eq:achiev_Proberror1}, we use the change of measure technique (see \cite[p.~4911]{Polyanskiy2011}).

We observe that, in the two-user case, the speed of convergence to the asymptotic limit is indeed slower than for the single-user case (the curve marked ``achievability (single)'' in Fig.~\ref{fig:num_bsc}, which is the point-to-point achievability bound reported in \cite[Th.~3]{Polyanskiy2011}). In particular, for $\ell \geq 1000$ and $K=2$, our converse bound is strictly below the single-user achievability bound, which implies that the maximum coding rate for the two-user case is strictly smaller than that for the single-user case. Additionally, the speed of convergence becomes slower as the number of users increases.

\subsection{Asymmetric Channels}
Next, we illustrate through an example that Theorem~\ref{thm:time_varying} indeed improves over Theorem~\ref{thm:asymp}. We consider the CM-DMBC depicted in Fig.~\ref{fig:W1W2}; we shall also assume that $\bscprob_{11}=0.01, \bscprob_{12}=0.40, \bscprob_{21} = 0.15$, and $\bscprob_{22}=0.10$. The two component channels, $W_1$ and $W_2$, are two BSCs with common outputs.
Let now $\mathcal{X}_1 = \{1,2\}$, $\mathcal{X}_2 = \{3,4\}$, and $W_{k,\sch}$ be a BSC with crossover probability $\bscprob_{k,\sch}$ for $k\in\{1,2\}$ and $\sch\in\{1,2\}$. One can verify that the condition $P^*(x) + C P^*_{\sch(x)}(x)\beta_{\sch(x)}'(w)\in[0,1]$ for $x\in\mathcal{X}$ and $w\in\mathbb{R}$ in Corollary~\ref{cor:tv_asymp_time_sharing} is satisfied using Lemma~\ref{lem:tv_opt_cond}.  Therefore, the asymptotic expansion in \eqref{eq:thm_tv_tight} holds, i.e., the converse bound in Theorem~\ref{thm:asymp} is tight up to second order.
The same is not true for the achievability bound in Theorem~\ref{thm:asymp}. Indeed, by computing \eqref{eq:constant_LHS_as} and \eqref{eq:constant_RHS_as}, we find that $\Xi_{\text{c}} = 0.2630$ but that $\Xi_{\text{a}} = 0.3175$.

\section{Conclusion}
In this paper, we considered the $K$-user CM-DMBC for the scenario where variable-length stop-feedback codes are used. We presented achievability and converse bounds on the maximum coding rate $\frac{1}{\ell}\log \Msf(\ell,\epsilon)$ for a fixed average blocklength $\ell$ and average error probability $\epsilon$. The main novelty in our nonasymptotic analysis is the converse bound, which relies on a nonstandard application of the meta-converse theorem \cite[Th.~26]{Polyanskiy2010b} to the variable-length setup. The achievability bound follows instead from a straightforward generalization of \cite[Th.~3]{Polyanskiy2011}. An asymptotic analysis of our bounds in the limit $\ell\rightarrow \infty$ reveals that, under mild technical conditions, the second-order asymptotic expansion of $\log \Msf(\ell,\epsilon)$ contains a square-root penalty. We provided upper and lower bounds on the second-order term in the asymptotic expansion of $\log \Msf(\ell,\epsilon)$ and derived necessary and sufficient conditions for our bounds on the second-order term to match. This occurs for example for the case when the component channels are two identical BSCs. For this case, we provide numerical evidence that the convergence to the asymptotic limit of the maximum coding rate is indeed slower than in the point-to-point case. Furthermore, our numerical results show that the first two terms in the asymptotic expansion of $\log \Msf(\ell,\epsilon)$ approximate $\log \Msf(\ell,\epsilon)$ accordingly.

Finally, we emphasize that our results are based on a setup in which the encoder output at time $n$ does not depend on the stop signals received before time $n$. 
A generalization of our analysis to the case when this dependency is allowed (which may result in a faster convergence to capacity) is left for future work. 
We recently showed that the dispersion is zero if full feedback is available~\cite{Trillingsgaard2018_VLF1}. 
It is then natural to ask how much feedback is needed for the dispersion to vanish.


\appendices

\section{Proof of Theorem~\ref{thm:simple_achiev}}
\label{app:achievability_proof} 
The proof follows closely \cite[Th.3~]{Polyanskiy2011}. Let $S$ be a Bernoulli RV with $\pr{S=1}=q$ and let its probability mass function be given by $P_S$. We start by specifying $U, f_n, \{g_{k,n}\}_{k\in\mathcal{K}}$, $\{\tau_k^*\}_{k\in\mathcal{K}}$. 
The RV $U$ has the following domain and probability mass function
\begin{align}
  \mathcal{U} &\triangleq \{0,1\} \times \underbrace{\mathcal{X}^\infty \times \cdots \times \mathcal{X}^\infty}_{M \text{ times}} \\
  P_U &\triangleq P_{S} \times\underbrace{P_{X^\infty} \times \cdots \times P_{X^\infty} }_{M \text{ times}}.
\end{align}
Note that the cardinality of $\mathcal{U}$ is unbounded. 
As remarked after Definition~\ref{def:VLSFcode}, the cardinality of $\mathcal{U}$ can always be reduced to $K+1$.

As in \cite{Polyanskiy2011}, the realization $u$ of $U=u$ defines a codebook $\{\mathbf{C}^{(u)}_1,\ldots,\mathbf{C}^{(u)}_M\}$ consisting of $M$ infinite dimensional codewords $\mathbf{C}^{(u)}_j \in \mathcal{X}^\infty$, $j\in\mathcal{M}$. Differently from \cite{Polyanskiy2011}, it also defines the RV $S$, which we shall use as a source of additional common randomness among the $K$ decoders. The encoder operates as follows:
\begin{align}
f_n(u,j) \triangleq \mathbf{C}_{j,n}^{(u)}, \qquad u\in\mathcal{U}, j\in \mathcal{M}.
\end{align}
Here, $\mathbf{C}_{j,n}^{(u)}$ denotes the $n$th entry of $\mathbf{C}_j^{(u)}$. To keep the notation compact, we shall omit the superscript $(u)$ in the remaining part of the proof. 
At time $n$, decoder $k$ computes the information densities
\begin{align}
  A_{k,n}(j) \triangleq i_{P_{X^n},W_k^n}(\mathbf{C}_j^n ; Y_k^n), \qquad j\in\mathcal{M}\label{eq:Skn_def}
\end{align}
where the vector $\mathbf{C}_j^n$ contains the first $n$ entries of $\mathbf{C}_j$. Define the stopping times
\begin{align}
\tau_{k}(j) &\triangleq \indi{S = 0}\inff{n \geq 0: A_{k,n}(j)\geq \gamma}
\end{align}
and let $\tau_k^*$ be the time at which decoder $k$ makes the final decision:
\begin{align}
  \tau_k^* \triangleq  \min_{j\in\mathcal{M}} \tau_k(j).\label{eq:tau_min}
\end{align}
The output of decoder $k$ at time $\tau^*_k$ is
\begin{align}
  g_{k,\tau^*_k}(U,Y_k^{\tau_k^*}) \triangleq \maxx{j \in\mathcal{M} : \tau_k(j) = \tau^*_k }.\label{eq:simple_achiev_decoder}
\end{align}
When $S=1$, we have $\tau^*_k = 0$, and hence the decoder $g_{k,\tau^*_k}(U,Y_k^{\tau_k^*})$ outputs $M$.
The average blocklength is then given by
\begin{IEEEeqnarray}{rCl}
\IEEEeqnarraymulticol{3}{l}{\EBig{\max_k \tau^*_k}}\nonumber\\
\quad &=& (1-q)\EBig{\max_k \tau^*_k | S= 0} \\
 &\leq& (1-q)\frac{1}{M}\sum_{j=1}^M \EBig{\max_k\tau_k(j)|J=j,S=0} \label{eq:simple_achiev_Emax_1}\\
&=& (1-q)\E{\max\tau_k(1)|J=1,S=0}\label{eq:simple_achiev_Emax_sym}\\
&=& (1-q)\EBig{\max_k\tau_k^{(0)}}\label{eq:simple_achiev_Emax_3}
\end{IEEEeqnarray}
where the expectation is over $J$, $Y_1^{\infty}$, $Y_2^{\infty}$, and $U$. Here, \eqref{eq:simple_achiev_Emax_1} follows from \eqref{eq:tau_min}; \eqref{eq:simple_achiev_Emax_sym} follows from symmetry; and \eqref{eq:simple_achiev_Emax_3} follows from the definition of $\tau_k^{(0)}$ in \eqref{eq:tau_k_def}.
For the error probability, we have that
\begin{IEEEeqnarray}{rCl}
  \IEEEeqnarraymulticol{3}{l}{\pr{ g_{k,\tau^*_k}(U, Y^{\tau^*_k}_k)\not= J} }\nonumber\\ 
   &\leq& \sprob + (1-\sprob)\pr{ g_{k,\tau^*_k}(U, Y^{\tau^*_k}_k) \not=  J| S=0} \\
   \quad&\leq& \sprob + (1-\sprob) \pr{ g_{k,\tau^*_k}(U, Y^{\tau^*_k}_k)\not= 1 | J=1,S=0}\label{eq:simple_achiev_error_1} \IEEEeqnarraynumspace\\
&\leq& \sprob + (1-\sprob)\pr{\tau_k(1) \geq \tau^*_k|S=0}\\
&=&  \sprob + (1-\sprob)\prBigg{ \bigcup_{j=2}^M \{\tau_k(1)\geq\tau_k(j)  \} \Bigg|S=0}\\
&\leq& \sprob + (1-\sprob) (M-1)\pr{\tau_k(1)\geq\tau_k(2)|S=0 }\\
&=& \sprob + (1-\sprob) (M-1)\pr{\tau_k^{(0)}\geq\bar \tau_k^{(0)}}.\label{eq:simple_achiev_error_final}
\end{IEEEeqnarray}
Here, \eqref{eq:simple_achiev_error_1} follows from \eqref{eq:simple_achiev_decoder} and \eqref{eq:simple_achiev_error_final} follows by noting that, given $J=1$, the RVs $\left(A_{k,n}(1), A_{k,n}(2), \tau_k(1),\tau_k(2)\right)$ (where the RVs $\{A_{k,n}(j)\}$ are defined in \eqref{eq:Skn_def}) have the same joint distribution as $\left(i_{P_{X^n},W_k^n}(X^n; Y_k^n), i_{P_{X^n},W_k^n}(\bar X^n ;  Y_k^n), \tau_k^{(0)},\bar \tau_k^{(0)}\right)$.
We conclude the proof by noting that, by Definition~\ref{def:VLSFcode}, the tuple $(U,f_n, \{g_{k,n}\}, \{\tau_k^*\})$ defines an $(\ell,M,\epsilon)$-VLSF code satisfying \eqref{eq:achiev_Emax} and \eqref{eq:achiev_Proberror1}.

The upper bound in \eqref{eq:achiev_Proberror2} follows from the same steps as in~\cite[Eqs. (111)--(118)]{Polyanskiy2011}.

\section{Proof of Lemma~\ref{lem:vlsf_metaconverse}}
\label{proof:vlsf_metaconverse}
Let $\epsilon_{k}^{(u)} \triangleq \pr{J\not= g_{k,\tau_k}(U, Y_k^{\tau_k})|U=u}$, $u\in\mathcal{U}$, be the conditional error probability at decoder $k$ given $U=u$
%
and define  the following probability measure on $\mathcal{X}^\infty \times \preffree{\overline Y}_1^{(u)} \times\cdots \times\preffree{\overline Y}_2^{(u)}$:
\begin{align}
  \mathds{Q}^{(u)}_{\X,\vect{\overline Y}_1,\cdots,\vect{\overline Y}_K}\farg{\x,\vect{\overline y}_1,\cdots,\vect{\overline y}_K} \triangleq P_{\X}^{(u)}(\x)\prod_{k=1}^K Q_k(\vect{\overline y}_{k}).\label{eq:Qk_def_eval}
\end{align}
For notational convenience, we shall indicate the two probability measures in \eqref{eq:Pk_def_eval} and \eqref{eq:Qk_def_eval} simply as  $\mathds{P}^{(u)}$ and $\mathds{Q}^{(u)}$, respectively.
For each decoder $k$, the average error probability is equal to $\epsilon^{(u)}_k$ under $\mathds{P}^{(u)}$, and it is no larger than $1-1/M$ under $\mathds{Q}^{(u)}$.
Hence, by an application of the meta-converse theorem \cite[Th.~26]{Polyanskiy2010b}, we conclude that for every $u\in\mathcal{U}$ and $k\in\mathcal{K}$
\begin{IEEEeqnarray}{rCl}
  M &\leq& \frac{1}{\beta_{1-\epsilon_k^{(u)}}(\mathds{P}_{\vect{X},\vect{\overline Y_k}}^{(u)}, \mathds{Q}_{\vect{X},\vect{\overline Y_k}}^{(u)})}.\label{eq:meta_conv_conclusion}
\end{IEEEeqnarray}
Here, $\beta_{\alpha}(P,Q)$ denotes the Neyman-Pearson function which is the minimum type-II error probability of a binary hypothesis test between two probability distributions $P$ and $Q$ on a common measurable space subject to the constraint that the type-I error probability does not exceed $1-\alpha$.
In order to obtain an information spectrum-type bound, we apply the following inequality \cite[Eq. (102)]{Polyanskiy2010b}
\begin{IEEEeqnarray}{rCl}
 \alpha \leq P\mathopen{}\Big[\frac{\dd P}{\dd Q}\geq \gamma\Big] + \gamma \beta_{\alpha}(P,Q)
\end{IEEEeqnarray}
to the left-hand side of \eqref{eq:meta_conv_conclusion}. Here, $\frac{\dd P}{\dd Q}$ denotes the Radon-Nikodym derivative. 
By doing so, we find that for every $ \gamma_k^{(u)} >0$ and for every $k\in\mathcal{K}$
\begin{multline}
  \log M \leq\\ \log  \gamma^{(u)}_k - \log\Big|\mathds{P}^{(u)}\mathopen{}\left[ i_k^{(u)}( \X ; \vect{\overline Y}_k )  \leq \log  \gamma^{(u)}_k\right]-\epsilon_k^{(u)} \Big|^+. \label{eq:converse_logM2_eval}
\end{multline}
Here,
\begin{IEEEeqnarray}{rCl}
 \IEEEeqnarraymulticol{3}{l}{i_k^{(u)}( \x;\vect{\bar  y}_k )}\nonumber\\
 \quad &\triangleq& \log \frac{\mathds{P}^{(u)}_{\X,\vect{\overline Y}_k}(\x,\vect{\overline y}_k)}{\mathds{Q}^{(u)}_{\X,\vect{\overline Y}_k}(\x,\vect{\overline y}_k)} =\log \frac{\prod_{i=1}^{\len{\vect{\overline y}_k}}{W_k(\overline y_{k,i}| x_i)}}{Q_k(\vect{\overline y}_k)}\IEEEeqnarraynumspace\\
 &=& i_k\farg{\x ; \mathbf{\overline y}_k}, \qquad  \x\in\mathcal{X}^\infty, \vect{\overline y}_k \in\preffree{\overline Y}_k^{(u)}
\end{IEEEeqnarray}
where $i_k(\x; \mathbf{\overline y}_k)$ was defined in \eqref{eq:mismatched_inf_def}.  Note that $i_k(\x; \vect{\overline y}_k)$ depends on $\x\in\mathcal{X}^\infty$ only through its first $\len{\vect{\overline y}_k}$ entries.
Set now
\begin{align}
  \gamma_k^{(u)} \triangleq \supp{\nu\in\mathbb{R}: \mathds{P}^{(u)}\mathopen{}\left[i_k(\X; \vect{\overline Y}_k) \leq \log \nu \right] \leq \epsilon_{k}^{(u)}+ \eta}\label{eq:converse_gamma2_eval}
\end{align}
where $\eta>0$ is arbitrary.
Using \eqref{eq:converse_gamma2_eval}, we conclude that 
\begin{multline}
   \mathds{P}^{(u)}\mathopen{}\left[i_k(\X;\vect{\overline Y}_k)<\log \gamma_k^{(u)} \right]\\
   \leq \epsilon_k^{(u)} + \eta \leq  \mathds{P}^{(u)}\mathopen{}\left[i_k(\X; \vect{\overline Y}_k) \leq \log \gamma_k^{(u)}\right].\label{eq:converse_delta_properties2_eval}
\end{multline}
Substituting the right-hand side of \eqref{eq:converse_delta_properties2_eval} in \eqref{eq:converse_logM2_eval}, we obtain
\begin{IEEEeqnarray}{rCl}
 \IEEEeqnarraymulticol{3}{l}{\log M }\nonumber\\
   &\leq&  \log \gamma^{(u)}_k- \log\left(\mathds{P}^{(u)}\mathopen{}\left[i_k(\X; \vect{\overline Y}_k) \leq \log \gamma^{(u)}_k\right]-\epsilon_k^{(u)} \right)\IEEEeqnarraynumspace \\
 &\leq&  \log \gamma_k^{(u)} - \log \eta.\label{eq:conve_logM_loglogM_eval}
\end{IEEEeqnarray}
Finally, the lemma is establishes by substituting \eqref{eq:conve_logM_loglogM_eval} into \eqref{eq:converse_delta_properties2_eval}, which yields
\begin{IEEEeqnarray}{rCl}
   \IEEEeqnarraymulticol{3}{l}{\mathds{P}^{(u)}\mathopen{}\left[i_k(\X ; \vect{\overline Y}_k) < \log(M\eta) \right]}\nonumber\\
   \qquad &\leq& \mathds{P}^{(u)}\mathopen{}\left[i_k(\X ; \vect{\overline Y}_k) < \log  \gamma_k^{(u)}  \right] \\
  &\leq&  \epsilon_{k}^{(u)}+\eta
  \label{eq:conve_const_before_tildeinf_eval}.
\end{IEEEeqnarray}

\section{Proof of Theorem~\ref{thm:converse_bound} (cardinality bound)}
\label{sec:converse_bound}

We simplify the minimization problem in \eqref{eq:conve_final_bound1} by showing that the minimum is also attained under the additional constraint that $|\mathcal{\overline U}|\leq K+1$. Define the $(K+1)$-dimensional region
\begin{multline}
  \mathcal{R} \triangleq \bigg\{ (\varepsilon_{1},\cdots,\varepsilon_{K}, L)\in [0,1]^K \times \mathbb{R}_+ : \\ L \geq \sum_{t=0}^\infty \left(1-L_t\left(\varepsilon_{1},\cdots,\varepsilon_{K}\right)\right) \bigg\}.\IEEEeqnarraynumspace\label{eq:region}
\end{multline}
Furthermore, let $\mathcal{R}_{\text{convex}}$ be the convex hull of $\mathcal{R}$.  Suppose that $\vect{p}_0$ lies on the lower convex envelope of $\mathcal{R}_{\text{convex}}$. Then we can write $\vect{p}_0$ as a convex combination of $I\in\mathbb{N}$ points in $\mathcal{R}$:
\begin{IEEEeqnarray}{rCl}
  \vect{p}_0 = \sum_{i=1}^I \alpha_i \vect{p}_i
\end{IEEEeqnarray}
where $\vect{p}_i \in\mathcal{R}$, $\alpha_i>0$, and $\sum_{i=1}^I \alpha_i=1$. Since $\vect{p}_0$ is a boundary point of $\mathcal{R}_{\text{convex}}$, there exists a supporting hyperplane $\{\vect{p}\in\mathbb{R}^{K+1} : \vect{a}^T \vect{p}  = \vect{a}^T \vect{p}_0 \}$ for some $\vect{a} \not= \vect{0}_{K+1}$, $\vect{a}\in\mathbb{R}^{K+1}$, with the property that $\vect{a}^T \vect{p} \leq \vect{a}^T \vect{p}_0$ for every $\vect{p} \in \mathcal{R}_{\text{convex}}$ \cite[pp.~50--51]{Boyd2004}. 
Now we note that the points  $\{\vect{p}_i\}$, $i\in\{1,\cdots,I\}$, must be on the supporting hyperplane $\{\vect{p}\in\mathbb{R}^{K+1} : \vect{a}^T \vect{p}  = \vect{a}^T \vect{p}_0 \}$ for every $i\in\{1,\cdots,I\}$. Indeed, suppose on the contrary that $\vect{a}^T \vect{p}_i < \vect{a}^T \vect{p}_0$ for some $i\in\{1,\cdots,I\}$. Then we have a contradiction:
\begin{IEEEeqnarray}{rCl}
  \vect{a}^T \vect{p}_0 =\sum_{i=1}^I \alpha_i \vect{a}^T \vect{p}_i < \sum_{i=1}^I \alpha_i \vect{a}^T \vect{p}_0 = \vect{a}^T \vect{p}_0.
\end{IEEEeqnarray}
Now, define the region
\begin{multline}
  \mathcal{R}_0 \triangleq \mathcal{R} \cap \Big\{\vect{p} =(p_1,\ldots,p_{K+1})\in\mathbb{R}^{K+1}: \\ \vect{a}^T \vect{p}  = \vect{a}^T \vect{p}_0    \text{ and } p_{K+1} \leq \max_{1 \leq i\leq I} p_{i,K+1}\Big\} 
\end{multline}
and the convex hull $\mathcal{R}_{0,\text{convex}}$ of $\mathcal{R}_0$. 
Here, $p_{i,K+1}$ denotes the $(K+1)$th entry of $\vect{p}_i$.
The region $\mathcal{R}_0$ is closed because $L_t(\cdot)$ is continuous in $\{\varepsilon_k^{(u)}\}$ and, since the $(K+1)$th entry of every point in $\mathcal{R}_0$ is bounded from above by $\max_{1\leq i \leq I} p_{i,K+1}$, the region is bounded in $\mathbb{R}^{K+1}$. 
This implies that $\mathcal{R}_0$ is compact. Moreover, $\vect{p}_0\in\mathcal{R}_{0,\text{convex}}$ because $\vect{p}_1,\cdots,\vect{p}_I \in\mathcal{R}_0$, and $\mathcal{R}_0$ lies in a $K$-dimensional affine subspace of $\mathbb{R}^{K+1}$. 
Hence, Caratheodory theorem \cite[Th.~15.3.5]{Cover2012} implies that $\vect{p}_0$ can be written as a convex combination of at most $K+1$ points in $\mathcal{R}_0$. But since $\mathcal{R}_0 \subseteq \mathcal{R}$, we can also  write $\vect{p}_0$ as a convex combination of at most $K+1$ points in $\mathcal{R}$.

Now, observe that the point
\begin{multline}
\Bigg(\E{\varepsilon_1^{(\bar U)}},\cdots,\E{\varepsilon_K^{(\bar U)}},\\ \E{\sum_{t=0}^\infty \left(1-L_t\left( \varepsilon_1^{(\bar U)},\cdots,\varepsilon_K^{(\bar U)}\right)\right)} \Bigg)
\end{multline}
evaluated for the RV $\overline U$ supported on the set $\mathcal{\bar U}$ and distributed as $P_{\bar U}$,  and for the $\{\varepsilon_k^{(\bar u)}\}$ that minimize \eqref{eq:conve_final_bound1}, is a boundary point of $\mathcal{R}_{\text{convex}}$.
By the above argument, we conclude that \eqref{eq:conve_final_bound1} is equal to
\begin{IEEEeqnarray}{rCl}
 \min_{\substack{P_{\bar U}\in\mathcal{P}(\mathcal{\bar U}), \varepsilon^{\bar u}_{k}\in[0,1]:\\ \E{\varepsilon_{k}^{(\bar U)}} \leq \epsilon + \eta}} \E{\sum_{t=0}^\infty \left(1-L_t\left(\varepsilon_{1}^{(\bar U)},\cdots,\varepsilon_{K}^{(\bar U)}\right)\right)}\label{eq:conve_final_bound}\IEEEeqnarraynumspace
 \end{IEEEeqnarray}
 where the RV $\bar U$ is supported on the set $\mathcal{\bar U}$ with $| \mathcal{\bar U}|\leq K+1$.

\section{Proof of Theorem~\ref{thm:asymp} (converse)}\label{sec:converse_asymptotics}
\begin{sloppypar}Fix a family of $(\ell, M,\epsilon)$-VLSF codes parameterized by the blocklength $\ell$. We shall assume that $\lim \inf_{\ell\rightarrow \infty} \log(M)/\ell>0$, that is, $M$ grows at least exponentially with $\ell$. If this does not occur, then the rightmost inequality in \eqref{eq:asymp_expansion} holds trivially. To establish the desired result, we analyze the nonasymptotic converse bound in Theorem~\ref{thm:converse_bound} in the limit $\log M \rightarrow \infty$. We shall set $\eta = (\log M)^{-1}$ and choose the auxiliary distributions $\{Q_k^{(\infty)}\}$ as follows. Let $x^t$, $t\in\mathbb{N}$, be an arbitrary $t$-dimensional vector in $\mathcal{X}^t$ and let $P_{x^t}\in\mathcal{P}(\mathcal{X})$ denote its type \cite[Def.~2.1]{Csiszar}. Furthermore, let $Q_{k,x^t}^{(\infty)}$ be the product distribution on $\mathcal{Y}_k^{\infty}$ generated by the marginal distribution $P_{x^t}W_k$. Finally, let $\mathcal{P}_t(\mathcal{X}) \subseteq \mathcal{P}(\mathcal{X})$ be the set of types of $t$-dimensional sequences. We choose $Q_k^{(\infty)}$ as follows:\end{sloppypar}
\begin{align}
 Q_k^{(\infty)}\farg{\vect{ y}} = \sum_{t=1}^{\lfloor \frac{2}{C}\log M \rfloor} \sum_{\substack{x^t\in\mathcal{X}^t:\\ P_{x^{t}} \in \mathcal{P}_{t}(\mathcal{X})}}\frac{Q_{k,x^t}^{(\infty)}(\vect{y})}{\lfloor \frac{2}{C}\log M \rfloor|\mathcal{P}_{t}(\mathcal{X})|} .\label{eq:Qinf_def}
\end{align}
Here, the inner sum is taken over the set of types of $t$-dimensional sequences and $C$ is the channel capacity given in \eqref{eq:capacity_def}. 
To keep notation compact, we set
\begin{IEEEeqnarray}{rCl}
  \tildeinf(x^n; y^n) \triangleq i_{P_{x^t},W_k}(x^n;y^n)\label{eq:asymp_conve_compact_inf_not}
\end{IEEEeqnarray}
which is defined for $n\leq t$ and for every $x^t\in\mathcal{X}^t$ and $y^n\in\mathcal{Y}_k^n$. 

We shall next summarize the key steps of the proof. These steps are analyzed in details in Sections~\ref{sec:disposing_of_maximum}--\ref{sec:lem63_first_order}.
\paragraph*{Step 1} We obtain an upper bound on $L_t(\vect{\varepsilon})$ in Theorem~\ref{thm:converse_bound} that does not involve any maximization over $n$ (the inner maximization in \eqref{eq:Lt_def}). Specifically, we show in Appendix~\ref{sec:disposing_of_maximum} that, whenever $t\leq \lfloor \frac{2}{C}\log M\rfloor$, we can dispose of this inner maximization by adding an error term of order $1/\log M$:
\begin{multline}
L_t(\vect{\varepsilon}) \\\leq   \max_{x^t\in\mathcal{X}^t} \prod_k \minn{1,\pr{\tildeinf(x^t;Y_k^t)\geq \lambda} +  \varepsilon_{k}} + \frac{2^K-1}{\lambda}.\label{eq:asymp_conve_key_step}
\end{multline}
Here,
\begin{multline}
\lambda \triangleq \log M - 2\log \log M -|\mathcal{X}| \log\mathopen{}\left(\frac{2}{C}\log M+1\right).\label{eq:lambda_def}
\end{multline}
and we denoted $L_t(\varepsilon_1,\cdots,\varepsilon_K)$ by $L_t(\vect{\varepsilon})$ with $\vect{\varepsilon}=[\varepsilon_1,\cdots,\varepsilon_K]$. 


\paragraph*{Step 2} We use \eqref{eq:asymp_conve_key_step} to lower-bound the right-hand side of \eqref{eq:conve_l_lower_bound} in Theorem~\ref{thm:converse_bound}. Since \eqref{eq:asymp_conve_key_step} holds only for $t\leq \lfloor \frac{2}{C} \log M\rfloor$, we must truncate the infinite sum in \eqref{eq:conve_l_lower_bound} as follows:
\begin{align}
 \sum_{t=0}^{\infty} (1- L_t(\vect{\varepsilon})) \geq \sum_{t=0}^{ \beta_{\vect{\varepsilon}}}(1- L_t(\vect{\varepsilon})) \label{eq:conve_proof_l_lower_bound_beta}
\end{align}
where $\beta_{\vect{\varepsilon}}\in\mathbb{N}$ is given by
\begin{IEEEeqnarray}{rCl}
\beta_{\vect{\varepsilon}}&\triangleq& \bigg\lfloor\frac{\lambda}{C} + \sqrt{\frac{\lambda V}{C^3}}\nu_{\vect{\varepsilon}}\bigg\rfloor.\label{eq:beta_eps_def}
\end{IEEEeqnarray}
Here, $V$ is defined in Theorem~\ref{thm:asymp} and $\nu_{\vect{\varepsilon}}$ 
is the solution of
\begin{align}
 \prod_{k} \left[Q\farg{ -\varrho_k   \nu_{\vect{\varepsilon}}}+ (1-2\delta_1)\varepsilon_k + \delta_1\right] = 1\label{eq:nu_eps_eq}
\end{align}
with $\delta_1\in (0,1/2)$ being an arbitrary constant that does not depend on $\lambda$. The role of $\delta_1$ is to ensure that $\nu_{\vect{\varepsilon}}$ is bounded from above and from below for all $\vect{\varepsilon}\in[0,1]^K$. We note that $\beta_{\vect{\varepsilon}} \leq \frac{2}{C}\log M$ for sufficiently large $M$ and for every $\vect{\varepsilon} \in [0,1]^K$. Note that we define $\beta_{\vect{\varepsilon}}$ as in \eqref{eq:beta_eps_def} instead of setting it equal $2/C \log M$ in order to control the error term originating from central limit theorem as we shall see later (see~\eqref{eq:asymp_conve_summing1}). Hence, since $\beta_{\vect{\varepsilon}} \leq \frac{2}{C}\log M$, we can use \eqref{eq:asymp_conve_key_step} to further lower-bound \eqref{eq:conve_proof_l_lower_bound_beta}. Note that $M \rightarrow \infty$ implies $\lambda \rightarrow \infty$. 

\paragraph*{Step 3} Next, we characterize the asymptotic behavior of the upper bound \eqref{eq:asymp_conve_key_step} in the limit $M\rightarrow \infty$. This will be used to provide an asymptotic lower bound on the right-hand side of \eqref{eq:conve_proof_l_lower_bound_beta}. 
It turns out convenient to subdivide the interval $[0,\beta_{\vect{\varepsilon}}]$ into  $K+2$ subintervals and to perform a different asymptotic analysis on each of the subintervals. Specifically, we set $[0,\beta_{\vect{\varepsilon}}] = \bigcup_{i=1}^{K+1} \mathcal{T}_i$ where $\mathcal{T}_i = [t_i,t_{i+1})$, $i\in\{0,\cdots,K\}$, and $\mathcal{T}_{K+1} = [t_{K+1},\beta_{\vect{\epsilon}}]$ with $t_0 = 0$ and 
\begin{IEEEeqnarray}{rCl}
t_i &\triangleq& \bigg\lfloor \frac{\lambda}{C_{i}} - \sqrt{\frac{V\lambda}{C^3}}\log \lambda\bigg\rfloor, \quad i\in\{1,\cdots,K\}\label{eq:ti_def}\\
t_{K+1}&\triangleq& \bigg\lfloor\frac{\lambda}{C}-\sqrt{\frac{V\lambda}{C^3}} \log \lambda\bigg\rfloor.
\end{IEEEeqnarray}
Recall that since $C_1 \geq C_2 \geq\cdots \geq C_K \geq C$ by assumption, we have that $t_0 \leq t_1 \leq \cdots \leq t_{K+1}$. Additionally, for sufficiently large $M$, we also have that $t_{K+1}<\beta_{\vect{\varepsilon}}$.

\begin{figure}[!t]
\begin{center}
\setlength\smallfigureheight{8.1cm}
\setlength\smallfigurewidth{8.1cm} 
\includegraphics{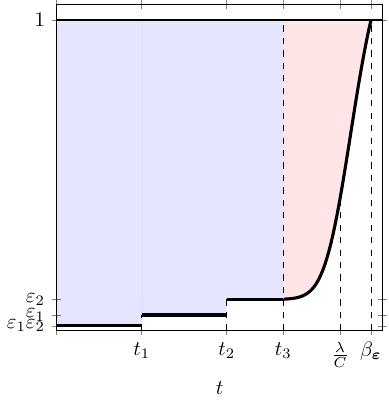}
\end{center} 
\caption{Our approach to lower-bounding \eqref{eq:conve_proof_l_lower_bound_beta}. For the case $K=2$ and $\vect{\varepsilon} = (0.05,0.10)$, the range of $t$ is divided into four subintervals: $[0,t_1)$, $[t_1,t_2)$, $[t_2,t_3)$, and $[t_3, \beta_{\vect{\varepsilon}}]$. The area of the blue shaded region depicts $\sum_{t=0}^{t_3-1} (1-L_t(\vect{\varepsilon}))$ while the area of the red shaded region depicts $\sum_{t=t_3}^{\beta_{\vect{\varepsilon}}} (1-L_t(\vect{\varepsilon}))$. The plotted curve represents a nonasymptotic upper bound on $L_t(\vect{\varepsilon})$ that is provided in Appendix~\ref{sec:large_deviation} and Appendix~\ref{sec:central_regime}.}
\label{fig:asymp_conve_regimes}
\end{figure}

In the first $K+1$ subintervals, we upper-bound \eqref{eq:asymp_conve_key_step} by means of a large-deviation analysis based on Hoeffding's inequality (see Appendix~\ref{sec:large_deviation}). In the last interval, our upper bound relies on Chebyshev's inequality and the Berry-Esseen central limit theorem (see Appendix~\ref{sec:central_regime}). These bounds are used to further lower-bound the right-hand side of \eqref{eq:conve_proof_l_lower_bound_beta}, as illustrated in Fig.~\ref{fig:asymp_conve_regimes}. 

We next summarize the asymptotic behavior of the bounds obtained in Sections~\ref{sec:large_deviation}--\ref{sec:central_regime}. 

To do so, it is convenient to introduce some notation that will allow us to keep our expressions compact. First, let $\rho>0$ and $\delta_2\in(0,1)$ be constants. For reasons that will become apparent later, we need $\rho$ to satisfy 
\begin{IEEEeqnarray}{rCl}
    \frac{1}{C_i} + \frac{i}{C_K} - \frac{(i+1) \rho}{C_1} > \frac{i}{C}, \quad i\in\{1,\cdots,K-1\}.\label{eq:rho_cond}\IEEEeqnarraynumspace
\end{IEEEeqnarray}
Note that the assumption \eqref{eq:thm_asymp_cond} ensures that one can find a $\rho$ that satisfies \eqref{eq:rho_cond}. 
Let 
\begin{IEEEeqnarray}{rCl}
d_0 &\triangleq& \frac{1}{C}-\frac{1}{C_{K}} + \frac{\rho}{C_1}\label{eq:d0_def}\\
d_1 &\triangleq& \frac{1}{C_1}(1-\rho)\\
d_i &\triangleq& \frac{1}{C_{i}}-\frac{1}{C_{i-1}}, \qquad  i\in \{2,\cdots,K\}\label{eq:di_def}
\end{IEEEeqnarray}
and define, for $\vect{\varepsilon}\in[0,1]^K$, the function\footnote{We use the convention that $\prod_{k\in\emptyset} a_k = 1$ for an arbitrary sequence $\{a_k\}$.}
\begin{IEEEeqnarray}{rCl}
f(\vect{\varepsilon})
&\triangleq&  \frac{1}{C}-d_0  \left(\max_k\varepsilon_{k}\right)-\sum_{i=1}^{K}d_i \left(\prod_{k \in \{i,\cdots,K\} }\varepsilon_k\right). \label{eq:feps_def}\IEEEeqnarraynumspace
\end{IEEEeqnarray}
%
%
This function is continuous in $\vect{\varepsilon}\in[0,1]^K$ and satisfies $f(\vect{0}_K)=1/C$ and $f(\vect{1}_K) = 0$. We shall also need the function $g_{\delta_1,\delta_2}(\vect{\varepsilon})$ given in \eqref{eq:geps_def} (its exact expression is not important for the level of detail provided in this section). 
This function is continuous in $\vect{\varepsilon}\in[0,1]^K$, $\delta_1>0$, and $\delta_2>0$, and has the following limits:
\begin{IEEEeqnarray}{rCl}
    \lim_{\delta_1 \rightarrow 0,\delta_2 \rightarrow 0}g_{\delta_1,\delta_2}(\vect{0}_K) &=& \sqrt{\frac{V}{C^3}}\E{\max_k H_k}\label{eq:limg0}\\
    \lim_{\delta_1 \rightarrow 0,\delta_2 \rightarrow 0} g_{\delta_1,\delta_2}(\vect{1}_K) &=& 0.\label{eq:limg1}
\end{IEEEeqnarray}
In Appendix~\ref{sec:large_deviation}, we prove the following asymptotic bound, which holds for every $\vect{\varepsilon}\in[0,1]^K$ and for sufficiently large $\lambda$:
\begin{multline}
 \sum_{t=0}^{t_{K+1}-1} \left(1-L_t(\vect{\varepsilon})\right)  \\\geq \lambda f(\vect{\varepsilon}) -\sqrt{\frac{\lambda V}{C^3}} \log(\lambda)\left(1-\max_k \varepsilon_k\right) - \const.\label{eq:conve_final_large2}
\end{multline}
This bound holds for every $\vect{\varepsilon}\in[0,1]^K$. Furthermore, in Appendix~\ref{sec:central_regime}, we provide the following asymptotic bound:
\begin{IEEEeqnarray}{rCl}
\IEEEeqnarraymulticol{3}{l}{\sum_{t= t_{K+1} }^{ \beta_{\vect{\varepsilon}} } (1-L_t(\vect{\varepsilon}))}\nonumber\\
 \quad&\geq& \sqrt{\lambda}g_{\delta_1,\delta_2}(\vect{\varepsilon})+\sqrt{\frac{\lambda V}{C^3}} \log(\lambda)\left(1-\max_k \varepsilon_k\right) \nonumber\\
 &&{} + \mathcal{O}(\log \lambda).\label{eq:asymp_conve_final3}\IEEEeqnarraynumspace
\end{IEEEeqnarray}
Here, the $\mathcal{O}(\log \lambda)$ term is uniform in $\vect{\varepsilon}\in[0,1]^K$.



%
By combining \eqref{eq:conve_proof_l_lower_bound_beta}, \eqref{eq:conve_final_large2}, and \eqref{eq:asymp_conve_final3}, we obtain for all $\vect{\varepsilon}\in[0,1]^K$ and for all sufficiently large $\lambda$ 
\begin{IEEEeqnarray}{rCl}
\IEEEeqnarraymulticol{3}{l}{\sum_{t=0}^{\infty}  \left(1-L_t(\vect{\varepsilon}) \right)}\nonumber\\
\quad &\geq& \sum_{t=0}^{\beta_{\vect{\varepsilon}}} \left(1-L_t(\vect{\varepsilon})\right)\\
 &\geq& \lambda f(\vect{\varepsilon}) - \sqrt{\frac{\lambda V}{C^3}} \log(\lambda)(1-\max_k \varepsilon_k)-\const \nonumber\\
 &&{}+ \sqrt{\lambda}g_{\delta_1,\delta_2}(\vect{\varepsilon}) + \sqrt{\frac{\lambda V}{C^3}} \log (\lambda)\left(1-\max_k \varepsilon_k\right) \nonumber\\
 &&{}+ \mathcal{O}(\log \lambda) \\
  &=& \lambda f(\vect{\varepsilon}) +\sqrt{\lambda}g_{\delta_1,\delta_2}(\vect{\varepsilon}) + \mathcal{O}(\log \lambda).\label{eq:asymp_conve_sum_t0_inf_bound}
    \end{IEEEeqnarray}
Again, we note that the $\mathcal{O}(\log \lambda)$ term in \eqref{eq:asymp_conve_sum_t0_inf_bound} is uniform in $\vect{\varepsilon}$.

\paragraph*{Step 4} We are left with solving the minimization in \eqref{eq:conve_l_lower_bound}. Specifically, we need to evaluate $\min \Big\{ \lambda f(\vect{\varepsilon}^{(U)}) + \sqrt{\lambda} g_{\delta_1,\delta_2}(\vect{\varepsilon}^{(U)}) \Big\}$, where the minimization is over all $\{\vect{\varepsilon}^{(u)}\}_{u\in\mathcal{U}}$ and all probability distributions $P_U\in\mathcal{P}(\mathcal{U})$ subject to $\EE{U}{\vect{\varepsilon}^{(u)}} \leq \epsilon + (\log M)^{-1}$. To do so, we rely on \cite[Lem.~63]{Polyanskiy2010b} which is repeated here for convenience.
\begin{lemma}[{{\cite[Lem.~63]{Polyanskiy2010b}}}]
\label{lem:lemma63}
    Let $D$ be a compact metric space. Suppose $f: D \mapsto \mathbb{R}$ and $g: D \mapsto \mathbb{R}$ are continuous. Define 
    \begin{IEEEeqnarray}{rCl}
        f^* \triangleq \max_{x\in D} f(x)
    \end{IEEEeqnarray}
    and 
    \begin{IEEEeqnarray}{rCl}
        g^* \triangleq \sup_{x: f(x) = f^*} g(x).
    \end{IEEEeqnarray}
    Then,
    \begin{IEEEeqnarray}{rCl}
        \max_{x\in D} \left[n f(x) + \sqrt{n} g(x)\right] = n f^* + \sqrt{n}g^* +o(\sqrt{n}).\IEEEeqnarraynumspace
    \end{IEEEeqnarray}
\end{lemma}
  As a first step, we show in Appendix~\ref{sec:lem63_first_order} that
\begin{multline}
\min_{\substack{P_U\in\mathcal{P}(\mathcal{U}),\vect{\varepsilon}^{(u)}\in[0,1]^K:\\ \EE{U}{\varepsilon_{k}^{(U)}}\leq \epsilon+(\log M)^{-1} }} \EE{U}{f(\vect{\varepsilon}^{(U)})} 
\\= \frac{1-\epsilon- (\log M)^{-1}}{C}.\label{eq:asymp_conve_first_order_opt} 
\end{multline}
and that the set of minimizers of the left-hand side of \eqref{eq:asymp_conve_first_order_opt} is given by
\begin{IEEEeqnarray}{rCl}
  \mathcal{G} &\triangleq& \bigg\{ \big(P_U,\{\vect{\varepsilon}^{(u)}\}\big) : \EE{U}{\varepsilon_k^{(U)}} = \epsilon+(\log M)^{-1} \text{ and }\nonumber\\
  && \quad P_U(u)>0 \Rightarrow \vect{\varepsilon}^{(u)} \in\{ \vect{0}_K\} \cup \{\vect{1}_K\} \text{ for } u\in\mathcal{U} \bigg\}.\IEEEeqnarraynumspace\label{eq:asymp_conve_set_of_minimizers}
\end{IEEEeqnarray}
Next, it follows from \eqref{eq:asymp_conve_set_of_minimizers} that
\begin{IEEEeqnarray}{rCl}
   \IEEEeqnarraymulticol{3}{l}{\min_{(P_U,\{\vect{\varepsilon}^{(u)}\})\in\mathcal{G}} \EE{U}{g_{\delta_1,\delta_2}(\vect{\varepsilon}^{U})}}\nonumber\\
    &=&  \min_{(P_U,\{\vect{\varepsilon}^{(u)}\})\in\mathcal{G}}\sum_{u\in\mathcal{U}: P_U(u)>0 } P_U(u)g_{\delta_1,\delta_2}(\vect{\varepsilon}^{u})\\
    &=&\min_{(P_U,\{\vect{\varepsilon}^{(u)}\})\in\mathcal{G}} \Bigg[g_{\delta_1,\delta_2}(\vect{0}_K)\sum_{\substack{u\in\mathcal{U}:\\ P_U(u)>0, \vect{\varepsilon}^{(u)}=\vect{0}_K }} P_U(u)\nonumber\\
    &&{}\quad\qquad\qquad\quad+g_{\delta_1,\delta_2}(\vect{1}_K)\sum_{\substack{u\in\mathcal{U}: \\ P_U(u)>0 , \vect{\varepsilon}^{(u)}=\vect{1}_K}} P_U(u)\Bigg]\IEEEeqnarraynumspace\\
    &=&   g_{\delta_1,\delta_2}(\vect{0}_K)\left(1-\epsilon-(\log M)^{-1} \right)\nonumber\\
    &&{}+ g_{\delta_1,\delta_2}(\vect{1}_K)(\epsilon + (\log M)^{-1}).\label{eq:asymp_conve_second_order_opt}
\end{IEEEeqnarray}
By \eqref{eq:conve_l_lower_bound} and \eqref{eq:asymp_conve_sum_t0_inf_bound}, we have that
\begin{IEEEeqnarray}{rCl}
\ell
&\geq& \min_{\substack{P_U,\vect{\varepsilon}^{(u)}:\\ \EE{U}{\varepsilon_{k}^{(U)}}\leq \epsilon+(\log M)^{-1} }}  \EE{U}{\lambda f(\vect{\varepsilon}^{(U)}) +  \sqrt{\lambda}g_{\delta_1,\delta_2}(\vect{\varepsilon}^{(U)}) } \nonumber\\
&&{}+ \mathcal{O}(\log \lambda) \IEEEeqnarraynumspace\label{eq:asymp_conve_optimization_problem}
\end{IEEEeqnarray}
where the $\mathcal{O}(\log\lambda)$ term is uniform in $\vect{\varepsilon}^{(u)}$.
We observe that the set of minimizers in \eqref{eq:asymp_conve_set_of_minimizers} and the Euclidean norm form a compact metric space. It follows from Lemma~\ref{lem:lemma63} that
\begin{IEEEeqnarray}{rCl}
\ell
&\geq& \min_{\substack{P_U,\varepsilon_{1}^{(u)},\varepsilon_{2}^{(u)}:\\ \EE{U}{\varepsilon_{k}^{(U)}}\leq \epsilon+(\log M)^{-1} }}  \EE{U}{  \lambda f(\vect{\varepsilon}^{(U)}) +  \sqrt{\lambda}g_{\delta_1,\delta_2}(\vect{\varepsilon}^{(U)}) }\nonumber\\
&&{} + \mathcal{O}(\log \lambda) \\
&=&  \frac{\lambda(1-\epsilon - (\log M)^{-1})}{C}\nonumber\\
&&{} + \sqrt{\lambda}\min_{(P_U, \{\vect{\varepsilon}\}) \in \mathcal{G}} \EE{U}{g_{\delta_1,\delta_2}(\vect{\varepsilon}^{(U)})}  +\mathcal{O}(\log \lambda) \label{eq:lem63_invoked}\\
&= & \frac{\lambda(1-\epsilon - (\log M)^{-1})}{C}+ \sqrt{\lambda} g_{\delta_1,\delta_2}(\vect{0}_K)\left(1-\epsilon-\frac{1}{\log M}\right)  \nonumber\\
& & {}+ \sqrt{\lambda}g_{\delta_1,\delta_2}(\vect{1}_K)\left(\epsilon+(\log M)^{-1}\right)+ \mathcal{O}(\log \lambda)\label{eq:second_order_opt}\\
  &=& \frac{\lambda (1-\epsilon)}{C} +  \sqrt{\lambda} g_{\delta_1,\delta_2}(\vect{0}_K)(1-\epsilon)  +\sqrt{\lambda}g_{\delta_1,\delta_2}(\vect{1}_K) \epsilon \nonumber\\
  &&{}+ \mathcal{O}(\log \lambda).\label{eq:last_opt_step}
\end{IEEEeqnarray}
Here, \eqref{eq:second_order_opt} follows from \eqref{eq:asymp_conve_first_order_opt} and in \eqref{eq:last_opt_step} we have used that $\lambda (\log M)^{-1} = \mathcal{O}(1)$.
Recall that $g_{\delta_1,\delta_2}(\vect{0}_K)$ and $g_{\delta_1,\delta_2}(\vect{1}_K)$ are continuous in $\delta_1>0$ and $\delta_2>0$, and that we have the limits \eqref{eq:limg0} and \eqref{eq:limg1}.
Hence, by choosing $\delta_1$ and $\delta_2$ arbitrarily small, we obtain the inequality
\begin{IEEEeqnarray}{rCl}
\ell \geq  \frac{\lambda (1-\epsilon)}{C} + (1-\epsilon)\sqrt{\frac{\lambda V}{C^3}} \EBig{\max_k \tvRV_k } + o(\sqrt{\lambda}).\IEEEeqnarraynumspace
\end{IEEEeqnarray}
Here, recall that $\{H_k\}$ have cumulative distribution function given in \eqref{eq:Wk_def}.
Finally, using the definition of $\lambda$ in \eqref{eq:lambda_def}, we conclude that
\begin{IEEEeqnarray}{rCl}
  \log M \leq \frac{\ell C}{1-\epsilon} -  \sqrt{\frac{\ell V}{1- \epsilon}} \E{\max_k \tvRV_k} + o(\sqrt{\ell})
\end{IEEEeqnarray}
which establishes the desired result.

\subsection{Disposing of the maximum in \eqref{eq:Lt_def}}
\label{sec:disposing_of_maximum}
By \eqref{eq:Qk_def1}, we have that for all $\vect{\bar y}\in\preffree{Y}_k$,
\begin{IEEEeqnarray}{rCl}
\IEEEeqnarraymulticol{3}{l}{Q_k(\vect{\bar y})}\nonumber\\
 &=& \sum_{\substack{ \vect{y}\in\mathcal{Y}_k^\infty:\\ \vect{\bar y} = [y_1,\cdots,y_{\len{\bar y}}]}} \!\!\!\!\! \sum_{t=1}^{\lfloor \frac{2}{C}\log M \rfloor} \!\!\!\!\! \sum_{P_{x^{t}} \in \mathcal{P}_{t}(\mathcal{X})} \frac{Q_{k,x^t}^{(\infty)}(\vect{y})}{\lfloor \frac{2}{C}\log M \rfloor|\mathcal{P}_{t}(\mathcal{X})|} \label{eq:Qk_bary_def}\IEEEeqnarraynumspace\\
&=&\!\sum_{t=1}^{\lfloor \frac{2}{C}\log M \rfloor} \!\!\!\!\!\sum_{P_{x^{t}} \in \mathcal{P}_{t}(\mathcal{X})}\!\frac{1}{\lfloor \frac{2}{C}\log M \rfloor|\mathcal{P}_{t}(\mathcal{X})|}\! \prod_{i=1}^{\len{\vect{\bar y}}} P_{x^t}W_k(\bar y_{i}).\IEEEeqnarraynumspace
\end{IEEEeqnarray}
Let now $\tilde t$ be an integer no larger than $\frac{2}{C}\log M$. Using that $Q_k$ is a convex combination of measures on $\preffree{Y}_k$, we obtain the following relation between $i_{k}(x^t;y_k^t)$ and $i_{P_{x^{\tilde t}}, W_k}(x^t;y_k^t)$
\begin{IEEEeqnarray}{rCl}
  \IEEEeqnarraymulticol{3}{l}{i_{k}(x^t; y_k^t)}\nonumber\\
   \quad &=&  \log \frac{W_k^t(y_k^t|x^t)}{Q_k(y_k^t)} \\
  &\leq& \log \frac{W_k^t(y_k^t|x^t)}{\frac{1}{\lfloor \frac{2}{C}\log M\rfloor |\mathcal{P}_{\tilde t}(\mathcal{X})|}P_{x^{\tilde t}}W_k^{t}(y_k^{ t}) }\label{eq:convex_imath_rel1}\\
  &\leq&  i_{P_{x^{\tilde t}},W_k}(x^t; y_k^t) + |\mathcal{X}| \log\mathopen{}\left(\frac{2}{C}\log M+1\right).\label{eq:convex_imath_rel} 
\end{IEEEeqnarray}
In \eqref{eq:convex_imath_rel1}, $P_{x^{\tilde t}} W_k^t(y_k^t) \triangleq (P_{x^{\tilde t}} W_k)^t(y_k^t)$ denotes the product distributions induced on $\mathcal{Y}_k^t$ by the probability distribution $(P_{x^{\tilde t}})^t$. The inequality in \eqref{eq:convex_imath_rel1} follows because the logarithm is monotonically increasing and because the $\{P_{x^{\tilde t}} W_k^t(y_k^t)\}$ are nonnegative. Finally, \eqref{eq:convex_imath_rel} follows because $|\mathcal{P}_{t}\farg{\mathcal{X}}|\leq(t+1)^{|\mathcal{X}|}$ \cite[Th.~11.1.1]{Cover2012}. 
We can now upper-bound $L_t(\vect{\varepsilon})$ in \eqref{eq:Lt_def} for $t\leq \frac{2}{C}\log M$, where $\vect{\varepsilon} \triangleq [\varepsilon_1,\cdots,\varepsilon_K]$, as follows. Let $\tilde \lambda$ be defined as
\begin{IEEEeqnarray}{rCl}
\tilde \lambda \triangleq \log M - \log \log M -|\mathcal{X}| \log\mathopen{}\left(\frac{2}{C}\log M+1\right).\label{eq:lambda_tilde_def}\IEEEeqnarraynumspace
\end{IEEEeqnarray} Then,
\begin{IEEEeqnarray}{rCl}
  \IEEEeqnarraymulticol{3}{l}{L_t(\vect{\varepsilon})}\nonumber\\
    &=&\max_{x^t\in\mathcal{X}^t} \prod_{k} \minnbigg{1,\nonumber\\
    &&{}\qquad\quad \pr{\max_{0\leq n \leq t}i_k(x^n; Y_k^n)\geq \log M + \log \eta} + \varepsilon_k }\IEEEeqnarraynumspace\\
&\leq& \max_{x^t\in\mathcal{X}^t} \prod_{k} \minnbigg{1,\nonumber\\
&&{}\qquad\quad\pr{\max_{0\leq n \leq t}i_{P_{x^t}, W_k}(x^n; Y_k^n)\geq \tilde \lambda} + \varepsilon_k }\label{eq:asymp_conve_Lt_max2}\\
&=& \max_{x^t\in\mathcal{X}^t} \prod_{k} \minn{1,\pr{\max_{0\leq n \leq t} \tilde \imath_k(x^n; Y_k^n)\geq \tilde \lambda} + \varepsilon_k }.\label{eq:asymp_conve_Lt_max}
\end{IEEEeqnarray}
In \eqref{eq:asymp_conve_Lt_max2}, we have used \eqref{eq:convex_imath_rel} and that $\eta = (\log M)^{-1}$, and in \eqref{eq:asymp_conve_Lt_max}, we have used \eqref{eq:asymp_conve_compact_inf_not}. 

Fix a positive constant $\nu$. We dispose of the inner maximization in \eqref{eq:asymp_conve_Lt_max} through the steps \eqref{eq:asymp_conve_max_cup0}--\eqref{eq:asymp_conve_max_cup}, shown in the top of the next page.\begin{figure*}[!t]
\normalsize
\setcounter{MYtempeqncnt}{\value{equation}}
\setcounter{equation}{203}
\begin{IEEEeqnarray}{rCl}
\pr{\max_{0\leq n \leq t}\tildeinf(x^n; Y_k^n)\geq \tilde \lambda}
 &=& \prBigg{\bigcup_{n = 0}^t \left\{\tildeinf(x^n; Y_k^n)\geq \tilde \lambda\right\}}\label{eq:asymp_conve_max_cup0}\\
 &\leq& \prBigg{\left\{\tildeinf(x^t; Y_k^t)\geq \tilde \lambda\right\}\cup\bigcup_{n=0}^{t-1} \bigg(\left\{\tildeinf(x^t; Y_k^t)\geq \tilde \lambda-\nu\right\} \cup\Big\{\tildeinf(x_{n+1}^t;Y_{k,n+1}^t) \leq -\nu\Big\}\bigg) }\label{eq:asymp_conve_max_cup1}\IEEEeqnarraynumspace\\
  &=& \prBigg{\left\{\tildeinf(x^t; Y_k^t)\geq \tilde \lambda-\nu\right\}\cup\bigcup_{n=1}^{t}  \left\{\tildeinf(x_n^t;Y_{k,n}^t) \leq -\nu\right\} }\\
&\leq& \pr{\tildeinf(x^t; Y_k^t)\geq \tilde \lambda - \nu}+ \pr{\bigcup_{n=1}^{t} \left\{ \tildeinf(x_{n}^t; Y_{k,n}^t) \leq -\nu \right\} }.\label{eq:asymp_conve_max_cup}\IEEEeqnarraynumspace
\end{IEEEeqnarray}

\setcounter{equation}{\value{MYtempeqncnt}}
\hrulefill
\vspace*{4pt}
\end{figure*}\addtocounter{equation}{4} In \eqref{eq:asymp_conve_max_cup1}, we denoted the last $t-n$ entries of $Y_k^t$ by $Y_{k,n+1}^t$. The inequality in \eqref{eq:asymp_conve_max_cup1} holds because $\tilde \imath_k(x^t; Y_{k}^t) \geq \tilde \lambda -\nu$ and $\tilde \imath_k(x_{n+1}^t; Y_{k,n+1}^t) \leq -\nu$ imply that $\tilde \imath_k(x^n; Y^n_k) = \tilde \imath_k(x^t; Y_{k}^t) - \tilde \imath_k(x_{n+1}^t; Y_{k,n+1}^t) \geq \tilde\lambda$ for $n\in\{0,\cdots,t-1\}$. Finally, we have used the union bound in \eqref{eq:asymp_conve_max_cup}.
Define now
\begin{multline}
   \tilde \tau_k(x^t, y^t)\\ \triangleq \maxx{ \{0\} \cup \{1\leq n \leq t: \tildeinf(x_{n}^t; y_{n}^t) \leq -\nu\} }.
\end{multline}
This definition implies the following: if $\tilde \tau(x^t, y^t)=n$, then for all sequences $\tilde y^t \in\mathcal{Y}_k^t$ such that $\tilde y_n^t = y_n^t$, we also have that $\tilde \tau_k(x^t, \tilde y^t)=n$.

Let $\bar y^t_k\in\mathcal{Y}_k^t$ be arbitrary vectors for $k\in\mathcal{K}$. We can now upper-bound the second term in \eqref{eq:asymp_conve_max_cup} as follows
\begin{IEEEeqnarray}{rCl}
\IEEEeqnarraymulticol{3}{l}{\pr{\bigcup_{n=1}^{t} \left\{ \tildeinf(x_{n}^t; Y_{k,n}^t) \leq -\nu \right\}} }\nonumber\\
 &=&\sum_{n=1}^t \sum_{\substack{y^t\in\mathcal{Y}_k^t:\\ \tilde \tau_k(x^t,y^t) = n}} W^t_k(y^t|x^t)  \\
&=& \sum_{n=1}^t \sum_{\substack{y^t\in\mathcal{Y}_k^t:\\ \tilde \tau_k(x^t,y^t) = n}}  \nonumber\\
&&\qquad{}\frac{ \prod_{i=n}^t W_{k}(y_i|x_i)}{\prod_{i=n}^t P_{x^t}W_k(y_i)} \frac{\prod_{i=n}^t P_{x^t}W_k(y_i)}{\prod_{i=n}^t W_{k}(y_i|x_i)}W_k^t(y^t|x^t)\IEEEeqnarraynumspace\\
&\leq&  \ee{-\nu}  \sum_{n=1}^t \sum_{\substack{y^t\in\mathcal{Y}_k^t:\\ \tilde \tau_k(x^t,y^t) = n}}\Bigg( \prod_{i=n}^t P_{x^t}W_{k}(y_i)\Bigg)\nonumber\\
&&\qquad\quad\qquad\qquad\qquad\qquad{}\times W_k^{n-1}(y^{n-1}|x^{n-1})\label{eq:asymp_conve_exp_nu_bound}\\
&=&  \ee{-\nu}  \sum_{n=1}^t \sum_{\substack{y^t\in\mathcal{Y}_k^t: \\  \bar y_k^{n-1}=y^{n-1} \\ \tilde \tau_k(x^t,y^t) = n}} \prod_{i=n}^t P_{x^t}W_{k}(y_i)\label{eq:asymp_conve_max_sum}\\
&=&  \ee{-\nu}  \sum_{n=1}^t  P_{x^t}W_k^t[ \tilde \tau_k(x^t, Y^t) = n]\label{eq:asymp_conve_measure}\\
&\leq& \ee{-\nu}.\label{eq:asymp_conve_exp_bound}
\end{IEEEeqnarray}
Here, \eqref{eq:asymp_conve_exp_nu_bound} holds because of \eqref{eq:asymp_conve_compact_inf_not} and because $\tilde \imath_k(x^t_n; y^t_n) \leq -\nu$ for every $y^t$ such that $\tilde \tau_k(x^t, y^t) = n$; \eqref{eq:asymp_conve_max_sum} holds because $\{\bar y^t_k\}_{k\in\mathcal{K}}$ are fixed arbitrary vectors and because all sequences $\tilde y^t$ whose last $t-n+1$ entries coincide with the ones of $y^t$ also satisfy $\tilde \tau_k(x^t, \tilde y^t)=n$.

By substituting \eqref{eq:asymp_conve_exp_bound} into \eqref{eq:asymp_conve_max_cup}, by choosing $\nu =\log \log M$, and by recalling the definition of $\lambda$ in \eqref{eq:lambda_def}, we conclude that 
\begin{multline}
\pr{\max_{0\leq n \leq t}\tilde \imath_k(x^n; Y_k^n)\geq \tilde \lambda} \\\leq \pr{\tilde \imath_k(x^t; Y_k^t)\geq  \lambda} + \frac{1}{\log M}.\label{eq:asymp_conve_max_bound_conclusion}
\end{multline}
It follows from \eqref{eq:asymp_conve_Lt_max}, from \eqref{eq:asymp_conve_max_bound_conclusion}, and from the inequality $(\log M)^{-1} \leq 1/\lambda$ that for all $t\leq \frac{2}{C} \log M$, we have
\begin{IEEEeqnarray}{rCl}
\IEEEeqnarraymulticol{3}{l}{L_t(\vect{\varepsilon})}\nonumber\\
 &\leq& \max_{x^t \in\mathcal{X}^t} \prod_k  \minn{1,\pr{\tilde \imath_k(x^t; Y_k^t)\geq  \lambda } + \frac{1}{\log M} + \varepsilon_k} \\
 &\leq& \max_{x^t \in\mathcal{X}^t} \prod_k \left( \minn{1,\pr{\tilde \imath_k(x^t; Y_k^t)\geq  \lambda }  + \varepsilon_k}+\frac{1}{\lambda}\right) \label{eq:Ltvareps_second_to_last_step}\\
&\leq& \max_{x^t\in\mathcal{X}^t} \prod_k \minn{1,\pr{\tildeinf(x^t;Y_k^t)\geq \lambda} +  \varepsilon_{k}} + \frac{2^K-1}{\lambda}.\label{eq:Ltvareps_last_step}\IEEEeqnarraynumspace
\end{IEEEeqnarray}
In \eqref{eq:Ltvareps_last_step}, we have expanded the product in \eqref{eq:Ltvareps_second_to_last_step} into $2^K$ terms and used that $\minn{1,\pr{\tildeinf(x^t;Y_k^t)\geq \lambda}}\leq 1$. This establishes \eqref{eq:asymp_conve_key_step}.

\subsection{Large-deviation analysis}\label{sec:large_deviation}
To apply Hoeffding's inequality for all $t$ within the first $K+1$ intervals, we use that 
\begin{IEEEeqnarray}{rCl}
b \triangleq \max_k\max_{x\in\mathcal{X}} \max_{y\in\mathcal{Y}_k(x)} |\tildeinf(x,y)|\label{eq:b_def}
\end{IEEEeqnarray}
is finite. Here, $\mathcal{Y}_k(x)$ denotes the support of $W_k(\cdot|x)$. We shall first treat the first $K$ intervals and shortly return to the interval $\mathcal{T}_K$ for which we need the additional property that $\min_k I_k(P)\leq C$ for every $P\in\mathcal{P}(\mathcal{X})$.  We first obtain the following large-deviation bound, which holds for all $t\in\mathcal{T}_i$, $i\in\{0,\cdots,K-1\}$ and $k \in\{i+1,\cdots,K\}$: 
\begin{IEEEeqnarray}{rCl}
 \IEEEeqnarraymulticol{3}{l}{\max_{x^t \in\mathcal{X}^t}\pr{\tildeinf(x^t ; Y_{k}^t) \geq \lambda}}\nonumber\\
  &=& \max_{x^t \in\mathcal{X}^t}\pr{\frac{\tildeinf(x^t;Y^t_{k})}{t}-  I_{k}(P_{x^t})\geq \frac{\lambda}{t} - I_{k}(P_{x^t}) } \label{eq:hoeffding_step0}\\
  &\leq& \max_{x^t \in\mathcal{X}^t} \ee{ -2t^2 \frac{\left(\lambda/t - I_k(P_{x^t})\right)^2}{4 t b^2} }\label{eq:hoeffding_step}\\
    &=& \max_{x^t \in\mathcal{X}^t} \ee{ -\frac{1}{2b^2} \left(\frac{\lambda -t I_k(P_{x^t})}{\sqrt{t}}  \right)^2 }\\
    &\leq& \max_{x^t \in\mathcal{X}^t} \ee{ -\frac{1}{2b^2} \left(\frac{\lambda -t_{i+1} I_k(P_{x^t})}{\sqrt{t_{i+1}}}  \right)^2 }\label{eq:hoeffding_step2}\\
    &\leq&  \ee{ -\frac{1}{2b^2} \left(\frac{\lambda -t_{i+1} C_{i+1}}{\sqrt{t_{i+1}}}  \right)^2 }\label{eq:hoeffding_step3}\\
    &\leq&   \ee{ -\frac{1}{2b^2} \left(\frac{C_{i+1}\sqrt{V \lambda/C^3}\log \lambda }{\sqrt{\lambda/C_{i+1} - \sqrt{V \lambda/C^3}\log \lambda}}  \right)^2 }\label{eq:hoeffding_step5}\IEEEeqnarraynumspace\\
        &\leq&  \eeBigg{ -\frac{1}{2b^2} \left(\frac{C_{i+1}\sqrt{V \lambda/C^3}\log \lambda }{\sqrt{\lambda/C_{i+1}}}  \right)^2 }\label{eq:hoeffding_step6}\\
        &=& \eeBig{ -\frac{1}{2b^2} \left(C_{i+1}^{3/2}\sqrt{V/C^3}\log \lambda \right)^2 }\label{eq:hoeffding_step7}\\
                &=&  \eeBig{ -\underbrace{\frac{V C_{i+1}^{3}}{2b^2 C^3}}_{\triangleq c_{1i}} \log^2 \lambda  }\\
  &=&  \left(\frac{1}{\lambda}\right)^{c_{1i}  \log \lambda}.\label{eq:hoeffding_bound}
\end{IEEEeqnarray}
Here, \eqref{eq:hoeffding_step} follows from Hoeffding's inequality \cite[Th.~2]{Hoeffding1963} and from \eqref{eq:b_def}, \eqref{eq:hoeffding_step2} follows because $(\lambda - t I_k(P_{x^t}))/\sqrt{t}$, for a fixed distribution $P_{x^t}$, is a nonincreasing function of $t$ and because $t< t_{i+1}$, \eqref{eq:hoeffding_step3} follows because $I_k(P_{x^t})$ is uniformly upper-bounded by $C_{i+1}$ for $k\in\{i+1,\cdots,K\}$ (recall that we assumed $C_1 \geq \cdots \geq C_K$), \eqref{eq:hoeffding_step5} follows from \eqref{eq:ti_def}, and \eqref{eq:hoeffding_step6}--\eqref{eq:hoeffding_bound} follow from algebraic manipulations.

Next, we consider the interval $\mathcal{T}_K$. Fix a probability distribution $P\in\mathcal{P}(\mathcal{X})$, and let
\begin{IEEEeqnarray}{rCl}
\kappa(P)\triangleq \argmin_k I_k(P).\label{eq:kappa_def}
\end{IEEEeqnarray}
Note that $I_{\kappa(P)}(P)\leq C$ for every $P\in \mathcal{P}(\mathcal{X})$. Let also $c_{1K} \triangleq V/(2b^2)$. We have the following bound for all $t\in\mathcal{T}_K$
\begin{IEEEeqnarray}{rCl}
\IEEEeqnarraymulticol{3}{l}{ \max_{x^t \in\mathcal{X}^t} \min_k \pr{\tildeinf(x^t ; Y_k^t) \geq \lambda}}\nonumber\\
\quad &\leq& \max_{x^t \in\mathcal{X}^t} \min_k \ee{ -\frac{1}{2b^2} \left(\frac{\lambda -t_{K+1} I_k(P_{x^t})}{\sqrt{t_{K+1}}}  \right)^2 }\label{eq:hoeffding2_step1}\IEEEeqnarraynumspace\\ 
 &\leq& \max_{x^t \in\mathcal{X}^t} \ee{ -\frac{1}{2b^2} \left(\frac{\lambda -t_{K+1} I_{\kappa(P_{x^t})}(P_{x^t})}{\sqrt{t_{K+1}}}  \right)^2 }\\
  &\leq& \max_{x^t \in\mathcal{X}^t}  \ee{ -\frac{1}{2b^2} \left(\frac{\lambda -t_{K+1} C}{\sqrt{t_{K+1}}}  \right)^2 }\label{eq:hoeffding2_step2}\\
 &\leq& \left(\frac{1}{\lambda}\right)^{ c_{1K} \log \lambda}.\label{eq:hoeffding_bound2}
\end{IEEEeqnarray}
Here, \eqref{eq:hoeffding2_step1} follows from steps similar to \eqref{eq:hoeffding_step0}--\eqref{eq:hoeffding_step2}, \eqref{eq:hoeffding2_step2} holds because $I_{\kappa(P)}(P) \leq C$ for every $P\in\mathcal{P}(\mathcal{X})$, and \eqref{eq:hoeffding_bound2} follows from steps similar to \eqref{eq:hoeffding_step3}--\eqref{eq:hoeffding_bound}.
Using \eqref{eq:hoeffding_bound}, we conclude that for all $i\in\{0,\cdots,K-1\}$
\begin{IEEEeqnarray}{rCl}
 \IEEEeqnarraymulticol{3}{l}{\sum_{t=t_i}^{t_{i+1}-1} L_t(\vect{\varepsilon}) }\nonumber\\
 \quad&\leq&\sum_{t= t_{i} }^{ t_{i+1}-1  } \max_{x^t \in\mathcal{X}^t}\prod_k \minn{1, \pr{\tildeinf(x^t;Y_k^t)\geq\lambda} +  \varepsilon_{k}} \nonumber\\
 &&{}+ \mathcal{O}(1) \\
 &\leq&\sum_{t= t_{i} }^{ t_{i+1}-1  } \max_{x^t \in\mathcal{X}^t}\prod_{k=i+1}^K\left( \pr{\tildeinf(x^t;Y_k^t)\geq\lambda} +  \varepsilon_{k}\right) \nonumber\\
 &&{}+ \mathcal{O}(1)\\
  &\leq&\sum_{t= t_{i} }^{ t_{i+1}-1  }\prod_{k=i+1}^K\left(  \lambda^{-c_{1i} \log \lambda} +  \varepsilon_{k}\right) + \mathcal{O}(1)\label{eq:asymp_conve_key_step0}\\
 &\leq& (t_{i+1} - t_{i})\prod_{k\in\{i+1,\cdots,K\}}\varepsilon_{k}+ \mathcal{O}(1)\label{eq:asymp_conve_key_step1}
\end{IEEEeqnarray}
as $\lambda\rightarrow \infty$. Here, \eqref{eq:asymp_conve_key_step1} follows because $(t_{i+1}-t_i )\lambda^{-c_{1i}\log \lambda} \leq \const \lambda^{-c_{1i}\log \lambda+1} = o(1)$ as $\lambda\rightarrow \infty$. 
Similarly, it follows from \eqref{eq:hoeffding_bound2} that
\begin{IEEEeqnarray}{rCl}
\IEEEeqnarraymulticol{3}{l}{\sum_{t=t_K}^{t_{K+1}-1} L_t(\vect{\varepsilon}) }\nonumber\\
&\leq&\sum_{t=t_{K}}^{ t_{K+1}-1 }\max_{ x^t\in\mathcal{X}^t} \prod_{k} \minn{1,\pr{\tildeinf(x^t;Y_k^t)\geq \lambda} +  \varepsilon_{k} }\nonumber\\
&&{}+ \mathcal{O}(1)\IEEEeqnarraynumspace\\
&\leq& \sum_{t=t_{K}}^{ t_{K+1}-1 }\max_{ x^t\in\mathcal{X}^t} \min_{k}\left\{\pr{\tildeinf(x^t;Y_k^t)\geq \lambda} +  \varepsilon_{k} \right\}\nonumber\\
&&{}+ \mathcal{O}(1) \label{eq:asymp_conve_last_interval_sum1}\\
&\leq& \sum_{t=t_{K}}^{ t_{K+1}-1 }\left(\max_{ x^t\in\mathcal{X}^t} \min_{k}\pr{\tildeinf(x^t;Y_k^t)\geq \lambda}+\max_k \varepsilon_{k} \right)\nonumber\\
&&{}+ \mathcal{O}(1)\label{eq:asymp_conve_last_interval_sum2}\\
&\leq&  (t_{K+1} - t_{K})  \left( \lambda^{-c_{1K} \log \lambda} +\max_k\varepsilon_{k}\right)+ \mathcal{O}(1)\IEEEeqnarraynumspace \label{eq:asymp_conve_last_interval_sum4}\\
&\leq& (t_{K+1}-t_{K}) \max_k\varepsilon_{k} + \mathcal{O}(1)\label{eq:asymp_conve_key_step2}
\end{IEEEeqnarray}
as $\lambda\rightarrow \infty$. Here, \eqref{eq:asymp_conve_last_interval_sum1} follows because $\prod_k \minn{1,a_k} \leq \min_k a_k$ for all nonnegative constants $\{a_k\}$, \eqref{eq:asymp_conve_last_interval_sum2} holds because $\min_k \{a_k +\varepsilon_k\} \leq \min_k a_k + \max_k \varepsilon_k$, and \eqref{eq:asymp_conve_last_interval_sum4} follows from \eqref{eq:hoeffding_bound2}.
Finally, by adding \eqref{eq:asymp_conve_key_step1} and \eqref{eq:asymp_conve_key_step2}, we obtain
\eqref{eq:conve_final_large}--\eqref{eq:asymp_conve_di_subst}, shown in the top of the next page.\begin{figure*}[!t]
\normalsize
\setcounter{MYtempeqncnt}{\value{equation}}
\setcounter{equation}{243}
\begin{IEEEeqnarray}{rCl}
 \IEEEeqnarraymulticol{3}{l}{\sum_{t=0}^{t_{K+1}-1} L_t(\vect{\varepsilon})}\nonumber\\
    \quad&\leq&\sum_{i=1}^{K} \bigg[(t_{i}-t_{i-1}) \prod_{k\in\{i,\cdots,K\}} \varepsilon_{k}\bigg] + (t_{K+1}-t_{K})\max_k\varepsilon_{k}+\mathcal{O}(1)\label{eq:conve_final_large}\\
        &=& \left(\frac{\lambda}{C_1} - \sqrt{\frac{\lambda V}{C^3}}\log \lambda \right) \prod_k \varepsilon_k+ \sum_{i=2}^{K} \bigg[\left(\frac{\lambda}{C_i} - \frac{\lambda}{C_{i-1}}\right) \prod_{k\in\{i,\cdots,K\}} \varepsilon_{k}\bigg]+ \left(\frac{\lambda}{C}-\frac{\lambda}{C_K}\right)\max_k\varepsilon_{k}+\mathcal{O}(1)\label{eq:asymp_conve_rho_loosening_before}\IEEEeqnarraynumspace\\
        &\leq& \left(\frac{\lambda}{C_1} - \frac{\lambda \rho}{C_1}\right) \prod_k \varepsilon_k+ \sum_{i=2}^{K} \bigg[\left(\frac{\lambda}{C_i} - \frac{\lambda}{C_{i-1}}\right) \prod_{k\in\{i,\cdots,K\}} \varepsilon_{k}\bigg] + \left(\frac{\lambda}{C}-\frac{\lambda}{C_K}+\frac{\lambda \rho}{C_1}-\sqrt{\frac{\lambda V}{C^3}}\log\lambda\right)\max_k\varepsilon_{k}+\mathcal{O}(1)\label{eq:asymp_conve_rho_loosening}\IEEEeqnarraynumspace\\
&=&  \sum_{i=1}^{K} \bigg[ d_i \prod_{k\in\{i,\cdots,K\}} \varepsilon_{k}\bigg]+ \left(d_0-\sqrt{\frac{\lambda V}{C^3}}\log\lambda\right)\max_k\varepsilon_{k}+\mathcal{O}(1).\label{eq:asymp_conve_di_subst}
\end{IEEEeqnarray}

\setcounter{equation}{\value{MYtempeqncnt}}
\hrulefill
\vspace*{4pt}
\end{figure*}\addtocounter{equation}{4}
Here, \eqref{eq:asymp_conve_rho_loosening} follows by adding to \eqref{eq:asymp_conve_rho_loosening_before} the term $\left(\max_k \varepsilon_k - \prod_k\varepsilon_k\right) \left(\lambda \rho/C_1 - \sqrt{\lambda V/C^3}\log \lambda\right)$, which is positive for all sufficiently large $\lambda$; and \eqref{eq:asymp_conve_di_subst} follows by the definition of the $\{d_i\}_{i\in\{0,\cdots,K\}}$ (see \eqref{eq:d0_def}--\eqref{eq:di_def}). Finally, \eqref{eq:conve_final_large2} follows from \eqref{eq:asymp_conve_di_subst} and by the definition of $f(\vect{\varepsilon})$ in \eqref{eq:feps_def}.

\subsection{Central-limit analysis}\label{sec:central_regime}
Within the interval $[t_{K+1},\beta_{\vect{\varepsilon}}]$, we use Chebyshev's inequality and the Berry-Esseen central limit theorem to obtain a bound on \eqref{eq:asymp_conve_key_step}. Fix a constant $\delta_2\in (0,1)$ and let $\mathcal{A}$ be a compact convex neighborhood of $P^*$ such that for all $P\in\mathcal{A}$, we have both $V_k(P)> 0$ and
\begin{align}
  \left| \sqrt{\frac{V_k(P)}{I_k(P)^3}} - \sqrt{\frac{V_k}{C^3}} \right| \leq \sqrt{\frac{V_k}{C^3}} \delta_2.\label{eq:VkIk3_approx}
\end{align}
The existence of such a set $\mathcal{A}$ follows from the continuity of $I_k(\cdot)$ and $V_k(\cdot)$ at $P^*$. By the definition of $\mathcal{A}$, it follows that
\begin{align}
   I_{\kappa(P) }(P) < C - \delta_3\label{eq:not_in_A_Ik_upper_bound}
\end{align}
for every $P \not \in \mathcal{A}$ and for some $C > \delta_3>0$. This is a consequence of the uniqueness of $P^*$.
The inequality \eqref{eq:VkIk3_approx} enables us to approximate $\sqrt{V_k(P)/I_k(P)^3}$ by $\sqrt{V_k/C^3}$ (which does not depend on $P$) as long as $P$ belongs to the set $\mathcal{A}$. In other words, we can eliminate the dependency on $P$ by introducing an error term proportional to $\delta_2$, which can be made arbitrarily small.

We shall use the following upper bound on $L_t(\vect{\varepsilon})$:
\begin{IEEEeqnarray}{rCl} 
   \IEEEeqnarraymulticol{3}{l}{L_t(\vect{\varepsilon})}\nonumber\\
    \quad &\leq&  \max_{x^t\in\mathcal{X}^t} \prod_k \minn{1,\pr{\tildeinf(x^t;Y_k^t)\geq \lambda} +  \varepsilon_{k}}\nonumber\\
    &&{} + \frac{2^K-1}{\lambda}\label{eq:asymp_conve_key_step_repeat}\IEEEeqnarraynumspace\\
    &\leq& \maxxbigg{ \max_{\substack{x^t\in\mathcal{X}^t:\\ P_{x^t}\not\in\mathcal{A}}} \min_k \pr{\tildeinf(x^t;Y_k^t)\geq \lambda} + \max_k \varepsilon_{k},\nonumber\\
    &&{}\quad \qquad \max_{\substack{x^t\in\mathcal{X}^t:\\ P_{x^t}\in\mathcal{A}}} \prod_k \left(\pr{\tildeinf(x^t;Y_k^t)\geq \lambda} +  \varepsilon_{k}\right)}  \nonumber\\
    &&{}+ \frac{2^K-1}{\lambda}.\label{eq:asymp_conve_key_step_central1}\IEEEeqnarraynumspace
\end{IEEEeqnarray}
Here, \eqref{eq:asymp_conve_key_step_central1} follows because $\prod_k (a_k+b_k) \leq \min_k (a_k + b_k) \leq \min_k a_k + \max_k b_k$ for all constants $\{a_k\}$ and $\{b_k\}$. 


For all $x^t\in\mathcal{X}^t$ for which $P_{x^t}\not\in \mathcal{A}$, we use Chebyshev's inequality to obtain the estimate
\begin{multline}
 \pr{\tildeinf(x^t;Y_k^t)\geq \lambda} \\\leq \left\{\begin{array}{ll}
 \frac{t V_{k}\farg{P_{x^t}}}{(\lambda - t I_{k}\farg{P_{x^t}})^2}  & \text{ if } \lambda > t I_{k}\farg{P_{x^t}}\\
 1 & \text{otherwise}
 \end{array}
\right..
 \label{eq:conve_chebyshev}
\end{multline}
It follows from \eqref{eq:not_in_A_Ik_upper_bound} and from the condition $t\leq \beta_{\vect{\varepsilon}}$ (see definition of $\beta_{\vect{\varepsilon}}$ in \eqref{eq:beta_eps_def}) that $\lambda> t  I_{\kappa(P)}(P)$ for every $P \not\in \mathcal{A}$ and for all sufficiently large $\lambda$ (recall the definition of $\kappa(\cdot)$ in \eqref{eq:kappa_def}). 
Using \eqref{eq:conve_chebyshev}, we obtain the following upper bound on the first term on the right-hand side of \eqref{eq:asymp_conve_key_step_central1}, which holds for $t\in[t_{K+1},\beta_{\vect{\varepsilon}}]$ and for all sufficiently large $\lambda$:
\begin{IEEEeqnarray}{rCl}
\IEEEeqnarraymulticol{3}{l}{\max_{\substack{x^t\in\mathcal{X}^t:\\ P_{x^t}\not\in\mathcal{A}}} \min_{k} \pr{\tilde  \imath_k(x^t; Y_k^t) \geq \lambda}}\nonumber\\
\qquad &\leq&\max_{P \not \in \mathcal{A}}\frac{t V_{\kappa(P)}\farg{P}}{(\lambda - t I_{\kappa(P)}\farg{P})^2}\label{eq:chebyshev_bound_start}\\
&\leq&  \frac{ V_{\text{max}} t }{(\lambda - t C + t\delta_3 )^2}\label{eq:conve_chebyshev_cap_upper}\\
&\leq&  \frac{ 2V_{\text{max}}  \lambda  }{(\lambda \delta_3/C - \const\sqrt{\lambda}\log \lambda - \const)^2}\label{eq:conve_chebyshev2}\\
&\leq&  \frac{1}{\lambda}\underbrace{\frac{4V_{\text{max}} C^2}{\delta_3}}_{\triangleq c_2}.\label{eq:chebyshev_bound_final}
\end{IEEEeqnarray}
Here, \eqref{eq:conve_chebyshev_cap_upper} follows because there exists a constant $V_{\text{max}}>0$ such that $V_k\farg{P}\leq V_{\text{max}}$  for every $P\in\mathcal{P}(\mathcal{X})$ \cite[p.~7048]{Tomamichel2013} and because of \eqref{eq:not_in_A_Ik_upper_bound}; \eqref{eq:conve_chebyshev2} follows because $t \leq \beta_{\vect{\varepsilon}}$ and $\delta_3 < C$, which imply that, for all sufficiently large $\lambda$,
\begin{IEEEeqnarray}{rCl} 
\lambda - tC + t\delta_3 &\geq& \lambda - (C - \delta_3)\beta_{\vect{\varepsilon}} \\
&=& \lambda - (C - \delta_3)(\lambda/C + \const \sqrt{\lambda}\log \lambda)\IEEEeqnarraynumspace\\
&=& \lambda \delta_3/C - \const \sqrt{\lambda}\log\lambda - \const > 0.
\end{IEEEeqnarray}
Finally, \eqref{eq:chebyshev_bound_final} holds for all sufficiently large $\lambda$.
We see that \eqref{eq:chebyshev_bound_final} can be made arbitrarily close to zero by choosing $\lambda$ sufficiently large. Now, we continue the chain of inequalities in \eqref{eq:asymp_conve_key_step_central1} as follows:
\begin{IEEEeqnarray}{rCl}
    \IEEEeqnarraymulticol{3}{l}{L_t(\vect{\varepsilon})}\nonumber\\
     &\leq&  \maxx{  \frac{c_2}{\lambda}+\max_k \varepsilon_k  , \max_{\substack{x^t\in\mathcal{X}^t:\\ P_{x^t}\in\mathcal{A}}} \prod_{k} \left(\pr{\tilde  \imath_k(x^t; Y_k^t) \geq \lambda} + \varepsilon_k\right) }\nonumber\\
     &&{} + \frac{2^K-1}{\lambda}\IEEEeqnarraynumspace\\
       &\leq&  \maxx{  \frac{c_2}{\lambda}  , \max_{\substack{x^t\in\mathcal{X}^t:\\ P_{x^t}\in\mathcal{A}}} \prod_{k} \left(\pr{\tilde  \imath_k(x^t; Y_k^t) \geq \lambda} + \varepsilon_k\right) -\prod_k \varepsilon_k }\nonumber\\
       &&{} + \max_k \varepsilon_k + \frac{2^K-1}{\lambda}\label{eq:asymp_conve_plus_maxeps_minprodeps}\\
     &\leq&   \max_{\substack{x^t\in\mathcal{X}^t:\\ P_{x^t}\in\mathcal{A}}} \prod_{k} \left(\pr{\tilde  \imath_k(x^t; Y_k^t) \geq \lambda} + \varepsilon_k\right) - \prod_k \varepsilon_k +\max_k \varepsilon_k \nonumber\\
     &&{} + \frac{1}{\lambda}\underbrace{\Big(2^K-1+ c_2\Big)}_{\triangleq c_3}.\label{eq:central_Lt_final}
\end{IEEEeqnarray}
Here, in \eqref{eq:asymp_conve_plus_maxeps_minprodeps}, we used that $\max_k \varepsilon_k \geq \prod_k \varepsilon_k$ because $\varepsilon_k\in[0,1]$ for all $k\in\mathcal{K}$. Note that the upper bound \eqref{eq:central_Lt_final} allows us to consider only the $x^t\in\mathcal{X}^t$ for which $P_{x^t}\in\mathcal{P}(\mathcal{X})$. This in turn allows us to make use of \eqref{eq:VkIk3_approx}. Specifically, for all $x^t\in\mathcal{X}^t$ for which $P_{x^t}\in \mathcal{A}$, the Berry-Esseen central limit theorem \cite[Th.~V.2.3]{V.V.Petro} yields the following estimate:
\begin{IEEEeqnarray}{rCl}
\IEEEeqnarraymulticol{3}{l}{\pr{\tildeinf(x^t ; Y_k^t) \geq \lambda}}\nonumber\\
 \quad&\leq& Q\farg{\frac{\lambda-t I_{k}(P_{x^t})}{\sqrt{t V_{k}(P_{x^t})}}}+ \frac{6 t T_{k}(P_{x^t}) }{(t V_{k}(P_{x^t}))^{3/2}}\\
&\leq& Q\farg{\frac{\lambda-t I_{k}(P_{x^t})}{\sqrt{t V_{k}(P_{x^t})}}}\nonumber\\
&&{}+ \frac{1}{\sqrt{2 C t}}\underbrace{\frac{6 \sqrt{2 C} \max_{P\in \mathcal{A}} T_{k}(P) }{ \min_{P\in \mathcal{A}} V_{k}(P)^{3/2}}}_{\triangleq c_4} \label{eq:conve_berry1}\\
&\leq& Q\farg{\frac{\lambda/I_{k}(P_{x^t})-t  }{\sqrt{\lambda V_{k}(P_{x^t})/I_k(P_{x^t})^3}}}+ \frac{c_4}{\sqrt{2C t}}\label{eq:conve_berry2}\\
&\leq& Q\farg{\min_{\nu_k \in\{-1,1\}} \frac{\lambda/I_{k}(P_{x^t})-t }{\sqrt{\lambda V_{k}/C^3}\left( 1 + \delta_2 \nu_k  \right)}  }+ \frac{c_4}{\sqrt{\lambda}}.\label{eq:conve_berry}\IEEEeqnarraynumspace
\end{IEEEeqnarray}
\begin{sloppypar}In \eqref{eq:conve_berry1}, $c_4$ is a well-defined positive constant because $T_k(P)<\const$ uniformly \cite[Lem.~46]{Polyanskiy2010b} and because the condition $V_k(P)>0$ for $P\in\mathcal{A}$ combined with the compactness of $\mathcal{A}$ imply that $\min_{P\in\mathcal{A}} V_k(P)$ is well-defined and positive; \eqref{eq:conve_berry2} follows by the inequality (proven in Appendix~\ref{app:lam_property})
\end{sloppypar}
\begin{IEEEeqnarray}{rCl}
\frac{a - t b}{\sqrt{t}} \geq \frac{a - t b}{\sqrt{a/b}}\label{eq:converse_property}
\end{IEEEeqnarray}
which holds for all positive $a,b$ and $t$; and \eqref{eq:conve_berry}, which holds for all sufficiently large $\lambda$, follows from \eqref{eq:VkIk3_approx} (recall that $P_{x^t}\in \mathcal{A}$), which is equivalent to
\begin{IEEEeqnarray}{rCl}
\sqrt{\frac{V_k}{C^3}}(1-\delta_2) \leq \sqrt{\frac{V_k(P)}{I_k(P)^3}} \leq \sqrt{\frac{V_k}{C^3}}(1+\delta_2).\label{eq:lower_and_upper_bound_VkIk3}
\end{IEEEeqnarray}
In \eqref{eq:conve_berry}, the role of $\nu_k$ is to select the upper or the lower bound in \eqref{eq:lower_and_upper_bound_VkIk3}. The choice depends on the sign of $\left(\lambda/I_k(P_{x^t})-t\right)$.
The bound \eqref{eq:conve_berry} implies that
\begin{IEEEeqnarray}{rCl}
\IEEEeqnarraymulticol{3}{l}{\prod_{k} \left(\pr{\tilde  \imath_k(x^t; Y_k^t) \geq \lambda} + \varepsilon_k\right)}\nonumber\\
 &\leq&\prod_{k } \left( Q\farg{\min_{\nu_k \in\{-1,1\}} \frac{\lambda/I_{k}(P_{x^t})-t }{\sqrt{\lambda V_{k}/C^3}\left( 1 + \delta_2\nu_k  \right) }  } + \varepsilon_k\right)\nonumber\\
 &&{} + \frac{c_5}{\sqrt{\lambda}}\label{eq:conve_berry_esseen_estimate_Q2}
\end{IEEEeqnarray}
where $c_5 \triangleq (2^K-1)c_4$.

We shall next eliminate the dependency of the first term of the right-hand side of \eqref{eq:conve_berry_esseen_estimate_Q2} on $x^t$ by further upper-bounding this term. Let $P\in\mathcal{A}$; we have
\begin{IEEEeqnarray}{rCl}
\IEEEeqnarraymulticol{3}{l}{Q\farg{\min_{\nu_k \in\{-1,1\}} \frac{\lambda/I_k(P) - t   }{\sqrt{\lambda V_{k}/C^3}\left( 1 + \delta_2 \nu_k  \right) } }}\nonumber\\
\quad &\leq& Q\farg{\min_{\nu_k \in\{-1,1\}}\frac{\lambda/(C+\diff I_k(P - P^*) )  - t   }{\sqrt{\lambda V_{k}/C^3}\left( 1 + \delta_2 \nu_k  \right) } }\label{eq:Qtaylor_expansion0}\IEEEeqnarraynumspace\\
&\leq& Q\farg{\min_{\nu_k \in\{-1,1\}}\frac{ \frac{\lambda}{C} - \frac{\lambda}{C^2} \diff I_k(P - P^*)  - t   }{\sqrt{\lambda V_{k}/C^3}\left( 1 + \delta_2 \nu_k  \right) } }.\label{eq:Qtaylor_expansion}
\end{IEEEeqnarray}
Here, \eqref{eq:Qtaylor_expansion0} follows because $I_k(P)$ is concave in $P$ and because the $Q$ function is monotonically decreasing; \eqref{eq:Qtaylor_expansion} follows from the inequality $\frac{a}{b+c} \geq \frac{a}{b} - \frac{a}{b^2}c$ which holds for all $a>0, b>0$, and $b+c > 0$. Indeed, $C +\diff I_k(P - P^*) > 0 $ for sufficiently small $\delta_2$ since $P\in \mathcal{A}$.
Using \eqref{eq:Qtaylor_expansion} in \eqref{eq:conve_berry_esseen_estimate_Q2}, we obtain the steps \eqref{eq:before_in_terms_of_FWk0}--\eqref{eq:before_in_terms_of_FWk}, shown in the top of the next page.\begin{figure*}[!t]
\normalsize
\setcounter{MYtempeqncnt}{\value{equation}}
\setcounter{equation}{271}
\begin{IEEEeqnarray}{rCl}
\prod_{k} \left(\pr{\tilde  \imath_k(x^t; Y_k^t) \geq \lambda} + \varepsilon_k\right) 
&\leq&  \prod_{k} \Bigg(Q\fargBigg{\min_{\nu_k \in\{-1,1\}} \frac{ \frac{\lambda}{C} - \frac{\lambda}{C^2} \diff I_k(P_{x^t} - P^*) - t   }{ \sqrt{\lambda V_{k}/C^3}\left( 1 + \delta_2 \nu_k  \right) }  }+\varepsilon_{k}\Bigg) + \frac{c_5}{\sqrt{\lambda}} \label{eq:before_in_terms_of_FWk0}\\
&\leq& \max_{P\in \mathcal{A}} \prod_k \left( Q\farg{\min_{\nu_k \in\{-1,1\}}\frac{ \frac{\lambda}{C} - \frac{\lambda}{C^2} \diff I_k(P - P^*)  - t   }{\sqrt{\lambda V_{k}/C^3}\left( 1 + \delta_2 \nu_k \right)}  } +\varepsilon_{k}\right)+\frac{c_5}{\sqrt{\lambda}}\IEEEeqnarraynumspace\\
&\leq& \max_{\vect{v}\in\mathbb{R}_0^{|\mathcal{X}|}  }\prod_k \left(Q\farg{\min_{\nu_k \in\{-1,1\}} \frac{ \frac{\lambda/C  - t}{\sqrt{\lambda V/C^3}} -  \diff I_k(\vect{v})   }{ \varrho_{k}(1 + \delta_2 \nu_k )}  }+\varepsilon_{k}\right)+\frac{c_5}{\sqrt{\lambda}}.\label{eq:before_in_terms_of_FWk}
\end{IEEEeqnarray}

\setcounter{equation}{\value{MYtempeqncnt}}
\hrulefill
\vspace*{4pt}
\end{figure*}\addtocounter{equation}{3}
Here, in \eqref{eq:before_in_terms_of_FWk}, we used that $\varrho_k=\sqrt{V_k/V}$.
Let now $\{\tvRV_{\delta_2,k}\}$ be i.i.d. RVs with cumulative distribution function
\begin{multline}
  F_{\tvRV_{\delta_2,k}}(w) \triangleq Q\farg{ \min_{\nu_k \in \{-1,1\}} \frac{ -w - \diff I_k(\optx_{\delta_2}(w))    }{ \varrho_{ k}(1 + \delta_2\nu_k)}}\label{eq:FtvRV_delta2_def}
\end{multline}
where\footnote{If the maximizer of \eqref{eq:optx_delta2} is not unique, $\optx_{\delta_2}(w)$ is chosen arbitrarily from the set of maximizers.}
\begin{multline}
\optx_{\delta_2}(w) \\ \triangleq \argmax_{\vect{v}\in\mathbb{R}_0^{|\mathcal{X}|} } \prod_{k} \left( Q\farg{\min_{\nu_k \in\{-1,1\}} \frac{ -w -  \diff I_k(\vect{v})    }{\varrho_{ k}(1 + \delta_2 \nu_k) }} +\varepsilon_{k} \right).\label{eq:optx_delta2}
\end{multline}
We also denote by 
\begin{IEEEeqnarray}{rCl}
  \preffree{K} &\triangleq& \{\mathcal{\tilde K}: \mathcal{\tilde K}\subseteq \mathcal{K}\}\setminus \{\emptyset\}\label{eq:frakK}
\end{IEEEeqnarray}
the set of all nonempty subsets of $\mathcal{K}$. Using these definitions and \eqref{eq:before_in_terms_of_FWk}, we have that for every $x^t\in \mathcal{X}^t$ satisfying $P_{x^t}\in \mathcal{A}$,\footnote{In \eqref{eq:conve_final_central_bound2}, we use the convention that $\prod_{k\in\emptyset} a_k = 1$ for every $a_k \in\mathbb{R}$.}
\begin{IEEEeqnarray}{rCl}
\IEEEeqnarraymulticol{3}{l}{L_t(\vect{\varepsilon})}\nonumber\\
 \quad&\leq& \max_{\substack{x^t\in\mathcal{X}^t:\\ P_{x^t}\in\mathcal{A}}} \prod_{k} \left(\pr{\tilde  \imath_k(x^t; Y_k^t) \geq \lambda} + \varepsilon_k\right) - \prod_k \varepsilon_k\nonumber\\
 &&{} +\max_k \varepsilon_k  + \frac{c_3}{\lambda} \\
  &\leq& \prod_{k} \left(  F_{\tvRV_{\delta_2,k}}\farg{-\frac{\lambda/C  - t}{\sqrt{\lambda V/C^3}} }+ \varepsilon_{k}\right)  - \prod_k \varepsilon_k \nonumber\\
  &&{}+ \max_k \varepsilon_k+  \underbrace{2 c_5}_{\triangleq c_6}\frac{1}{\sqrt{\lambda}}\label{eq:asymp_conve_central_Fdelta2k} \\
  &=& \sum_{\mathcal{\bar K} \in \preffree{K}} \left( \prod_{k\in\mathcal{K}\setminus \{\mathcal{\bar K}\}} \varepsilon_k \right) \left(  \prod_{k\in \mathcal{\bar K}} F_{\tvRV_{\delta_2,k}}\farg{-\frac{\lambda/C  - t}{\sqrt{\lambda V/C^3}} }\right) \nonumber\\
  &&{}+ \max_k \varepsilon_k + \frac{c_6}{\sqrt{\lambda}}.\IEEEeqnarraynumspace\label{eq:conve_final_central_bound2}
\end{IEEEeqnarray}
Here, \eqref{eq:asymp_conve_central_Fdelta2k}, which holds for sufficiently large $\lambda$, follows from \eqref{eq:before_in_terms_of_FWk} and \eqref{eq:FtvRV_delta2_def}; \eqref{eq:conve_final_central_bound2} follows from \eqref{eq:frakK} and by expanding the product in the first term of \eqref{eq:asymp_conve_central_Fdelta2k} into $2^K$ terms.

We now evaluate \eqref{eq:conve_final_central_bound2}. For every nonempty subset $\mathcal{\bar K}\subseteq \mathcal{K}$ and for sufficiently large $\lambda$, we have
\begin{IEEEeqnarray}{rCl}
\IEEEeqnarraymulticol{3}{l}{\sum_{t= t_{K+1}}^{  \beta_{\vect{\varepsilon}} } \prod_{k \in\mathcal{\bar K}}  F_{\tvRV_{\delta_2,k}}\farg{-\frac{\lambda/C  - t}{\sqrt{\lambda V/C^3}} }}\nonumber\\
&\leq & \int_{-\infty}^{\beta_{\vect{\varepsilon}}} \prod_{k \in\mathcal{\bar K}}  F_{\tvRV_{\delta_2,k}}\farg{-\frac{\lambda/C  - t}{\sqrt{\lambda V/C^3}} }  \intdiff t + 1\label{eq:FWk_increasing}\\
&=& \int_{-\infty}^{\beta_{\vect{\varepsilon}}} \pr{ \max_{k\in\mathcal{\bar K}}\left\{ \tvRV_{\delta_2,k} + \frac{\lambda/C  - t}{\sqrt{\lambda V/C^3}} \right\}\leq 0 }\intdiff t +1 \\
  &= & \int_{-\infty}^{\beta_{\vect{\varepsilon}}}\pr{ \max_{k\in\mathcal{\bar K}}\left\{ \frac{\lambda}{C}  +\sqrt{\frac{\lambda V}{C^3}} \tvRV_{\delta_2,k}   \right\}\leq t }\intdiff t +1 \label{eq:delta3k_dek}\\
 &= & \beta_{\vect{\varepsilon}} - \E{\min\mathopen{} \bigg\{\beta_{\vect{\varepsilon}}, \max_{k\in\mathcal{\bar K}}\bigg\{ \frac{\lambda}{C}+ \sqrt{\frac{\lambda V}{C^3}} \tvRV_{\delta_2,k}  \bigg\} \bigg\}}  +1\label{eq:asymp_conve_Eprob_property}\IEEEeqnarraynumspace\\
 &= & \sqrt{\frac{\lambda V}{C^3}}\Big( \nu_{\vect{\varepsilon}}-\EBig{\min\mathopen{}\Big\{\nu_{\vect{\varepsilon}}, \max_{k\in\mathcal{\bar K}} \tvRV_{\delta_2,k} \Big\}}\Big)  +1.\label{eq:asymp_conve_prodQ_final}
\end{IEEEeqnarray}
In \eqref{eq:FWk_increasing}, we have used that $F_{\tvRV_{\delta_2,k}}(\cdot)$ is a monotonically increasing function upper-bounded by one; \eqref{eq:asymp_conve_Eprob_property} holds because for every RV $X$,
\begin{align}
  \E{\minn{a, X}} = a - \int_{-\infty}^a \pr{X \leq t} \intdiff t.
\end{align}
Finally, \eqref{eq:asymp_conve_prodQ_final} follows from \eqref{eq:beta_eps_def}.

Next, we substitute \eqref{eq:asymp_conve_prodQ_final} into \eqref{eq:conve_final_central_bound2} and obtain a lower bound on $\sum_{t=t_{K+1}}^{\beta_{\vect{\varepsilon}}} (1-L_t(\vect{\varepsilon}))$ through the steps \eqref{eq:asymp_conve_summing1}--\eqref{eq:asymp_conve_final23}, shown in the top of the next page. This lower bound holds for all sufficiently large $\lambda$ and for all $\vect{\varepsilon} \in [0,1]^K$\begin{figure*}[!t]
\normalsize
\setcounter{MYtempeqncnt}{\value{equation}}
\setcounter{equation}{286}
\begin{IEEEeqnarray}{rCl}
\IEEEeqnarraymulticol{3}{l}{\sum_{t= t_{K+1} }^{ \beta_{\vect{\varepsilon}} } (1-L_t(\vect{\varepsilon})) } \nonumber\\
\quad&\geq& \sqrt{\frac{\lambda V}{C^3}}(\log \lambda + \nu_{\vect{\varepsilon}})\left(1-\max_k \varepsilon_k\right)-\sum_{t= t_{K+1} }^{ \beta_{\vect{\varepsilon}} }\left[ \sum_{\mathcal{\bar K} \in \preffree{K}} \left( \prod_{k\in\mathcal{K}\setminus \{\mathcal{\bar K}\}} \varepsilon_k \right) \left(  \prod_{k\in \mathcal{\bar K}} F_{\tvRV_{\delta_2,k}}\farg{-\frac{\lambda/C  - t}{\sqrt{\lambda V/C^3}} }\right)\right]  \nonumber\IEEEeqnarraynumspace\\
&& {} - 2^K -\sqrt{\frac{V}{C^3}} c_6 (\log \lambda + \nu_{\vect{\varepsilon}}) \label{eq:asymp_conve_summing1}\\
&=& \sqrt{\frac{\lambda V}{C^3}}(\log \lambda + \nu_{\vect{\varepsilon}})\left(1-\max_k \varepsilon_k\right)-\sum_{\mathcal{\bar K} \in \preffree{K}} \left( \prod_{k\in\mathcal{K}\setminus \{\mathcal{\bar K}\}} \varepsilon_k \right) \left( \sum_{t= t_{K+1} }^{ \beta_{\vect{\varepsilon}} } \prod_{k\in \mathcal{\bar K}} F_{\tvRV_{\delta_2,k}}\farg{-\frac{\lambda/C  - t}{\sqrt{\lambda V/C^3}} }\right) -c_7 \log \lambda\label{eq:asymp_conve_summing2} \\
 &\geq&\sqrt{\frac{\lambda V}{C^3}} \Bigg( (\log \lambda+\nu_{\vect{\varepsilon}})(1-\max_k\varepsilon_{k})-\sum_{\mathcal{\bar K} \in \preffree{K}} \Bigg(\prod_{k \in \mathcal{K}\setminus\{\mathcal{\bar K}\}}\varepsilon_k  \Bigg)\Big(\nu_{\vect{\varepsilon}}- \EBig{\min\mathopen{}\Big\{\nu_{\vect{\varepsilon}}, \max_{k\in\mathcal{\bar K}} \tvRV_{\delta_2,k}  \Big\}}\Big)\Bigg)- c_7 \log \lambda \IEEEeqnarraynumspace\label{eq:asymp_conve_final22}\\
&\geq& \sqrt{\lambda}g_{\delta_1,\delta_2}(\vect{\varepsilon})+\sqrt{\frac{\lambda V}{C^3}}(1- \max_k \varepsilon_k)\log \lambda -c_7 \log \lambda.\label{eq:asymp_conve_final23}
\end{IEEEeqnarray}

\setcounter{equation}{\value{MYtempeqncnt}}
\hrulefill
\vspace*{4pt}
\end{figure*}\addtocounter{equation}{4}
and the function $g_{\delta_1,\delta_2}(\vect{\varepsilon})$ is defined in \eqref{eq:geps_def}, shown in the top of the next page.\begin{figure*}[!t]
\normalsize
\setcounter{MYtempeqncnt}{\value{equation}}
\setcounter{equation}{290}
\begin{IEEEeqnarray}{rCl}
g_{\delta_1,\delta_2}(\vect{\varepsilon}) &\triangleq& \sqrt{\frac{V}{C^3}} \Bigg(\nu_{\vect{\varepsilon}} (1-\max_k\varepsilon_{k}) -\sum_{\mathcal{\bar K} \in \preffree{K}} \Bigg(\prod_{k \in \mathcal{K}\setminus\{\mathcal{\bar K}\}}\varepsilon_k  \Bigg)\left( \nu_{\vect{\varepsilon}}- \E{\min\mathopen{}\left\{ \nu_{\vect{\varepsilon}}, \max_{k\in\mathcal{\bar K}} \tvRV_{\delta_2,k}  \right\}}\right)  \Bigg).\label{eq:geps_def}
\end{IEEEeqnarray}

\setcounter{equation}{\value{MYtempeqncnt}}
\hrulefill
\vspace*{4pt}
\end{figure*}\addtocounter{equation}{1}
In \eqref{eq:asymp_conve_summing1}, we have used \eqref{eq:conve_final_central_bound2} and \eqref{eq:asymp_conve_summing2} follows by interchanging the order of summations. We also used that $\nu_{\vect{\varepsilon}}$ is bounded from above, which implies that there exists a constant $c_7$ that does not depend on  $\vect{\varepsilon}\in[0,1]^K$ such that $\sqrt{V/C^3} c_6(\nu_{\vect{\varepsilon}}+\log \lambda) + 2^K \leq c_7 \log \lambda$ for all sufficiently large $\lambda$ (recall that $|\preffree{K}| = 2^K-1$). Finally, \eqref{eq:asymp_conve_final22}follows from \eqref{eq:asymp_conve_prodQ_final}.
This establishes \eqref{eq:asymp_conve_final3}.

\subsection{Optimization over $f(\cdot)$}\label{sec:lem63_first_order}
We observe that \eqref{eq:rho_cond} implies that, for every $i\in\{1,\cdots,K-1\}$,
\begin{IEEEeqnarray}{rCl}
    \sum_{j=1}^{i} (d_j-d_0) > 0.\label{eq:dj_d0_ineq}
\end{IEEEeqnarray}
In turn, these inequalities imply that there exist constants $\{\zeta_i\}_{i\in\{1,\cdots,K-2\}}$ such that
\begin{IEEEeqnarray}{rCl}
  d_i - d_0 + \zeta_{i-1} - \zeta_i &>& 0
\label{eq:asymp_conve_consts_zeta}
\end{IEEEeqnarray}
for every $i\in\{1,\cdots,K-1\}$.
Here, we have set $\zeta_0 \triangleq \zeta_{K-1} \triangleq 0$ for convenience. A proof of this claim can be found in Lemma~\ref{lem:zeta_constants} in Appendix~\ref{app:lam_property}.

 We use \eqref{eq:asymp_conve_consts_zeta} to show that $f(\vect{\varepsilon})$ is lower-bounded by an affine function through the steps \eqref{eq:asymp_conve_f_bound_1}--\eqref{eq:asymp_conve_f_bound3}, shown in the top of the next page.\begin{figure*}[!t]
\normalsize
\setcounter{MYtempeqncnt}{\value{equation}}
\setcounter{equation}{293}
\begin{IEEEeqnarray}{rCl}
  f(\vect{\varepsilon}) &\geq& \frac{1}{C}-\sum_{i=1}^{K} d_i \min_{k \in \{i,\cdots,K\} }\varepsilon_k-d_0  \max_k\varepsilon_{k}\label{eq:asymp_conve_f_bound_1}\\
    &=& \frac{1}{C}- \sum_{i=1}^{K-1} (d_i-d_0) \min_{k \in \{i,\cdots,K\} }\varepsilon_k  - d_0\left(\max_k\varepsilon_{k} +\sum_{i=1}^{K-1} \min_{k\in\{i,\cdots,K\}} \varepsilon_k\right)- d_{K}\varepsilon_K\IEEEeqnarraynumspace\\
     &\geq& \frac{1}{C}- \sum_{i=1}^{K-1} (d_i-d_0+ \zeta_{i-1} - \zeta_i) \min_{k \in \{i,\cdots,K\} }\varepsilon_k   - d_0\left(\varepsilon_1 + \cdots +\varepsilon_K\right) - d_{K}\varepsilon_K\label{eq:asymp_conve_f_bound2}\\
        &\geq& \frac{1}{C}- \sum_{i=1}^{K-1} (d_i-d_0+ \zeta_{i-1} - \zeta_i) \frac{\varepsilon_i+\cdots+\varepsilon_K}{K - i + 1}   - d_0\left(\varepsilon_1 + \cdots +\varepsilon_K\right) - d_{K}\varepsilon_K.\label{eq:asymp_conve_f_bound3}
\end{IEEEeqnarray}

\setcounter{equation}{\value{MYtempeqncnt}}
\hrulefill
\vspace*{4pt}
\end{figure*}\addtocounter{equation}{4}
Here, \eqref{eq:asymp_conve_f_bound_1} holds because $\prod_{k\in\{i,\cdots,K\}} \varepsilon_k \leq \min_{k\in\{i,\cdots,K\}} \varepsilon_k$; in \eqref{eq:asymp_conve_f_bound2}, we used that $\min_{k\in \{i,\cdots,K\}} \varepsilon_k \leq \min_{k\in \{i+1,\cdots,K\}} \varepsilon_k$ for $i\in\{1,\cdots,K-1\}$, which implies that
\begin{IEEEeqnarray}{rCl}
\IEEEeqnarraymulticol{3}{l}{  \sum_{i=1}^{K-1}( \zeta_{i-1}- \zeta_i) \min_{k\in\{i,\cdots,K\}} \varepsilon_k}\nonumber\\
 &\geq&  \sum_{i=1}^{K-1} \zeta_{i-1}\min_{k\in\{i,\cdots,K\}} \varepsilon_k-\sum_{i=1}^{K-1} \zeta_i \min_{k\in\{i+1,\cdots,K\}} \varepsilon_k\\
  &=&  \sum_{i=1}^{K-1} \zeta_{i-1}\min_{k\in\{i,\cdots,K\}} \varepsilon_k-\sum_{i=2}^{K} \zeta_{i-1} \min_{k\in\{i,\cdots,K\}} \varepsilon_k\IEEEeqnarraynumspace\\
  &=& \zeta_0 \varepsilon_1 - \zeta_{K-1}\varepsilon_K  = 0.
\end{IEEEeqnarray}
Furthermore, \eqref{eq:asymp_conve_f_bound3} holds because $\min_{k\in\{i,\cdots,K\}}\varepsilon_k \leq (\varepsilon_i + \cdots + \varepsilon_K)/(K-i+1)$. It follows from \eqref{eq:asymp_conve_consts_zeta} that $(d_i-d_0+ \zeta_{i-1} - \zeta_i)$ is positive for all $i \in\{1,\cdots,K-1\}$. Hence, the inequality \eqref{eq:asymp_conve_f_bound3} holds with equality if and only if $\varepsilon_1 =\cdots = \varepsilon_K$. This implies that a necessary and sufficient condition for the chain of inequalities \eqref{eq:asymp_conve_f_bound_1}--\eqref{eq:asymp_conve_f_bound3} to hold with equality is that $\varepsilon_1=\cdots=\varepsilon_K$ and that $\varepsilon_1\in\{0,1\}$. 
This implies that
\begin{IEEEeqnarray}{rCl}
\IEEEeqnarraymulticol{3}{l}{\min_{\substack{P_U,\vect{\varepsilon}^{(u)}\in[0,1]^K:\\ \EE{U}{\varepsilon_{k}^{(U)}}\leq \epsilon+(\log M)^{-1} }} \EE{U}{f(\vect{\varepsilon}^{(U)})}}\nonumber\\
 \qquad \qquad\qquad\qquad&=& (1-\epsilon - (\log M)^{-1}) f(\vect{0}_K) \nonumber\\
 &&{}+ (\epsilon +(\log M)^{-1})f(\vect{1}_K)\IEEEeqnarraynumspace
\end{IEEEeqnarray}
which is equivalent to \eqref{eq:asymp_conve_first_order_opt}.

\section{Proof of Theorem~\ref{thm:asymp} (achievability) and of Theorem~\ref{thm:time_varying}}
\label{sec:achievability_asymptotics}

First, we prove Theorem~\ref{thm:time_varying}. Then, we show that the achievability part of Theorem~\ref{thm:asymp} follows as a special case of Theorem~\ref{thm:time_varying}.
To establish Theorem~\ref{thm:time_varying}, we make use of the following lemma. 
\begin{lemma}
\label{lem:time_varying}
Under the conditions of Theorem~\ref{thm:time_varying}, there exists a joint probability distribution $P_{X^\infty}$ on $\mathcal{X}^\infty$ such that the stopping times $\{ \tau_k(\gamma)\}$
\begin{IEEEeqnarray}{rCl}
\tau_k(\gamma) &\triangleq& \inff{n \geq 0: i_{P_{X^n},W_k^n}(X^n ; Y_{k}^n)\geq \gamma}\label{eq:achiev_xn} 
\end{IEEEeqnarray}  
satisfy
  \begin{IEEEeqnarray}{rCl}
    \EBig{\max_k \tau_k(\gamma)} \leq \frac{\gamma}{C} +\sqrt{\frac{\gamma V}{C^3}}\EBig{\max_k \tvRVbar_k} + o(\sqrt{\gamma}).\label{eq:lem_Emaxtau}\IEEEeqnarraynumspace
  \end{IEEEeqnarray}
  Here, the independent RVs $\{\tvRVbar_k\}$ have cumulative distribution functions given in \eqref{eq:FtvRVk_def} and $P_{X^n}$ in \eqref{eq:achiev_xn} denotes the joint  probability distribution of the first $n$ entries of $X^\infty \sim P_{X^\infty}$.
\end{lemma}
\begin{IEEEproof}
    See Appendix~\ref{proof:tv_asymp_time_sharing}.
\end{IEEEproof}
Define now the function
\begin{align}
g\farg{x}\triangleq \sqrt{\frac{x V}{C^3}} \EBig{   \max_k   \tvRVbar_k }
\end{align}
and let $P_{X^\infty}$ be distributed according to Lemma~\ref{lem:time_varying}. In view of Theorem~\ref{thm:simple_achiev}, let for all integers $\bar\ell>0$ and for all $\delta>0$
 \begin{IEEEeqnarray}{rCl}
  \gamma_{\bar\ell} &\triangleq& C \left( \bar\ell - (1+\delta)g(C \bar\ell) \right)\label{eq:gamma_def}\\
  \sprob_{\bar\ell} &\triangleq& \frac{\bar\ell \epsilon -1}{ \bar\ell - 1}\\
  M_{\bar\ell} &\triangleq& \left\lfloor \ee{\gamma_{\bar\ell}- \log \bar\ell }\right\rfloor.
\end{IEEEeqnarray}
Then, we have (cf.~\eqref{eq:achiev_Proberror2})
\begin{multline}
  \sprob_{\bar\ell}+(1-\sprob_{\bar\ell})( M_{\bar\ell}-1)\exp\left\{-\gamma_{\bar\ell}\right\} \\\leq \frac{\bar\ell \epsilon -1}{ \bar\ell - 1} + \frac{\bar\ell(1-\epsilon) }{\bar\ell-1} \frac{1}{\bar\ell} = \epsilon.\label{eq:asymp_achiev_prob_error}
\end{multline}
Additionally, suppose that there exists an integer $\ell_0 \geq 0$ such that, for all $\bar \ell > \ell_0$,
\begin{align}
  \EBig{\max_k \tau_k(\gamma_{\bar\ell})} \leq \bar\ell\label{eq:Emaxtau_U0}
\end{align}
and $ M_{\bar\ell}\geq 2$. Then, for $\bar\ell\geq \ell_0$, we have 
\begin{align}
(1-\sprob_{\bar\ell})\EBig{\max_k \tau_k(\gamma_{\bar\ell}) }\leq  \frac{\bar\ell(1-\epsilon) }{\bar\ell-1}\bar\ell \triangleq \ell_{\bar\ell}.\label{eq:asymp_achiev_Emax}
\end{align}
By invoking Theorem~\ref{thm:simple_achiev} with $q=q_{\bar\ell}$, $\gamma=\gamma_{\bar\ell}$, and $M =  M_{\bar \ell}$, and by using \eqref{eq:achiev_Proberror2} along with the inequalities \eqref{eq:asymp_achiev_prob_error} and \eqref{eq:asymp_achiev_Emax}, we conclude that there exists a sequence of $(\ell_{\bar\ell}, M_{\bar\ell}, \epsilon)$-VLSF codes for all $\bar\ell\geq \ell_0$. Consequently, we have that for all $\bar \ell \geq \ell_0$,
\begin{IEEEeqnarray}{rCl}
  \IEEEeqnarraymulticol{3}{l}{\log \Msf(\ell_{\bar\ell},\epsilon)}\nonumber\\
   \quad&\geq& \log  M_{\bar\ell} \label{eq:asymp_achiev_Msf_lower_bound0}\\
  & \geq& C \left( \bar\ell - (1+\delta)g(C \bar\ell) \right) - \log \bar\ell -1\label{eq:asymp_achiev_Msf_lower_bound1}\\
  &=&\frac{C \ell_{\bar\ell}}{1-\epsilon} - (1+\delta) \sqrt{\frac{ V \ell_{\bar\ell}}{1-\epsilon}} \EBig{   \max_k   \tvRVbar_k } +\mathcal{O}(1).\label{eq:asymp_achiev_last_step}\IEEEeqnarraynumspace
\end{IEEEeqnarray}
Here, in \eqref{eq:asymp_achiev_Msf_lower_bound1}, we used that $\log(\lfloor x\rfloor) \geq \log(x-1) \geq \log x - 1$ for $x\geq 2$; furthermore, \eqref{eq:asymp_achiev_last_step} follows because
\begin{align}
  \ell_{\bar\ell} = \frac{(\bar\ell)^2 (1-\epsilon)}{\bar\ell-1} \leq \bar\ell (1-\epsilon) + o(1).
\end{align}
Since we can choose $\delta$ arbitrarily small, we conclude that \eqref{eq:asymp_achiev_last_step} implies \eqref{eq:TheoremTV_asymp}.

\paragraph*{Proof of \eqref{eq:Emaxtau_U0}} By Lemma~\ref{lem:time_varying}, there exists an integer $\ell_0$ such that, for all $\bar\ell\geq \ell_0$, we have
\begin{IEEEeqnarray}{rCl}
  \EBig{\max_k \tau_k(\gamma_{\bar\ell})}
  &\leq& \frac{\gamma_{\bar\ell}}{C}+(1+\delta)g(\gamma_{\bar\ell})\label{eq:asymp_Emax_final1}  \\
    &=&\bar\ell -  (1+\delta)g(C \bar\ell) \nonumber\\
    &&{}+ (1+\delta)g(C \bar\ell- C (1+\delta)g(C \bar\ell))\label{eq:asymp_Emax_final15}\IEEEeqnarraynumspace\\
    &\leq& \bar\ell.\label{eq:asymp_Emax_final2}
\end{IEEEeqnarray}
Here, \eqref{eq:asymp_Emax_final15} follows by the definition of $\gamma_{\bar\ell}$ in \eqref{eq:gamma_def}, and \eqref{eq:asymp_Emax_final2} holds because $g(x)$ is nonnegative and nondecreasing, which implies that $g(C\bar \ell) - g(C\bar\ell - C(1+\delta)g(C\bar\ell)) \geq 0$.

\paragraph*{Proof of the achievability part of Theorem~\ref{thm:asymp}} The achievability bound in Theorem~\ref{thm:asymp} follows by setting $\optbar(w)$ in Theorem~\ref{thm:time_varying} equal to the constant vector
\begin{IEEEeqnarray}{rCl}
\optbar_{\text{const}} \triangleq -\argmin_{\vect{v}\in\mathbb{R}_0^{|\mathcal{X}|}}\Ebigg{   \max_k   \diff I_k( \vect{v} ) +  \varrho_k Z_k }
\end{IEEEeqnarray}
where $Z_k \stackrel{\text{i.i.d.}}{\sim} \mathcal{N}(0,1)$. This implies that $\optbar'(w) = \vect{0}_{|\mathcal{X}|}$. Hence, $E_k(s) = 0$. In this case, we have that for every $w\in\mathbb{R}$,
\begin{IEEEeqnarray}{rCl}
F_{\tvRVbar_k}(w) &=& \Phi\farg{ \frac{w + \diff I_k(\optbar_{\text{const}})}{\varrho_k} }\\
  &=& \pr{-\diff I_k(\optbar_{\text{const}}) + \varrho_k Z_k \leq w}.
\end{IEEEeqnarray}
Hence, the $\{\tvRVbar_k\}$ have the same distribution as $\{-\diff I_k(\optbar_{\text{const}}) + \varrho_k Z_k\}$. The achievability part of Theorem~\ref{thm:asymp} is established by noting that 
\begin{IEEEeqnarray}{rCl}
\EBig{\max_k \tvRVbar_k} &=& \E{\max_k\mathopen{}\left\{ -\diff I_k(\optbar_{\text{const}}) + \varrho_k Z_k\right\}} \\
&=& \min_{\vect{v} \in\mathbb{R}_0^{|\mathcal{X}|}}\EBig{ \max_k\mathopen{}\left\{\diff I_k(\vect{v})  +\varrho_k Z_k\right\}}.\IEEEeqnarraynumspace
\end{IEEEeqnarray}

\section{Proof of Lemma~\ref{lem:tv_opt_cond}}
\label{app:tv_opt_cond}
  Our objective is to show that, if the conditions in Lemma~\ref{lem:tv_opt_cond} are satisfied, then $\vect{\beta}(w)$ given in \eqref{eq:betaw_def} satisfies 
  \begin{IEEEeqnarray}{rCl}
  P^*(x) + C P_{\sch(x)}^*(x) \beta'_{\sch(x)}(w)\in[0,1]\label{eq:tv_opt_cond_keycond}
  \end{IEEEeqnarray}
  for every $x\in\mathcal{X}$ and every $w\in\mathbb{R}$.
    First, given $w\in\mathbb{R}$, we shall analyze the function
  \begin{IEEEeqnarray}{rCl}
   v(w) \triangleq \argmax_{v\in\mathbb{R}} \prod_{k=1}^2 \Phi\farg{\frac{1}{\varrho_{ k}} \left(w +  v \Delta_{k}\right)}.\label{eq:cond_lem_opt}
  \end{IEEEeqnarray}
The objective function in \eqref{eq:cond_lem_opt} is differentiable everywhere in $v$. Furthermore, it is strictly log-concave in $v$ because $\Phi(\cdot)$ is strictly log-concave, and it tends to $0$ as $|v|\rightarrow \infty$ (recall that $\Delta_1 >0$ and $\Delta_2 < 0$). This implies that \eqref{eq:cond_lem_opt} has exactly one maximum which is also the unique stationary point.

By \eqref{eq:betaw_def}, we have that $\vect{\beta}(w) = [v(w), -v(w)]\transpose$. Therefore \eqref{eq:tv_opt_cond_keycond} is equivalent to
\begin{IEEEeqnarray}{rCl}
    P^*(x) +(-1)^{\sch(x)+1} C P^*_{\sch(x)} v'(w) \in[0,1]\label{eq:cond_lem_vdiff}
\end{IEEEeqnarray}
for every $x\in\mathcal{X}$ and every $w\in\mathbb{R}$.
We characterize $v'(w)$ in two steps. First, we show that $v'(w) > D$ if $\Delta_1/\varrho_1 + \Delta_2/\varrho_2 > 0$, that $v'(w) < D$ if $\Delta_1/\varrho_1 + \Delta_2/\varrho_2 < 0$, and that $v'(w) = D$ if $\Delta_1/\varrho_1 + \Delta_2/\varrho_2 = 0$. Next, we show that $v(\cdot)$ is monotonic. Specifically, we demonstrate that \eqref{eq:lem_cond2} implies that $v'(w) < 0$ if $\Delta_1/\varrho_1 + \Delta_2/\varrho_2 > 0$ and $v'(w) > 0$ if $\Delta_1/\varrho_1 + \Delta_2/\varrho_2 < 0$. By combining the two steps, we find that $\Delta_1/\varrho_1 + \Delta_2/\varrho_2 > 0$ implies $D < v'(w) < 0$ and that $\Delta_1/\varrho_1 + \Delta_2/\varrho_2 < 0$ implies $0 < v'(w) < D$. This argument establishes \eqref{eq:tv_opt_cond} because $P^*(x) \in[0,1]$ for all $x\in\mathcal{X}$. 

\paragraph*{First step}
  Let $\psi(x)\triangleq \phi(x)/\Phi(x)$. We characterize the stationary point of the objective function in \eqref{eq:cond_lem_opt} by taking its logarithm, by differentiating it with respect to $v$, and by equating the resulting expression to zero:
  \begin{IEEEeqnarray}{rCl}
  \sum_{k=1}^2 \psi\farg{\frac{w +  v \Delta_{k}}{\varrho_{ k}} } \frac{\Delta_k}{\varrho_k} = 0.\label{eq:cond_lem_kkt}
  \end{IEEEeqnarray}
  One can readily verify that \eqref{eq:cond_lem_kkt} has exactly one solution, which we denote by $v(w)$ to emphasize its dependence on $w$.
 Our objective is to characterize $v'(w)$. Hence, we differentiate both sides of \eqref{eq:cond_lem_kkt} with respect to $w$ to obtain an implicit equation for $v'(w)$:
  \begin{IEEEeqnarray}{rCl}
    \sum_{k=1}^2 \psi'\farg{\frac{w +  v(w) \Delta_{k}}{\varrho_{ k}} }\frac{\Delta_k}{\varrho_k^2} \left(1 +  v'(w) \Delta_{k}\right)  = 0.\label{eq:lem_stationarity_cond}
  \end{IEEEeqnarray}
  Let $\bar v(w,v)$ be the solution to the following equation in $z$
  \begin{IEEEeqnarray}{rCl}
    \sum_{k=1}^2 \psi'\farg{\frac{w +  v \Delta_{k}}{\varrho_{ k}}}\frac{\Delta_k}{\varrho_k^2} \left(1 +  z \Delta_{k}\right)  = 0.\label{eq:lem_vwv_def}
  \end{IEEEeqnarray}
  Note that we must have $v'(w) = \bar v(w, v(w))$. Solving \eqref{eq:lem_vwv_def} for $z$, we obtain that
  \begin{IEEEeqnarray}{rCl}
      \bar v(w,v)  = - \frac{ \mathlarger{\sum_{k=1}^2 \frac{\Delta_k}{\varrho_{ k}^2}} \psi'\farg{\mathlarger{\frac{w+v\Delta_k}{\varrho_k}}} }{\mathlarger{\sum_{k=1}^2 \frac{\Delta_k^2}{\varrho_{ k}^2}}\psi'\farg{\mathlarger{\frac{w+v\Delta_k}{\varrho_k}}}}\label{eq:barv_wv_func}
      \end{IEEEeqnarray}
  for every $w\in\mathbb{R}$ and every $v\in\mathbb{R}$. 
  Since $\psi'(\cdot)\in(-1,0)$, since $\psi'(x)$ is an increasing function in $x$ (this result is proven in Lemma~\ref{prop:psi_property}(a)-(b) in Appendix~\ref{app:lam_property}), and since $\Delta_1 > 0 > \Delta_2$, we conclude that $\bar v(w,v)$ is an increasing function of $v$ for fixed $w$. 

We proceed by noting that the following equation in $\zeta$
\begin{IEEEeqnarray}{rCl}
  \frac{w + \zeta w \Delta_1}{\varrho_1}=\frac{w + \zeta w \Delta_2}{\varrho_2}\label{eq:invariant}
\end{IEEEeqnarray}
is solved by
\begin{IEEEeqnarray}{rCl}
\zeta \triangleq \frac{\varrho_1 - \varrho_2}{\Delta_1\varrho_2 - \Delta_2\varrho_1}.\label{eq:lemma_alpha}
\end{IEEEeqnarray}
For the case $\Delta_1/\varrho_1+\Delta_2/\varrho_2 = 0$, we observe that $\zeta = D$ (recall that $D$ is defined in \eqref{eq:chi}) and that $v(w) = D w$ solves \eqref{eq:cond_lem_kkt}.
Next, consider the case $\Delta_1/\varrho_1 + \Delta_2/\varrho_2 > 0$. Define \begin{IEEEeqnarray}{rCl}
  \kappa_w(a) \triangleq \sum_{k=1}^2\psi\farg{\frac{w+a \Delta_k}{\varrho_k}} \frac{\Delta_k}{\varrho_k}.
\end{IEEEeqnarray}
Note that $\kappa_w(a)$ is a decreasing function in $a$ because $\psi(\cdot)$ is a decreasing function (proved in Lemma~\ref{prop:psi_property}(a)) and because $\Delta_1 > 0 > \Delta_2$. Additionally, \eqref{eq:cond_lem_kkt} implies that $\kappa_w(v(w)) = 0$. Now, we use \eqref{eq:invariant}, the positivity of $\psi(\cdot)$, and that $\Delta_1/\varrho_1 + \Delta_2/\varrho_2 > 0$,  to conclude that 
\begin{IEEEeqnarray}{rCl}
    \kappa_w(\zeta w) &=& \sum_{k=1}^2\psi\farg{\frac{w+\zeta w \Delta_k}{\varrho_k}} \frac{\Delta_k}{\varrho_k}\\
    &=& \psi\farg{\frac{w+\zeta w \Delta_1}{\varrho_1}}\left(\frac{\Delta_1}{\varrho_1}+\frac{\Delta_2}{\varrho_2}\right)\\
    &>& 0 \\
    &=& \kappa_w(v(w)).\label{eq:kappa_zetaw_vw_ineq}
\end{IEEEeqnarray}
Since $\kappa_w(\cdot)$ is a decreasing function, \eqref{eq:kappa_zetaw_vw_ineq} implies that $v(w) > \zeta w$. 
Thus, using that $\bar v(w,v)$ is increasing in $v$ for fixed $w$, we conclude that 
\begin{IEEEeqnarray}{rCl}
v'(w) = \bar v(w,v(w)) > \bar v(w,\zeta w) = D.
\end{IEEEeqnarray}
Following a similar line of reasoning, one can show that $\Delta_1/\varrho_1 + \Delta_2/\varrho_2 < 0$ implies $v'(w) < D$.

\paragraph*{Second step}
Next, we show that $v(w)$ is monotonic. First, note that the solutions $v^*$ and $w^*$ of the optimization problem
\begin{IEEEeqnarray}{rCl}
  \max_{v,w} \mathopen{}\left\{\e{-w\alpha} \prod_{k=1}^2 \Phi\farg{\frac{1}{\varrho_{ k}} \left(w +  v\Delta_{k}\right)}\right\}\label{eq:para_opt}
\end{IEEEeqnarray}
where $\alpha >0$, satisfy $v^* = v(w^*)$. The objective function in \eqref{eq:para_opt} is differentiable everywhere, strictly log-concave in $w$ and $v$, and it tends to zero as $|v|\rightarrow\infty$ or $|w|\rightarrow \infty$. These properties imply that \eqref{eq:para_opt} has exactly one maximum, which is also the unique stationary point.
We let $v^*(\alpha)$ and $w^*(\alpha)$ denote the solution of \eqref{eq:para_opt}. Then, for each $\tilde w \in\mathbb{R}$, there exists a constant $\tilde \alpha>0$ such that $v(\tilde w) = v^*(\tilde \alpha)$ and $\tilde w = w^*(\tilde \alpha)$. Hence, the set of points $\{(v(w),w)\}_{w\in\mathbb{R}}$ is equivalently parameterized by set of points $\{(v^*(\alpha),w^*(\alpha)\}_{\alpha>0}$.  We also point out that $w^*(\alpha)$ is nonincreasing in $\alpha$. By taking the logarithm of the objective function \eqref{eq:para_opt} and equating it to zero, we obtain the following stationarity conditions for $v^*(\alpha)$ and $w^*(\alpha)$
\begin{IEEEeqnarray}{rCl}
   \sum_{k=1}^2 \psi\farg{\frac{w^*( \alpha) +  v^*( \alpha) \Delta_{k}}{\varrho_{ k}} }\frac{1}{\varrho_k} &= \alpha\label{eq:para_kkt1}\\
   \sum_{k=1}^2 \psi\farg{\frac{w^*( \alpha) +  v^*( \alpha) \Delta_{k}}{\varrho_{ k}} }\frac{\Delta_k}{\varrho_k} &= 0.\label{eq:para_kkt2}
\end{IEEEeqnarray}
Solving \eqref{eq:para_kkt1} and \eqref{eq:para_kkt2} for $v^*(\alpha)$ and $w^*(\alpha)$, we find that
\begin{IEEEeqnarray}{rCl}
  v^*(\alpha) &=& \frac{1}{\Delta_1-\Delta_2} \bigg( \varrho_1 \psi^{-1}\farg{\frac{\alpha \varrho_1}{1-\Delta_1/\Delta_2}}\nonumber\\
  &&\qquad\qquad\quad{}-\varrho_2 \psi^{-1}\farg{\frac{\alpha \varrho_2}{1-\Delta_2/\Delta_1}} \bigg)\label{eq:para_vopt}\\
    w^*(\alpha) &=& \frac{\Delta_1\Delta_2}{\Delta_2-\Delta_1} \bigg( \frac{\varrho_1}{\Delta_1} \psi^{-1}\farg{\frac{\alpha \varrho_1}{1-\Delta_1/\Delta_2}}\nonumber\\
    &&\qquad\qquad\quad{}-\frac{\varrho_2}{\Delta_2}\psi^{-1}\farg{\frac{\alpha \varrho_2}{1-\Delta_2/\Delta_1}} \bigg).\IEEEeqnarraynumspace
\end{IEEEeqnarray}
Note that $\Delta_1 - \Delta_2 > 0$ due to the assumption $\Delta_1 > 0 > \Delta_2$.
Since $w^*(\alpha)$ is a nonincreasing function of $\alpha$, our objective is to show that $v^*(\alpha)$ is either nonincreasing or nondecreasing in $\alpha$. In particular, if $v^*(\alpha)$ is nonincreasing then $v(w)$ must be nondecreasing and vice versa. Let $g(x)\triangleq -\psi' ( \psi^{-1}\farg{ x }) $ for $x>0$. By taking the derivative of \eqref{eq:para_vopt} with respect to $\alpha$, we obtain
\begin{multline}
  \frac{\partial v^*}{\partial \alpha}\\ = \frac{1}{\Delta_1 - \Delta_2} \left( \frac{ \varrho_2^2/(1-\Delta_2/\Delta_1)}{g\farg{\frac{\alpha \varrho_2}{1-\Delta_2/\Delta_1}} }-\frac{ \varrho_1^2/(1-\Delta_1/\Delta_2)}{g\farg{\frac{\alpha \varrho_1}{1-\Delta_1/\Delta_2}}}\right).\label{eq:para_vopt_diff}
\end{multline}
Note that $g(x)$ is a positive function, which satisfies the following property: $\beta g(x) \geq g(\beta x) $ for $\beta\geq 1$ and $x>0$ (this is proved in Lemma~\ref{prop:psi_property}(c) in Appendix~\ref{app:lam_property}). This implies that, when $\beta>1$,
\begin{IEEEeqnarray}{rCl}
\frac{g(\beta x)}{g(x)} < \beta.\label{eq:gbeta_leq_beta}
\end{IEEEeqnarray}
Similarly, when $0<\beta<1$, one readily finds that \eqref{eq:gbeta_leq_beta} implies
\begin{IEEEeqnarray}{rCl}
\frac{g(\beta x)}{g(x)} > \frac{1}{\beta}.\label{eq:gbeta_leq_beta2}
\end{IEEEeqnarray}
To show monotonicity of $v^*(\alpha)$, we analyze the sign of \eqref{eq:para_vopt_diff}. Let
\begin{IEEEeqnarray}{rCl} 
    h(\alpha) &\triangleq& \frac{ \varrho_2^2 (1-\Delta_1/\Delta_2) g\farg{\frac{\alpha \varrho_1}{1-\Delta_1/\Delta_2}}}{{\varrho_1^2(1-\Delta_2/\Delta_1) g\farg{\frac{\alpha \varrho_2}{1-\Delta_2/\Delta_1}}}}\\
     &=& -\frac{ \varrho_2^2 \Delta_1 g\farg{\frac{\alpha \varrho_1}{1-\Delta_1/\Delta_2}}}{{\varrho_1^2 \Delta_2 g\farg{\frac{\alpha \varrho_2}{1-\Delta_2/\Delta_1}}}}.
\end{IEEEeqnarray}
Note that $h(\alpha)>1$ implies $\partial v^*/\partial \alpha > 0$. Furthermore, $h(\alpha)<1$ implies $\partial v^*/\partial \alpha < 0$.

Consider the case $\Delta_1/\varrho_1 +\Delta_2/\varrho_2 > 0$. By \eqref{eq:lem_cond2}, we must have that $\varrho_2 \geq \varrho_1$. But this implies that
\begin{IEEEeqnarray}{rCl}
   h(\alpha) &>& \frac{\varrho_2}{\varrho_1} \frac{  g\farg{\frac{\alpha \varrho_1}{1-\Delta_1/\Delta_2}}}{{ g\farg{\frac{\alpha \varrho_2}{1-\Delta_2/\Delta_1}}}}\label{eq:halpha_1}\\ 
   &\geq& -\frac{ \varrho_2 }{{\varrho_1}}\frac{ \varrho_2 \Delta_1 }{{ \varrho_1 \Delta_2}}\label{eq:halpha_2}\\
   &\geq&\frac{\varrho_2}{\varrho_1}\label{eq:halpha_3}\\
   &\geq& 1.\label{eq:halpha_4}
\end{IEEEeqnarray}
Here, \eqref{eq:halpha_1} follows from $\Delta_1\varrho_2/(\Delta_2\varrho_1) < -1$; \eqref{eq:halpha_2} follows from \eqref{eq:gbeta_leq_beta2} with $x = \alpha \varrho_2/(1-\Delta_2/\Delta_1)>0$ and $\beta = -\varrho_1\Delta_2/(\varrho_2 \Delta_1)\in(0,1)$, \eqref{eq:halpha_3}  holds because $\Delta_1\varrho_2/(\Delta_2\varrho_1) < -1$, and \eqref{eq:halpha_4}  holds because $\varrho_2/\varrho_1 \geq 1$.
Using a similar line of reasoning, we obtain for the case $\Delta_1/\varrho_1 +\Delta_2/\varrho_2 < 0$ that
\begin{IEEEeqnarray}{rCl}
   h(\alpha) &<&  1.
\end{IEEEeqnarray}
Using that $h(\alpha) > 1$ implies $\partial v^*/\partial \alpha > 0$ and that $h(\alpha) < 1$ implies $\partial v^*/\partial \alpha < 0$, we conclude that $D <v'(w) < 0$ if $\Delta_1/\varrho_1 +\Delta_2/\varrho_2 > 0$ and that $D > v'(w) > 0$ if $\Delta_1/\varrho_1 +\Delta_2/\varrho_2 < 0$.

\section{Proof of Lemma~\ref{lem:time_varying}}
\label{proof:tv_asymp_time_sharing}
To prove Lemma~\ref{lem:time_varying}, we shall first construct a suitable nonstationary joint probability distribution $P_{X^\infty}$ on $\mathcal{X}^\infty$. Then we shall set
\begin{IEEEeqnarray}{rCl}
  \beta_- \triangleq \bigg\lfloor \frac{\gamma}{C} - \sqrt{\frac{\gamma V}{C^3}} \log \gamma \bigg\rfloor 
  \end{IEEEeqnarray}
   and
   \begin{IEEEeqnarray}{rCl}
    \beta_+ \triangleq \bigg\lfloor \frac{\gamma}{C} + \sqrt{\frac{\gamma V}{C^3}} \log \gamma \bigg\rfloor
\end{IEEEeqnarray}
and compute an asymptotic upper bound on
\begin{IEEEeqnarray}{rCl}
  \EBig{\max_k \tau_k} &=& \sum_{t=0}^\infty \Big(1-\prBig{\max_k \tau_k \leq t}\Big)\\
                  &\leq& \beta_-+1 + \sum_{t=\beta_-+1}^{\beta_+} \Big(1-\prBig{\max_k \tau_k \leq t}\Big)\nonumber\\
                  &&{}+\sum_{t=\beta_++1}^\infty \Big(1-\prBig{\max_k \tau_k \leq t}\Big)\label{eq:tv_achiev_upper_bound_terms}
\end{IEEEeqnarray} 
that matches \eqref{eq:lem_Emaxtau} in the limit $\gamma\rightarrow \infty$. This is done by obtaining lower bounds on $\pr{\max_k \tau_k \leq t}$ for $t\geq \beta_- + 1$.
Similarly as in Appendix~\ref{sec:converse_asymptotics}, it turns out convenient to treat the two subintervals $[\beta_-+1, \beta_+]$ and $[\beta_++1,\infty)$ differently. In the former subinterval, our main tool is a multivariate version of the Berry-Esseen central limit theorem for sums of independent RVs. In the latter subinterval, we  apply Hoeffding's inequality.
To compute the desired lower bound on $\pr{\max_k \tau_k \leq t}$, we shall use the following relation between $\tau_k$ and $i_{P_{X^t},W_k^t}(X^t; Y^t_k)$, which follows from the definition of $\tau_k$ in \eqref{eq:achiev_xn}:
\begin{IEEEeqnarray}{rCl}
    \prBig{\max_k \tau_k \leq t} &\geq& \prBig{\min_k i_{P_{X^t},W_k^t}(X^t; Y^t_k)\geq \gamma }.\label{eq:tv_achiev_tau_inf_rel0}
\end{IEEEeqnarray}   

We next summarize the key steps of the proof. Details are provided in Appendix~\ref{sec:Coptbar_approx}--\ref{eq:achiev_central_regime}.
\paragraph*{Step 1}
First, we specify the probability distribution $P_{X^\infty}$ on $\mathcal{X}^\infty$ for which \eqref{eq:lem_Emaxtau} holds. Let 
\begin{IEEEeqnarray}{rCl}
\convnorm{t} &\triangleq& \frac{t - \gamma/C}{\sqrt{\gamma V/C^3}}\label{eq:convnorm_def}\\
\bar P^{(1)}(\gamma) &\triangleq& P^* +\sqrt{V C/\gamma} \optbar\farg{\convnorm{\beta_-}}\label{eq:tv_barP1}\\
P^{(2)}(w) &\triangleq& P^* + C \optbar'\farg{w}\label{eq:tv_barP2}\\
P^{(3)}(\gamma) &\triangleq& P^* +\sqrt{V C/\gamma} \optbar\farg{\convnorm{\beta_+}}\label{eq:tv_barP3}
\end{IEEEeqnarray}
and let $P^{(1)}(\gamma)\in\mathcal{P}_{\beta_-}(\mathcal{X})$ be the type that minimizes $\vectornorm{\bar P^{(1)}(\gamma)-P^{(1)}(\gamma)}$ (recall that $\mathcal{P}_n(\mathcal{X})$ denotes the set of types of $n$-dimensional sequences and that $\vectornorm{\cdot}$ denotes the Euclidean distance). For all sufficiently large $\gamma$, $\bar P^{(1)}(\gamma)$ and $\bar P^{(3)}(\gamma)$ are legitimate probability distributions, and  \eqref{eq:tv_input_dist_cond} implies that $P^{(2)}(w)$ is a valid probability distribution as well. The probability distribution $P_{X^\infty}$ is specified as follows. 
We let the distribution $P_{X^{\beta_-}}$ of $X^{\beta_-}$ be uniform over the set of all codewords of type $P^{(1)}(\gamma)$. For $t\in[\beta_-+1,\beta_+]$, the RVs $\{X_t\}$ are generated independently according to $P^{(2)}(\convnorm{t})$ and, for $t\geq \beta_++1$, the RVs $\{X_t\}$ are generated independently according to $P^{(3)}(\gamma)$. For notational convenience, the dependency of $P^{(1)}(\gamma)$ and $P^{(3)}(\gamma)$ on $\gamma$ is omitted in the remainder of the proof. 

We need $X^{\beta_-}$ to be of constant composition because the capacity-achieving input distributions of the components channels $\{W_k\}$ are generally not given by $P^*$. The above choice of the distribution of $X^{\beta_-}$ thereby parallels the achievability proof of the asymptotic expansion of the maximum coding rate for compound DMCs for the fixed blocklength case \cite{Polyanskiy}.

We note that $\vectornorm{\bar P^{(1)}-P^{(1)}}_1\leq \mathcal{O}(1/\gamma)$
as $\gamma\rightarrow \infty$. Furthermore, differentiability of $I_k(\cdot)$  implies that
\begin{IEEEeqnarray}{rCl}
I_k\farg{P^{(1)}} = I_k\farg{\bar P^{(1)}} + \mathcal{O}\farg{\frac{1}{\gamma}}.\label{eq:Ik_P_Pgam_approx}
\end{IEEEeqnarray}
\begin{sloppypar}\noindent Additionally, since $X^{\beta_-}$ are of constant composition, we cannot write $i_{P_{X^{\beta_-}}, W_k^{\beta_-}}(X^{\beta_-};Y_k^{\beta_-})$ as a sum of independent RVs since $P_{Y^{\beta_-}}$ is not a product distribution. Hence, to dispose of the dependency among the RVs $X^{\beta_-}$, we use the inequality \cite[Eq.~(4.49)]{Tan2014}\end{sloppypar}
\begin{multline}
 P_{Y_k^{\beta_-}}(\mathbf{y})\\= P_{X^{\beta_-}} W_k^{\beta_-}(\vect{y}) \leq |\mathcal{P}_{\beta_-}(\mathcal{X})|(P^{(1)} W_k)^{\beta_-}(\vect{y})
\end{multline}
which holds for all $\vect{y}\in\mathcal{Y}_k^{\beta_-}$,
and the inequality $|\mathcal{P}_n(\mathcal{X})|\leq (n+1)^{|\mathcal{X}|}$ \cite[Th.~11.1.1]{Cover2012} to conclude that, for all $t\geq \beta_-$,
\begin{IEEEeqnarray}{rCl}
  \IEEEeqnarraymulticol{3}{l}{i_{P_{X^{t}},W_k^{t}}(x^{t}; y_{k}^{t})}\nonumber\\
  \qquad &=& \log \frac{W_k^{\beta_-}(y_k^{\beta_-} | x^{\beta_-})}{P_{Y_k^{\beta_-}(y_k^{\beta_-})}} \nonumber\\
   &&{}+ \sum_{n = \beta_-+1}^t i_{P_{X_n},W_{k,n}}(x_n; y_{k,n})\\
  &\geq& \log \frac{\prod_{n=1}^{\beta_-} W_k(y_{k,n}|x_n)}{ |\mathcal{P}_{\beta_-}(\mathcal{X})| (P^{(1)} W_k)^{\beta_-}(y_k^{\beta_-})}\nonumber\\
  &&{}+ \sum_{n = \beta_-+1}^t i_{P_{X_n},W_{k,n}}(x_n; y_{k,n})\\
  &\geq& \sum_{n=1}^{t} i_{P_{X_n}, W_k}(x_n;y_{k,n}) -|\mathcal{X}| \log(\beta_-+1).\label{eq:inf_dens_constant_comp}
\end{IEEEeqnarray}
It follows from \eqref{eq:tv_achiev_tau_inf_rel0} and \eqref{eq:inf_dens_constant_comp} that
\begin{IEEEeqnarray}{rCl}
 \prBig{\max_k \tau_k \leq t}
  &\geq& \prBigg{\min_k \mathopen{}\bigg\{\sum_{n=1}^{t} i_{P_{X_n}, W_k}(X_n;Y_{k,n}) \nonumber\\
  &&{}\qquad\qquad-|\mathcal{X}| \log(\beta_-+1)\bigg\}\geq \gamma }.\label{eq:tv_achiev_tau_inf_rel}\IEEEeqnarraynumspace
\end{IEEEeqnarray}
We also note that the marginal probability distribution of $X_t$, for $t\leq \beta_-$, is given by $P^{(1)}$.

\begin{sloppypar}
\paragraph*{Step 2} We shall next estimate the expected value of $\sum_{n=1}^{t} i_{P_{X_n}, W_k}(X_n;Y_{k,n})$, which is given by $\sum_{i=1}^t I_k(P_{X_i})$. This is needed to lower-bound $\prBig{\min_k i_{P_{X^t}; W_k^t}(X^t, Y^t_k)\geq \gamma }$ using Hoeffding's inequality for $t\leq \beta_-$ and a multivariate version of the Berry-Esseen central limit theorem for the case $t\geq \beta_-+1$.
  %
%
We first treat the case $t\in[\beta_- +1,\beta_+]$. We will return to the case $t\in[\beta_++1,\infty)$ shortly. For $t\in[\beta_- +1,\beta_+]$, we have that\end{sloppypar}
\begin{IEEEeqnarray}{rCl}
  \IEEEeqnarraymulticol{3}{l}{\sum_{i=1}^t I_k(P_{X_i})}\nonumber\\
  \quad &=& \sum_{i=\beta_-+1}^t I_k\farg{P^{(2)}(\convnorm{i})}+\beta_- I_k\farg{P^{(1)}}\label{eq:improv_achiev_use_estimate00}\\
  &=& \sum_{i=\beta_-+1}^t  \Big[C - E_k(\convnorm{i} )+ C \diff I_k\farg{\optbar'(\convnorm{i})}\Big]\nonumber\\
  &&{}+\beta_-\left(C + \sqrt{\frac{VC}{\gamma}}\diff I_k(\optbar(\convnorm{\beta_-}) \right)+\mathcal{O}(1).\label{eq:improv_achiev_use_estimate01}\IEEEeqnarraynumspace
\end{IEEEeqnarray}
Here, in \eqref{eq:improv_achiev_use_estimate00} we used that the marginal distribution of $P_{X_t}$ for $t\in[1,\beta_-]$ is given by $P^{(1)}$. To obtain \eqref{eq:improv_achiev_use_estimate01}, we used \eqref{eq:Ik_P_Pgam_approx} and that $\beta_- = \mathcal{O}(\gamma)$. Furthermore, we performed a Taylor-expansion of $I_k(P^{(1)})$ around $P^*$, and used that $E_{k}(s) = C-  I_k(P^{(2)}(s)) + C \ \diff I_k(\optbar'(s))$ (recall the definition of $E_k(s)$ in \eqref{eq:Eks_cond} and of $P^{(2)}$ in \eqref{eq:tv_barP2}). We note that $C \diff I_k(\optbar'(s)) = \diff I_k(C\optbar'(s))$ because $\diff I_k(\cdot)$ is linear. To simplify \eqref{eq:improv_achiev_use_estimate01}, we shall apply the following two asymptotic expansions, which are proven in Appendix~\ref{sec:Coptbar_approx} and Appendix~\ref{sec:ts_error_bound}, respectively:
  \begin{IEEEeqnarray}{rCl}
 \IEEEeqnarraymulticol{3}{l}{C \diff I_k (\optbar'(\convnorm{t}))}\nonumber\\
  &=& t\sqrt{\frac{V C}{\gamma}} \diff I_k(\optbar(\convnorm{t}))- (t-1)\sqrt{\frac{V C}{\gamma}} \diff I_k(\optbar(\convnorm{t-1})) \nonumber\\
  &&{} + \mathcal{O}\farg{\frac{\log \gamma}{\sqrt{\gamma}}}\IEEEeqnarraynumspace\label{eq:Coptbar_approx}
 \end{IEEEeqnarray}
 and
\begin{IEEEeqnarray}{rCl}
  \sum_{i=\beta_-+1}^{t} E_k(w(i)) &=& \sqrt{\frac{\gamma V}{C^3}} \int_{w(\beta_-)}^{w(t)} E_k(s) \intdiff s  +  \mathcal{O}\farg{\log \gamma}\label{eq:ts_error_bound}\IEEEeqnarraynumspace
\end{IEEEeqnarray}
as $\gamma\rightarrow \infty$ for all $t\in[\beta_-+1,\beta_+]$.
By substituting \eqref{eq:Coptbar_approx} and \eqref{eq:ts_error_bound} into \eqref{eq:improv_achiev_use_estimate01}, we obtain \eqref{eq:improv_achiev_use_estimate}--\eqref{eq:improv_achiev_inductive}, shown in the top of the next page.\begin{figure*}[!t]
\normalsize
\setcounter{MYtempeqncnt}{\value{equation}}
\setcounter{equation}{371}
\begin{IEEEeqnarray}{rCl}
\sum_{i=1}^t I_k(P_{X_i})
 &=&   \sum_{i=\beta_-+1}^t  \Bigg(C + i \sqrt{\frac{V C}{\gamma}} \diff I_k(\optbar(\convnorm{i})) - (i-1)\sqrt{\frac{VC}{\gamma}}\diff I_k(\optbar(\convnorm{i-1})) \Bigg)\nonumber\\
 &&{}+ \beta_-\left(C + \sqrt{\frac{VC}{\gamma}}\diff I_k(\optbar(\convnorm{\beta_-}) \right) - \sqrt{\frac{\gamma V}{C^3}} \int_{w(\beta_-) }^{w(t)} E_k(s) \intdiff s + \mathcal{O}(\log\gamma)+ (t-\beta_-)\mathcal{O}\farg{\frac{\log \gamma}{\sqrt{\gamma}}}\label{eq:improv_achiev_use_estimate}\\
  &=&  t\left( C+\sqrt{\frac{V C}{\gamma}} \diff I_k(\optbar(\convnorm{t}))\right)-\sqrt{\frac{\gamma V}{C^3}}\int_{w(\beta_-) }^{w(t)} E_k(s) \intdiff s+ \mathcal{O}(\log^2 \gamma).\label{eq:improv_achiev_inductive}
\end{IEEEeqnarray}

\setcounter{equation}{\value{MYtempeqncnt}}
\hrulefill
\vspace*{4pt}
\end{figure*}\addtocounter{equation}{2}
Here, \eqref{eq:improv_achiev_inductive} follows because $(t-\beta_-)\leq (\beta_{+}-\beta_-)=\mathcal{O}(\sqrt{\gamma}\log \gamma)$.

We now move to the case $t \in[\beta_++1,\infty)$ for which, proceeding as in \eqref{eq:improv_achiev_use_estimate00}--\eqref{eq:improv_achiev_inductive}, we obtain that
\begin{IEEEeqnarray}{rCl}
\sum_{i=1}^t I_k(P_{X_i}) &=& \beta_+\left( C+\sqrt{\frac{V C}{\gamma}} \diff I_k(\optbar(\convnorm{\beta_+}))\right)\nonumber\\
&&{}-\sqrt{\frac{\gamma V}{C^3}}\int_{w(\beta_-) }^{w(\beta_+)} E_k(s) \intdiff s\nonumber\\
&&{} + \sum_{i=\beta_++1}^t I_k\farg{P^{(3)}} + \mathcal{O}(\log^2 \gamma)\label{eq:Ik_large_deviation01}\\
&=& t \left(C + \mathcal{O}\left(\frac{\log \gamma}{\sqrt{\gamma}}\right) \right). \label{eq:Ik_large_deviation02}
\end{IEEEeqnarray}
Here, \eqref{eq:Ik_large_deviation01} follows from \eqref{eq:improv_achiev_inductive} and  \eqref{eq:Ik_large_deviation02} follows because $\int_{-\infty}^\infty E_k(s)\intdiff s< \infty$ (see~\eqref{eq:int_Ek_bound}) and because $I_k\farg{P^{(3)}} = C + \mathcal{O}(1/\sqrt{\gamma})$. We have also used that  $\diff I_k(\optbar(\convnorm{\beta_+})) = \mathcal{O}(\log \gamma)$, which follows because $\vect{\bar v}'$ is bounded by  \eqref{eq:tv_input_dist_cond} and from $w(\beta_+) = \mathcal{O}(\log \gamma)$. 

\paragraph*{Step 3} We now use \eqref{eq:improv_achiev_inductive} and \eqref{eq:Ik_large_deviation02} to compute the second and the third term in \eqref{eq:tv_achiev_upper_bound_terms}. By applying Hoeffding's inequality and then using \eqref{eq:Ik_large_deviation02}, we demonstrate in Appendix~\ref{sec:tv_large_dev} that
\begin{IEEEeqnarray}{rCl}
  \sum_{t=\beta_++1}^\infty \left(1-\prBig{\max_k \tau_k \leq t} \right) = o(1)\label{eq:tv_large_dev_bound}
\end{IEEEeqnarray}
as $\gamma\rightarrow \infty$. Hence, third term in \eqref{eq:tv_achiev_upper_bound_terms} vanishes as $\gamma \rightarrow \infty$. 

\begin{sloppypar}\noindent Next, we analyze the second term in \eqref{eq:tv_achiev_upper_bound_terms}, which require bounds on  $\prBig{\min_k i_{P_{X^t}; W_k^t}(X^t, Y^t_k)\geq \gamma }$ for $t\in[\beta_-+1,\beta_+]$. Let $\delta$ be an arbitrary positive constant. In Appendix~\ref{eq:achiev_central_regime}, we show using \eqref{eq:improv_achiev_inductive} and a multivariate version of the Berry-Esseen central limit theorem  for sums of independent RVs \cite[Th.~1.8]{Tan2014}, \cite[Th~1.3]{Gotze1991} that\end{sloppypar}
\begin{multline}
    \prBig{\min_k i_{P_{X^t},W_k^t}(X^t; Y_k^t) \geq \gamma}\\\geq \prod_k F_{\tvRVbar_{\delta,k}}(\convnorm{t}) + \mathcal{O}\farg{\frac{\log^2 \gamma}{\sqrt{\gamma}}}\label{eq:tv_achiev_central_bound}
\end{multline}
as $\gamma\rightarrow \infty$.  
Here, the $\mathcal{O}(\cdot)$ term is uniform in $t\in[\beta_-+1,\beta_+]$. The RVs $\{\tvRVbar_{\delta,k}\}$ have the cumulative distribution function
\begin{multline}
  F_{\tvRVbar_{\delta,k}}(w) \\\triangleq \Phi\farg{\frac{1}{\varrho_k} \min_{\nu_k \in\{-1,1\}} \frac{w + \diff I_k(\optbar(w)) - \frac{1}{C}\int_{-\infty}^w E_k(s)\intdiff s  }{1-\delta \nu_k}}.
\end{multline}
Thus, we have that
\begin{IEEEeqnarray}{rCl}
 \IEEEeqnarraymulticol{3}{l}{\sum_{t = \beta_-+1}^{\beta_+} \prBig{\max_k \tau_k \leq  t }}\nonumber\\
  &\geq& \sum_{t = \beta_-+1}^{\beta_+} \prod_k F_{\tvRVbar_{\delta,k}}\farg{- \frac{\gamma/C-t}{\sqrt{\gamma V/C^3}}}  +\mathcal{O}\mathopen{}(\log^3 \gamma)\label{eq:achiev_three_sums_improv00}\\
  &=& \int_{\beta_-}^{\beta_+} \prod_k F_{\tvRVbar_{\delta,k}}\farg{- \frac{\gamma/C-t}{\sqrt{\gamma V/C^3}}    }  \intdiff t +\mathcal{O}\mathopen{}(\log^3 \gamma)\label{eq:achiev_three_sums_improv}\\
 &=& \int_{\beta_-}^{\beta_+} \prod_k \prbigg{ \frac{\gamma}{C}+ \sqrt{ \frac{\gamma V}{C^3} } \tvRVbar_{\delta,k} \leq t}\intdiff t + \mathcal{O}(\log^3 \gamma)\label{eq:achiev_Q_to_prob_improv}\\
 &=& \int_{\beta_-}^{\beta_+}  \prbigg{ \max_k\mathopen{}\bigg\{  \frac{\gamma}{C}+ \sqrt{ \frac{\gamma V}{C^3} }\tvRVbar_{\delta,k}  \bigg\} \leq t}\intdiff t \nonumber\\
 &&{}+ \mathcal{O}(\log^3 \gamma) \\
    &\geq& \Ebigg{ \minnbigg{\beta_+ - \beta_-, \beta_+ - \max_k\mathopen{}\bigg\{ \frac{\gamma}{C} + \sqrt{ \frac{\gamma V}{C^3} }\tvRVbar_{\delta,k}\bigg\} } } \nonumber\\
    &&{}+ \mathcal{O}(\log^3 \gamma).\label{eq:tv_int_to_Emin}
\end{IEEEeqnarray}
Here, \eqref{eq:achiev_three_sums_improv00} follows from \eqref{eq:convnorm_def}, from \eqref{eq:tv_achiev_central_bound}, and because $(\beta_+-\beta_-) = \mathcal{O}(\sqrt{\gamma}\log \gamma)$; \eqref{eq:achiev_three_sums_improv} follows because $F_{\tvRVbar_{\delta,k}}(w)$ is a nondecreasing function in $w$ upper-bounded by one; and \eqref{eq:tv_int_to_Emin} follows because, for a continuous RV $X$ with probability density function $p_X(x)$ and $b \geq a$, we have that
\begin{IEEEeqnarray}{rCl}
  \IEEEeqnarraymulticol{3}{l}{\int_{a}^b \pr{X \leq x} \intdiff x}\nonumber\\
  \qquad  &=& \int_a^b \int_{-\infty}^x p_X(s) \intdiff s \intdiff x \\
  &=& \int_{a}^b \int_{-\infty}^\infty p_X(s)\indi{x \geq s}  \intdiff s  \intdiff x\\
  &=&  \int_{-\infty}^\infty p_X(x) \int_{a}^b \indi{x \geq s} \intdiff x \intdiff s  \label{eq:tonelli}\\
  &=& \int_{-\infty}^\infty p_X(s) \minn{b-a, (b-s)^+} \intdiff s\\
  &\geq& \E{\minn{b-a, b-X}}.
\end{IEEEeqnarray}
In \eqref{eq:tonelli}, we used Tonelli's theorem \cite[Th.~15.8]{apostel}. Now, since  $\delta$ can be chosen arbitrarily small, we obtain from \eqref{eq:tv_int_to_Emin} the asymptotic bound
\begin{IEEEeqnarray}{rCl}
\IEEEeqnarraymulticol{3}{l}{\sum_{t = \beta_-+1}^{\beta_+} \prBig{\max_k \tau_k \leq  t }}\nonumber\\
  &\geq& \Ebigg{ \minnbigg{\beta_+ - \beta_-, \beta_+ - \max_k\mathopen{}\bigg\{ \frac{\gamma}{C}+\sqrt{ \frac{\gamma V}{C^3} }\tvRVbar_k\bigg\}}} \nonumber\\
  &&{}+ o(\sqrt{\gamma})\\
&=& \beta_+ - \Ebigg{ \maxxbigg{ \beta_-,   \max_k\mathopen{}\bigg\{ \frac{\gamma}{C} +\sqrt{ \frac{\gamma V}{C^3} }\tvRVbar_k\bigg\}}}\nonumber\\
&&{} + o(\sqrt{\gamma})\\
&=& \beta_+ - \frac{\gamma}{C} - \sqrt{\frac{\gamma V}{C^3}}\EBig{ \maxxBig{ {-\log \gamma},   \max_k \tvRVbar_k}}\nonumber\\
&&{} + o(\sqrt{\gamma})\\
&=& \beta_+ - \frac{\gamma}{C} - \sqrt{\frac{\gamma V}{C^3}}\EBig{\max_k \tvRVbar_k} + o(\sqrt{\gamma}).\label{eq:improv_achiev_final_central_bound}
\end{IEEEeqnarray}
Recall that the $\{\tvRVbar_k\}$ have cumulative distribution function given in \eqref{eq:FtvRVk_def}.

Finally, substituting \eqref{eq:tv_large_dev_bound} and \eqref{eq:improv_achiev_final_central_bound} in \eqref{eq:tv_achiev_upper_bound_terms}, we obtain the desired result \eqref{eq:lem_Emaxtau}.

\subsection{Proof of \eqref{eq:Coptbar_approx}}\label{sec:Coptbar_approx}
We prove \eqref{eq:Coptbar_approx}, for $t\in[\beta_-+1,\beta_+]$, through the chain of equalities \eqref{eq:ts_asymp_convnorm_used}--\eqref{eq:ts_asymp_conve_hatx_diff2}, shown in the top of the next page.\begin{figure*}[!t]
\normalsize
\setcounter{MYtempeqncnt}{\value{equation}}
\setcounter{equation}{392}
\begin{IEEEeqnarray}{rCl}
\IEEEeqnarraymulticol{3}{l}{ t\sqrt{\frac{V C}{\gamma}} \diff I_k(\optbar(\convnorm{t})) - (t-1)\sqrt{\frac{V C}{\gamma}} \diff I_k(\optbar(\convnorm{t-1}) ) }\nonumber\\
\qquad &=& \diff I_k\fargBigg{\left(\frac{\gamma}{C} +  \sqrt{\frac{\gamma V}{C^3}}\convnorm{t}\right) \sqrt{\frac{V C}{ \gamma}} \optbar\farg{\convnorm{t}}-\left(\frac{\gamma}{C} +\sqrt{\frac{\gamma V}{C^3}} \convnorm{t}-1\right) \sqrt{\frac{V C}{ \gamma}} \optbar\farg{\convnorm{t} -\sqrt{\frac{C^3}{\gamma V}}}}\label{eq:ts_asymp_convnorm_used}\IEEEeqnarraynumspace\\
&=& \diff I_k\farg{\left(\frac{\gamma}{C} + \sqrt{\frac{\gamma V}{C^3}}\convnorm{t}\right) \sqrt{\frac{V C}{ \gamma}} \optbar\farg{\convnorm{t}}-\left(\frac{\gamma}{C} +\sqrt{\frac{\gamma V}{C^3}} \convnorm{t}-1\right) \sqrt{\frac{V C}{\gamma}} \left(\optbar\farg{\convnorm{t}} - \sqrt{\frac{C^3}{\gamma V}}\optbar'\farg{\convnorm{t}}\right)}\nonumber\\
&&{} + \mathcal{O}\farg{\frac{1}{\sqrt{\gamma}}}\label{eq:ts_asymp_conve_hatx_diff}\\
&=& \diff I_k\farg{\sqrt{\frac{V C}{\gamma}} \optbar\farg{\convnorm{t}}+\left(\frac{\gamma}{C} +\sqrt{\frac{\gamma V}{C^3}} \convnorm{t}-1\right) \frac{C^2}{\gamma} \optbar'\left( \convnorm{t}\right)} + \mathcal{O}\farg{\frac{1}{\sqrt{\gamma}}}\IEEEeqnarraynumspace\\
&=& \sqrt{\frac{V C}{\gamma}} \diff I_k\farg{\optbar\farg{\convnorm{t}}}+\left(\sqrt{\frac{ V C}{ \gamma}} \convnorm{t}-\frac{C^2}{\gamma}\right) \diff I_k\farg{\optbar'\farg{\convnorm{t}}} + C \diff I_k(\optbar'(\convnorm{t}))+ \mathcal{O}\farg{\frac{1}{\sqrt{\gamma}}}\\
&=& C \diff I_k(\optbar'(\convnorm{t}))+ \mathcal{O}\farg{\frac{\log \gamma}{\sqrt{\gamma}}}.\label{eq:ts_asymp_conve_hatx_diff2}
\end{IEEEeqnarray}

\setcounter{equation}{\value{MYtempeqncnt}}
\hrulefill
\vspace*{4pt}
\end{figure*}\addtocounter{equation}{5}
Here, \eqref{eq:ts_asymp_convnorm_used} follows because $\gamma/C + \sqrt{\gamma V/C^3}\convnorm{t} = t$ (see \eqref{eq:convnorm_def}); moreover, \eqref{eq:ts_asymp_conve_hatx_diff} follows from a first-order Taylor expansion of $\optbar(\cdot)$ around $\convnorm{t}$,
 and \eqref{eq:ts_asymp_conve_hatx_diff2} follows because $\optbar'(\cdot)$ is bounded and because $t\in[\beta_-+1,\beta_+]$ implies that $\convnorm{t} = \mathcal{O}(\log \gamma)$.

\subsection{Proof of \eqref{eq:ts_error_bound}}\label{sec:ts_error_bound}
For $t\in[\beta_- + 1,\beta_+]$, we obtain \eqref{eq:ts_error_bound} through the following steps:
\begin{IEEEeqnarray}{rCl}
  \IEEEeqnarraymulticol{3}{l}{\int_{\convnorm{\beta_-} }^{\convnorm{t}} E_k(s) \intdiff s}\nonumber\\ &=& \sum_{i=\beta_-+1}^{t}\int_{\convnorm{i-1}}^{\convnorm{i}} E_k(s)\intdiff s \\
 &=& \sum_{i=\beta_-+1}^{t} [\convnorm{i}-\convnorm{i-1}] E_k(s_i)\label{eq:ts_error_bound1}\\
 &=& \sum_{i=\beta_-+1}^{t} [\convnorm{i}-\convnorm{i-1}] \Big(E_k(\convnorm{i}) \nonumber\\
 &&\qquad\qquad\qquad\quad\qquad\qquad{}- [\convnorm{i}-s_i] E_k'(s_i')\Big) \label{eq:ts_error_bound2}\\
  &=& \sum_{i=\beta_-+1}^{t} [\convnorm{i}-\convnorm{i-1}] E_k(\convnorm{i})\nonumber\\
  &&{} -\sum_{i=\beta_-+1}^{t} (\convnorm{i}-\convnorm{i-1})(\convnorm{i}-s_i) E_k'(s_i') \IEEEeqnarraynumspace \\
  &=& \sqrt{\frac{C^3}{\gamma V}}\sum_{i=\beta_-+1}^{t}  E_k(\convnorm{i}) +\mathcal{O}\farg{\frac{1}{\gamma}}\sum_{i=\beta_-+1}^{t} E_k'(s_i')\label{eq:ts_error_bound3}\IEEEeqnarraynumspace \\
  &=& \sqrt{\frac{C^3}{\gamma V}} \sum_{i=\beta_-+1}^{t} E_k(\convnorm{i}) +  \mathcal{O}\farg{\frac{\log \gamma}{\sqrt{\gamma}}}\label{eq:ts_error_bound4}
\end{IEEEeqnarray}
as  $\gamma \rightarrow \infty$. Here, \eqref{eq:ts_error_bound1} follows from the mean value theorem for definite integrals \cite[Th.~7.30]{apostel} which implies that there exist constants $s_i \in (\convnorm{i-1},\convnorm{i})$ satisfying \eqref{eq:ts_error_bound1}; the equality \eqref{eq:ts_error_bound2} follows from the mean value theorem \cite[Th.~5.11]{apostel}, which guarantees the existence of constants $s'_i \in(s_i,\convnorm{i})$ such that \eqref{eq:ts_error_bound2} is satisfied; \eqref{eq:ts_error_bound3} follows because $\convnorm{i}-\convnorm{i-1} = \sqrt{C^3/(\gamma V)}$; and \eqref{eq:ts_error_bound4} holds because $t\in[\beta_-+1,\beta_+]$, which implies that $(t-\beta_-) = \mathcal{O}(\sqrt{\gamma}\log \gamma)$, and because $\{|E'_k(w)|\}$ are bounded (see \eqref{eq:bounded_Eks}).

\subsection{Proof of \eqref{eq:tv_large_dev_bound}}\label{sec:tv_large_dev}
We shall first apply Hoeffding's inequality to obtain an upper bound on $1- \prBig{\max_k \tau_k \leq t }$ that holds for $t\in[\beta_++1,\infty)$. To obtain \eqref{eq:tv_large_dev_bound}, we then sum this upper bound over all integers larger than $\beta_+$.

Observe that \eqref{eq:Ik_large_deviation02} implies that there exists a constant $c_1>0$ such that, for all sufficiently large $\gamma$ and for all $k\in\mathcal{K}$, we have
\begin{IEEEeqnarray}{rCl}
  \sum_{i=1}^t I_k(P_{X_i})-|\mathcal{X}| \log(\beta_-+1) &\geq & t\left(C - \frac{c_1 \log \gamma}{\sqrt{\gamma}}\right).\label{eq:Ik_large_dev}\IEEEeqnarraynumspace
\end{IEEEeqnarray}
Choose an arbitrary $\vect{\tilde x}\in\mathcal{X}^{\beta_-}$ of type $P^{(1)}$. Then, we proceed with the steps \eqref{eq:tv_large_dev_inf_rel}--\eqref{eq:achiev_hoeffding}, shown in the top of the next page.\begin{figure*}[!t]
\normalsize
\setcounter{MYtempeqncnt}{\value{equation}}
\setcounter{equation}{404}
\begin{IEEEeqnarray}{rCl}
 1- \prBig{\max_k \tau_k \leq t }
 &\leq& 1-\prBigg{\min_k \mathopen{}\Bigg\{\sum_{n=1}^{t} i_{P_{X_n}, W_k}(X_n;Y_{k,n}) -|\mathcal{X}| \log(\beta_-+1)\Bigg\}\geq \gamma} \label{eq:tv_large_dev_inf_rel}\\
  &\leq& \sum_k \pr{ \sum_{n=1}^{t} i_{P_{X_n}, W_k}(X_n;Y_{k,n}) -|\mathcal{X}| \log(\beta_-+1)\leq \gamma}\label{eq:achiev_union}\\
    &=& \sum_k \prbigg{ \sum_{n=1}^{t} i_{P_{X_n}, W_k}(X_n;Y_{k,n}) -|\mathcal{X}| \log(\beta_-+1)\leq \gamma\bigg|X^{\beta_-}=\vect{\tilde x}}\label{eq:cond_Xfixed_large}\\
  &\leq& \sum_k \ee{-\const \bigg(\frac{ \sum_{i=1}^t I_k(P_{X_i})-|\mathcal{X}| \log(\beta_-+1)-\gamma}{\sqrt{\gamma}} \bigg)^2}\label{eq:achiev_hoeffding1}\\
    &\leq& K \ee{-\const \bigg(\frac{t (C-c_1 \log(\gamma)/\sqrt{\gamma})-\gamma}{\sqrt{\gamma}} \bigg)^2}.\label{eq:achiev_hoeffding}
\end{IEEEeqnarray}

\setcounter{equation}{\value{MYtempeqncnt}}
\hrulefill
\vspace*{4pt}
\end{figure*}\addtocounter{equation}{5}
Here, \eqref{eq:tv_large_dev_inf_rel} follows from \eqref{eq:tv_achiev_tau_inf_rel}; \eqref{eq:achiev_union} follows from the union bound; \eqref{eq:cond_Xfixed_large} follows because the distribution of $\sum_{n=1}^{\beta_-}  i_{P_{X_n},W_k}(X_n; Y_{k,n})$ depends only on $X^{\beta_-}$ through its type, since $X^{\beta_-}$ is of constant composition; \eqref{eq:achiev_hoeffding1} follows from Hoeffding's inequality \cite[Th.~2]{Hoeffding1963}, and because $\{i_{P_{X_n},W_k}(X_n; Y_{k,n})\}_{n=1}^{\beta_-}$ are conditionally independent given $X^{\beta_-}$; and  \eqref{eq:achiev_hoeffding}, for sufficiently large $\gamma$, holds because $t C > \gamma$ and  because of \eqref{eq:Ik_large_dev}.
Consequently, we have
\begin{IEEEeqnarray}{rCl}
\IEEEeqnarraymulticol{3}{l}{ \sum_{t= \beta_++1}^{\infty}\Big(1- \prBig{\max_k \tau_k \leq t }\Big) }\nonumber\\
   &\leq& K \sum_{t= \beta_+}^{\infty} \ee{-\const  \left(\frac{t (C - c_1\log(\gamma)/\sqrt{\gamma}) -\gamma}{\sqrt{\gamma}}\right)^2}\label{eq:achiev_large_dev_taylor}\\
   &=& K \sum_{i= 1}^{\infty} \sum_{t=\lfloor \gamma/C + i \sqrt{\gamma V/C^3}\log \gamma \rfloor}^{\lfloor \gamma/C + (i+1) \sqrt{\gamma V/C^3}\log \gamma \rfloor-1} \nonumber\\
   &&\qquad{}\ee{-\const  \left(\frac{t (C - c_1\log(\gamma)/\sqrt{\gamma}) -\gamma}{\sqrt{\gamma}}\right)^2}\label{eq:achiev_large_dev_taylor2}\\
      &\leq& \const\sqrt{\gamma} \log(\gamma)  \sum_{i= 1}^{\infty}\eebigg{\nonumber\\
      &&\quad{}{-}\const  \bigg[\bigg(\frac{\sqrt{\gamma}}{C} +i \const\log \gamma \bigg)\bigg(C - \frac{c_1\log\gamma}{\sqrt{\gamma}}\bigg) -\sqrt{\gamma}\bigg]^2}\label{eq:achiev_decreasing}\IEEEeqnarraynumspace\\
      &\leq& \const  \sqrt{\gamma}\log(\gamma) \sum_{i= 1}^{\infty}\ee{-\const  \left(i \log \gamma- \const  \right)^2}\label{eq:achiev_hoeffding_trick1}\\
  &\leq& \const \sqrt{\gamma} \log(\gamma)\sum_{i=1}^{\infty} \ee{-\const \left( i \log \gamma\right)^2 }\label{eq:achiev_log_sum}\\
    &=&  \const \sqrt{\gamma}\log(\gamma) \sum_{i=1}^{\infty}  \ee{-\const  \log^2 \gamma }^{i^2}\\
    &\leq& \const  \sqrt{\gamma}\log(\gamma)\sum_{i=1}^{\infty}  \ee{-\const  \log^2 \gamma }^i\\
    &=&  \const \sqrt{\gamma}\log(\gamma) \frac{\ee{-\const  \log^2 \gamma }}{1-\ee{-\const  \log^2 \gamma }} = o(1)\label{eq:achiev_final_large_deviations_bound}
\end{IEEEeqnarray}
\begin{sloppypar}\noindent as $\gamma\rightarrow \infty$.
Here, \eqref{eq:achiev_large_dev_taylor} follows by \eqref{eq:achiev_hoeffding}, \eqref{eq:achiev_decreasing} follows because $ \ee{-\const  \left((t (C - c_1 \log(\gamma)/\sqrt{\gamma}) -\gamma)/\sqrt{\gamma}\right)^2}$ decreases in $t$ for sufficiently large $\gamma$, and both \eqref{eq:achiev_hoeffding_trick1} and \eqref{eq:achiev_log_sum}  hold for sufficiently large $\gamma$.\end{sloppypar}

\subsection{Proof of \eqref{eq:tv_achiev_central_bound}}\label{eq:achiev_central_regime}
We shall apply a multivariate version of the Berry-Esseen central limit theorem for sums of independent RVs to $\sum_{n=1}^{t} i_{P_{X_n}, W_k}(X_n;Y_{k,n})$ in \eqref{eq:tv_achiev_tau_inf_rel}. 
To do so, we need to compute the variance of $\sum_{n=1}^{t} i_{P_{X_n}, W_k}(X_n;Y_{k,n})$. It turns out convenient to define the unconditional information variance 
\begin{IEEEeqnarray}{rCl}
U_k(P)\triangleq \Vaa{P\times W_k}{i_{P, W_k}(X; Y_k)}\label{eq:Uk_def}
\end{IEEEeqnarray}
and
\begin{IEEEeqnarray}{rCl}
V_k^t &\triangleq& \frac{1}{t}\Bigg(\beta_- V_k(P^{(1)}) + \sum_{n=\beta_-+1}^t U_k(P_{X_n})\Bigg).\label{eq:Vkt_def}
\end{IEEEeqnarray} 
Although $V_k^t$ depends on $\gamma$, we omit denoting this explicitly to make the notation more convenient. 
Then, for $t\in[\beta_-+1,\beta_+]$, we have that
\begin{IEEEeqnarray}{rCl}
\IEEEeqnarraymulticol{3}{l}{\text{Var}\mathopen{}\left[\sum_{n=1}^{t} i_{P_{X_n}, W_k}(X_n;Y_{k,n}) \right]-\sum_{n=\beta_-+1}^t U_k(P_{X_n})}\nonumber\\
\qquad &=& \EBigg{\text{Var}\mathopen{}\Bigg[\sum_{n=1}^{\beta_-} i_{P^{(1)}, W_k}(X_n;Y_{k,n})\bigg| X^{\beta_-}  \Bigg]} \nonumber\\
 &&{}+ \text{Var}\mathopen{}\Bigg[ \EBigg{  \sum_{n=1}^{\beta_-} i_{P^{(1)}, W_k}(X_n;Y_{k,n}) \bigg| X^{\beta_-}} \Bigg]\label{eq:law_of_total_variance}\\
  &=& \sum_{n=1}^{\beta_-}\Ebig{\text{Var}\mathopen{}\big[ i_{P^{(1)}, W_k}(X_n;Y_{k,n})\big| X_n  \big]}\nonumber\\
  &&{}+ \text{Var}\mathopen{}\Bigg[   \sum_{n=1}^{\beta_-} \E{ i_{P^{(1)}, W_k}(X_n;Y_{k,n}) | X_n} \bigg| X^{\beta_-} \Bigg]\label{eq:law_of_total_variance2}\IEEEeqnarraynumspace\\
   &=& \beta_- V_k(P^{(1)}).\label{eq:law_of_total_variance3}
\end{IEEEeqnarray}
Here, \eqref{eq:law_of_total_variance} follows from the law of total variance and from \eqref{eq:Uk_def}, \eqref{eq:law_of_total_variance2} follows because $\{i_{P^{(1)}, W_k}(X_n;Y_{k,n})\}_{n=1}^{\beta_-}$ are conditionally independent given $X^{\beta_-}$, and \eqref{eq:law_of_total_variance3} follows since the marginal distribution of $X_n$, for $n\in[1,\beta_-]$, is given by $P^{(1)}$ and since $X^{\beta_-}$ is of constant composition. 

Recall that $\vect{\tilde x}\in\mathcal{X}^{\beta_-}$ has the type $P^{(1)}$. By using \eqref{eq:inf_dens_constant_comp} and by invoking the multivariate version of the Berry-Esseen central limit theorem for sums of independent RVs reported in \cite[Th.~1.8]{Tan2014} and \cite[Th~1.3]{Gotze1991}, we obtain, for $t\in[\beta_-+1,\beta_+]$, the estimate 
\begin{IEEEeqnarray}{rCl}
\IEEEeqnarraymulticol{3}{l}{\pr{ \min_k  i_{P_{X^{t}},W_k^{t}}(X^{t}; Y_{k}^{t}) \geq \gamma}}\nonumber\\
  &\geq&\prBigg{ \min_k \mathopen{}\Bigg\{ \sum_{n=1}^t  i_{P_{X_n},W_k}(X_n; Y_{k,n})\Bigg\} \nonumber\\
  &&{}\qquad\qquad\qquad\geq \gamma+|\mathcal{X}| \log(\beta_-+1)}\\
   &=&\prBigg{ \min_k \mathopen{}\Bigg\{ \sum_{n=1}^t  i_{P_{X_n},W_k}(X_n; Y_{k,n})\Bigg\} \nonumber\\
  &&{}\qquad\qquad\qquad \geq \gamma+|\mathcal{X}| \log(\beta_-+1) \bigg| X^{\beta_-} = \vect{\tilde x}}\IEEEeqnarraynumspace\label{eq:cond_Xfixed}\\
&\geq& \prod_k Q\farg{ \frac{\gamma +|\mathcal{X}|\log(\beta_-+1)- \sum_{n=1}^t I_k(P_{X_n})}{\sqrt{V_k^t}}}\nonumber\\
&&{} + \frac{\const}{ \sqrt{\gamma}  }\label{eq:improv_achiev_berry_intermediate}\\
&=& \prod_k Q\farg{ \frac{\gamma- \sum_{n=1}^t I_k(P_{X_n})}{\sqrt{V_k^t}}} +  \mathcal{O}\farg{\frac{\log \gamma}{\sqrt{\gamma}}}. \label{eq:achiev_uniform_berry_improv}
\end{IEEEeqnarray}
Here, \eqref{eq:cond_Xfixed} follows because the distribution of $\sum_{n=1}^{\beta_-}  i_{P_{X_n},W_k}(X_n; Y_{k,n})$ depends only on $X^{\beta_-}$ through its type since $X^{\beta_-}$ is of constant composition and \eqref{eq:improv_achiev_berry_intermediate} follows, in addition to the central limit theorem, from \eqref{eq:Vkt_def}, from \eqref{eq:law_of_total_variance3}, because $\{i_{P_{X_n},W_k}(X_n; Y_{k,n})\}_{n\in\{1,\cdots,\beta_-\}, k\in\mathcal{K}}$ are conditional independent given $X^{\beta_-}$ and because the $\{T_k(\cdot)\}$ are uniformly upper-bounded \cite[Lem.~46]{Polyanskiy2010b}. Furthermore, we obtained \eqref{eq:achiev_uniform_berry_improv} by performing a first-order Taylor expansion of the $Q$ function around $\Big(\gamma- \sum_{n=1}^t I_k(P_{X_n})\Big)/\sqrt{V_k^t}$.

Next, we approximate $V_k^t$ in \eqref{eq:achiev_uniform_berry_improv} by $V_k$ defined in \eqref{eq:cond_inf_var} in a sense we shall make precise shortly. 
Recall that $\delta$ is an arbitrarily positive constant. Then, for sufficiently large $\gamma$, we have
\begin{align}
  \left| \sqrt{\frac{V_k^t}{(\frac{1}{t}\sum_{n=1}^t I_k(P_{X_n}) )^3}} - \sqrt{\frac{V_k}{C^3}} \right| \leq \sqrt{\frac{V_k}{C^3}} \delta\label{eq:VkIk3_approx_achiev_improv}
\end{align}
for every $t\in[\beta_-+1,\beta_+]$ (recall that $P_{X_i}$ and $V_k^t$ both depends on $\gamma$). This follows because  $U_k(P)$ and $I_k(P)$ are upper-bounded by 
\begin{IEEEeqnarray}{rCl}
U_{\text{max}}\triangleq \max_{\substack{P\in\mathcal{P}(\mathcal{X})\\k\in\mathcal{K}}} U_k(P)<\infty
\end{IEEEeqnarray}
and by $C_k$, respectively, and lower-bounded by $0$. Hence, we have that

\begin{multline}
\sqrt{\frac{V_k^t}{(\frac{1}{t}\sum_{n=1}^t I_k(P_{X_n}) )^3}}
\\\geq \frac{t}{\beta_-}\sqrt{\frac{ V_k(P^{(1)}) }{\left( I_k(P^{(1)})+ (t/\beta_- -1) C_k\right)^3}}\label{eq:VkPt_convergence}
\end{multline}
which converges to $\sqrt{\frac{V_k}{C^3}}$ as $\gamma\rightarrow \infty$. This convergence follows from \eqref{eq:Ik_P_Pgam_approx}, because $t/\beta_- \rightarrow 1$, and because $\sqrt{V_k(P^{(1)})/I_k(P^{(1)})^3} \rightarrow \sqrt{V_k/C^3}$  as $\gamma \rightarrow \infty$. Likewise, we have
\begin{multline}
\sqrt{\frac{V_k^t}{(\frac{1}{t}\sum_{n=1}^t I_k(P_{X_n}) )^3}}
\\\leq \frac{t}{\beta_-}\sqrt{\frac{ V_k(P^{(1)})+(t/\beta_- -1) U_{\text{max}}}{ I_k(P^{(1)})^3}}
\end{multline}
which also converges to $\sqrt{V_k/C^3}$ as $\gamma\rightarrow\infty$. Consequently, these arguments imply that \eqref{eq:VkIk3_approx_achiev_improv} is satisfied for sufficiently large $\gamma$. Similarly to the asymptotic analysis of the converse bound in Appendix~\ref{sec:central_regime}, \eqref{eq:VkIk3_approx_achiev_improv} allows us to approximate $\sqrt{V_k^t/(\frac{1}{t}\sum_{i=1}^t I_k(P_{X_i}) )^3}$ by $\sqrt{V_k/C^3}$ and, hence, eliminate the dependency on $\gamma$.

In order to further bound \eqref{eq:achiev_uniform_berry_improv} for $t\in[\beta_-+1,\beta_+]$, we shall now use \eqref{eq:VkIk3_approx_achiev_improv} together with the inequality (proved in Appendix~\ref{app:lam_property})
\begin{IEEEeqnarray}{rCl}
   \frac{\gamma - \xi a}{\sqrt{\xi b}} \leq   \frac{\gamma - \xi a}{\sqrt{\gamma b/a}} + \sqrt{\frac{b}{a \gamma}} \left(\frac{\gamma - \xi a}{\sqrt{\xi b}}\right)^2
\end{IEEEeqnarray}
which holds for all positive $a, b, \xi$, and $\gamma$. This implies the steps \eqref{eq:achiev_identity_improv}--\eqref{eq:achiev_identity_upper_bound2_improv}, shown in the top of this page.\begin{figure*}[!t]
\normalsize
\setcounter{MYtempeqncnt}{\value{equation}}
\setcounter{equation}{431}
\begin{IEEEeqnarray}{rCl} 
\IEEEeqnarraymulticol{3}{l}{\pr{ \min_k i_{P_{X^t},W_{k}^t}(X^t; Y_{k}^t) \geq \gamma}}\nonumber\\
\qquad &\geq& \prod_k Q\fargBigg{ \frac{\gamma - \sum_{n=1}^t I_k(P_{X_n})}{\sqrt{\gamma V_k^t/\big(\frac{1}{t} \sum_{n=1}^t I_k(P_{X_n})\big)}}   + \sqrt{\frac{ V_k^t}{\gamma \sum_{n=1}^t I_k(P_{X_n})}}\left( \frac{\gamma - \sum_{n=1}^t I_k(P_{X_n})}{\sqrt{V_k^t }}\right)^2  }+  \mathcal{O}\farg{\frac{\log \gamma}{\sqrt{\gamma}}}\IEEEeqnarraynumspace\label{eq:achiev_identity_improv}\\
    &\geq& \prod_k Q\farg{ \frac{\gamma/\big(\frac{1}{t}\sum_{n=1}^t I_k(P_{X_n})\big) - t }{\sqrt{ \gamma  V_k^t/\big(\frac{1}{t}\sum_{n=1}^t I_k(P_{X_n})\big)^3 }}   + \const \frac{\log^2 \gamma}{\sqrt{\gamma}}} +  \mathcal{O}\farg{\frac{\log \gamma}{\sqrt{\gamma}}}\label{eq:achiev_simplification}\\
    &\geq& \prod_k Q\farg{\max_{\nu_k\in\{-1,1\}} \frac{\gamma/\big(\frac{1}{t}\sum_{n=1}^t I_k(P_{X_n})\big) - t }{\sqrt{ \gamma V_k/C^3 }(1-\delta_2 \nu_k) }   + \const \frac{\log^2 \gamma}{\sqrt{\gamma}}  } +  \mathcal{O}\farg{\frac{\log \gamma}{\sqrt{\gamma}}}\IEEEeqnarraynumspace\label{eq:achiev_identity_upper_bound_improv}\\
    &\geq& \prod_k Q\farg{\max_{\nu_k\in\{-1,1\}} \frac{\gamma/\big(\frac{1}{t}\sum_{n=1}^t I_k(P_{X_n})\big) - t }{\sqrt{ \gamma V_k/C^3 }(1-\delta_2 \nu_k) }  } +\mathcal{O}\farg{\frac{\log^2 \gamma}{\sqrt{\gamma}}}.\label{eq:achiev_identity_upper_bound2_improv}
\end{IEEEeqnarray}

\setcounter{equation}{\value{MYtempeqncnt}}
\hrulefill
\vspace*{4pt}
\end{figure*}\addtocounter{equation}{4}
Here,  \eqref{eq:achiev_simplification} holds for sufficiently large $\gamma$, \eqref{eq:achiev_identity_upper_bound_improv} follows from \eqref{eq:VkIk3_approx_achiev_improv}, and \eqref{eq:achiev_identity_upper_bound2_improv} follows because the derivative of the $Q$ function is bounded. Finally, we substitute \eqref{eq:improv_achiev_inductive} into \eqref{eq:achiev_identity_upper_bound2_improv} and get the chain of inequalities \eqref{eq:improv_achiev_Q_expand1}--\eqref{eq:improv_achiev_Q_expand4}, shown in the top of the next page.\begin{figure*}[!t]
\normalsize
\setcounter{MYtempeqncnt}{\value{equation}}
\setcounter{equation}{435}
\begin{IEEEeqnarray}{rCl}
\IEEEeqnarraymulticol{3}{l}{\pr{ \min_k i_{P_{X^t},W_k}(X^t; Y_{k}^t) \geq \gamma}}\nonumber\\
\qquad&\geq& \prod_k Q\farg{ \max_{\nu_k \in \{-1,1\}} \frac{\frac{\gamma}{C} - \frac{\gamma}{C^2}\big(\frac{1}{t}\sum_{n=1}^t I_k(P_{X_n})-C\big) +\const - t }{\sqrt{ \gamma V_k/C^3 }(1 - \delta_2 \nu_k)}  }+ \mathcal{O}\farg{\frac{\log^2 \gamma}{\sqrt{\gamma}}}\label{eq:improv_achiev_Q_expand1}\\
&\geq& \prod_k Q\farg{ \max_{\nu_k \in \{-1,1\}}\frac{\frac{\gamma}{C} -\sqrt{\frac{\gamma V}{C^3}}\diff I_k\farg{\optbar\farg{\convnorm{t}}}+ \frac{\gamma}{t C^2} \sqrt{\frac{\gamma V}{C^3}}\int_{w(\beta_-)}^{w(t)}E_k(s)\intdiff s  +  \const \log^2 \gamma   - t }{\sqrt{ \gamma V_k/C^3 }(1 - \delta_2 \nu_k)}  }+ \mathcal{O}\farg{\frac{\log^2 \gamma}{\sqrt{\gamma}}}\label{eq:improv_achiev_Q_expand2}\IEEEeqnarraynumspace\\
&\geq& \prod_k Q\farg{ \frac{1}{ \varrho_k}\max_{\nu_k \in \{-1,1\}}\frac{-\convnorm{t} -\diff I_k\farg{\optbar\farg{\convnorm{t}}}+ \frac{1}{C}\int_{-\infty}^{w(t)} E_{k}(s)\intdiff s }{1 - \delta_2 \nu_k}  }+ \mathcal{O}\farg{\frac{\log^2 \gamma}{\sqrt{\gamma}}}\label{eq:improv_achiev_Q_expand3}\\
&\geq& \prod_k F_{\tvRVbar_{\delta,k}}\farg{\convnorm{t} } + \mathcal{O}\farg{\frac{\log^2 \gamma}{\sqrt{\gamma}}}.\label{eq:improv_achiev_Q_expand4}
\end{IEEEeqnarray}

\setcounter{equation}{\value{MYtempeqncnt}}
\hrulefill
\vspace*{4pt}
\end{figure*}\addtocounter{equation}{4} 
Here, \eqref{eq:improv_achiev_Q_expand1} follows because $\frac{1}{t}\sum_{n=1}^t I_k(P_{X_n}) = C + \mathcal{O}(1/\sqrt{\gamma})$. Furthermore, we applied a Taylor expansion of $\gamma/x$ around $x=C$, which implies that there exists a positive constant $c_2$ such that the following inequality holds for all sufficiently large $\gamma$:
\begin{IEEEeqnarray}{rCl}
\IEEEeqnarraymulticol{3}{l}{\frac{\gamma}{\frac{1}{t}\sum_{n=1}^t I_k(P_{X_n})}}\nonumber\\
 \qquad &\leq& \frac{\gamma}{C} - \frac{\gamma}{C^2} \Bigg(\frac{1}{t}\sum_{n=1}^t I_k(P_{X_n})-C\Bigg)\nonumber\\
 &&{} + c_2 \gamma \Bigg(\frac{1}{t}\sum_{n=1}^t I_k(P_{X_n})-C\Bigg)^2\\
&\leq&   \frac{\gamma}{C} - \frac{\gamma}{C^2} \Bigg(\frac{1}{t}\sum_{n=1}^t I_k(P_{X_n})-C\Bigg) + \const.
\end{IEEEeqnarray}
Moreover, \eqref{eq:improv_achiev_Q_expand2} follows from \eqref{eq:improv_achiev_inductive}, and \eqref{eq:improv_achiev_Q_expand3} follows because the derivative of the $Q$-function is bounded and because
\begin{IEEEeqnarray}{rCl} 
\sqrt{\gamma}\left|\frac{1}{C} - \frac{\gamma}{t C^2}\right| = \mathcal{O}(\log \gamma).\label{eq:tv_achiev_approx_1C_gamtCsqr}
\end{IEEEeqnarray} 
To prove \eqref{eq:tv_achiev_approx_1C_gamtCsqr}, we used that $t\in [\beta_-+1, \beta_+]$ and we applied a first-order Taylor expansion of the $Q$ function, and absorbed  the remainder term in the $\mathcal{O}\farg{\log^2(\gamma)/\sqrt{\gamma}}$ term. Finally, \eqref{eq:improv_achiev_Q_expand4} follows from \eqref{eq:ts_error_bound} and by the definition of the cumulative distribution functions of the RVs $\{\tvRVbar_{\delta,k}\}$:
\begin{multline}
  F_{\tvRVbar_{\delta,k}}(w)
  \\\triangleq \Phi\farg{\frac{1}{\varrho_k} \min_{\nu_k \in\{-1,1\}} \frac{w + \diff I_k(\optbar(w)) - \frac{1}{C}\int_{-\infty}^w E_k(s)\intdiff s  }{1-\delta_2 \nu_k}}.
\end{multline}

\section{Basic Lemmas}
\label{app:lam_property}

\begin{lemma}
Fix arbitrary $x\in\mathbb{R}$, $a>0, b>0$, and $\lambda >0$. Suppose that $\xi$ is the unique solution to the equation
\begin{align}
\frac{\lambda - \xi a}{\sqrt{b \xi}} = x.
\end{align}
Then we have:
\begin{align}
  0 \leq  \xi - \left(\frac{\lambda}{a}  - x \sqrt{\frac{ \lambda b}{a^3}} \right)\leq \frac{b}{a^2} x^2.\label{eq:sqrt_property0}
  \end{align}
   The inequalities in \eqref{eq:sqrt_property0} are equivalent to 
\begin{align}
    \frac{\lambda -\xi a}{\sqrt{\lambda b/a}}  \leq x \leq \frac{\lambda -\xi a}{\sqrt{\lambda b/a}}  +\sqrt{\frac{b}{a \lambda}} x^2.\label{eq:sqrt_property}
\end{align}
\end{lemma}
\begin{IEEEproof}
 For all $x \in\mathbb{R}$, we have that
\begin{align}
\xi =  \frac{\lambda}{a} + \frac{b }{2 a^2}x^2 - x\sqrt{ \frac{b^2 }{4 a^4}x^2+\frac{b \lambda }{a^3 }}.
\end{align}
When $x\geq 0$, 
\begin{IEEEeqnarray}{rCl}
\frac{\lambda}{a} - x\sqrt{ \frac{b \lambda }{a^3 }} &\leq&    \frac{\lambda}{a} + \frac{b }{2 a^2}x^2 - x\sqrt{ \frac{b^2 }{4 a^4}x^2+\frac{b \lambda }{a^3 }} \nonumber\\
&=& \xi\\
& \leq& \frac{\lambda}{a}+\frac{b}{2a^2}x^2 - x \sqrt{\frac{b \lambda}{a^3} }.
\end{IEEEeqnarray}
Furthermore, when $x\leq 0$, 
\begin{IEEEeqnarray}{rCl}
\frac{\lambda}{a}+ \frac{b}{2a^2}x^2 - x\sqrt{ \frac{b \lambda }{a^3 }} &\leq&  \frac{\lambda}{a} + \frac{b }{2 a^2}x^2 - x\sqrt{ \frac{b^2 }{4 a^4}x^2+\frac{b \lambda }{a^3 }} \IEEEeqnarraynumspace\\
&=& \xi\\
& \leq& \frac{\lambda}{a} +\frac{b}{a^2}x^2 - x \sqrt{\frac{b \lambda}{a^3} }.
\end{IEEEeqnarray}
This establishes \eqref{eq:sqrt_property}.
\end{IEEEproof}
\begin{lemma}
\label{lem:zeta_constants}
Fix an integer $K\geq 2$. Let $\{x_j\}_{j=1}^{K-1}$ be constants such that
\begin{IEEEeqnarray}{rCl}
    \sum_{j=1}^{i} x_j > 0 \qquad \text{for } i\in\{1,\cdots,K-1\}.\label{eq:lem_hyph}
\end{IEEEeqnarray}
Then, there exist positive constants $\{\zeta_i\}_{i=1}^{K-2}$ such that
\begin{IEEEeqnarray}{rCl}
  x_i+ \zeta_{i-1} - \zeta_i &>& 0 \qquad \text{for } i\in\{1,\cdots,K-1\}.\label{eq:lem_di_d0}
\end{IEEEeqnarray}
In \eqref{eq:lem_di_d0}, we set  $\zeta_0 \triangleq \zeta_{K-1} \triangleq 0$.
\end{lemma}
\begin{IEEEproof}
The lemma is obviously satisfied when $K=2$. Next, we consider the case $K\geq 3$. Define 
\begin{IEEEeqnarray}{rCl}
\delta \triangleq \min_{i\in \{1,\cdots,K-1\}} \frac{x_1+\cdots+x_i}{K-1}\label{eq:delta_def}
\end{IEEEeqnarray}
  and let $\zeta_i \triangleq x_1 + \cdots + x_i - i\delta$ for $i\in\{1,\cdots, K-2\}$. Note that $\zeta_i$ is positive for $i\in\{1,\cdots,K-2\}$. Then, we establish  \eqref{eq:lem_di_d0} for $i\in\{1,\cdots,K-2\}$ as follows
\begin{IEEEeqnarray}{rCl}
  \IEEEeqnarraymulticol{3}{l}{x_i + \zeta_{i-1} - \zeta_i}\nonumber\\
   \quad &=& x_i + (x_1+\cdots+x_i - (i-1)\delta)\nonumber\\
   &&{} - (x_1+\cdots+x_i - i\delta)\\
  &=& \delta \\
  &>& 0.\label{eq:last_step_delta_zero}
\end{IEEEeqnarray}
Here, \eqref{eq:last_step_delta_zero} follows from \eqref{eq:lem_hyph} and \eqref{eq:delta_def}. To prove \eqref{eq:lem_di_d0} for $i=K-1$, we proceed as follows
\begin{IEEEeqnarray}{rCl}
   \IEEEeqnarraymulticol{3}{l}{x_{K-1}+ \zeta_{K-2} - \zeta_{K-1}}\nonumber\\
    &=& x_1 + \cdots +x_{K-1} - (K-2)\delta \\
   &\geq& x_1 + \cdots + x_{K-1} - \frac{K-2}{K-1}(x_1+\cdots + x_{K-1})\\
   &>& 0.
\end{IEEEeqnarray}
\end{IEEEproof}

\begin{lemma}
\label{prop:psi_property}
  Define $\psi(x) \triangleq \phi(x)/\Phi(x)$. Then the following holds:
  \begin{enumerate}[label=\alph*)]
    \item $\psi'(x)\in (-1,0)$ for all $x\in\mathbb{R}$,
    \item $\psi''(x)$ is positive for all $x\in\mathbb{R}$,
    \item $\beta \psi'(\psi^{-1}(x)) < \psi'(\psi^{-1}(\beta x))$ for all $x>0$ and $\beta > 1$.
  \end{enumerate}
\end{lemma}
\begin{IEEEproof}
Define $\nu(x) \triangleq \psi(-x)$. Then,
\begin{IEEEeqnarray}{rCl}
    \frac{1}{\nu(x)} = \e{\frac{1}{2}x^2} \int_{x}^\infty \e{-\frac{1}{2}u^2}\intdiff u.
\end{IEEEeqnarray}
This quantity is known as Mill's ratio \cite{Sampford1953}. It follows from \cite[Eq.~(3)]{Sampford1953} that $\nu'(x) \in (0,1)$, which implies that $\psi'(x) \in (-1,0)$. 
Similarly, (b) follows from \cite[Eq.~(4)]{Sampford1953}, which states that $\nu''(x) > 0$, thereby implying that $\psi''(x)>0$.

To establish (c), we use the identity
\begin{IEEEeqnarray}{rCl}
\psi'(\psi^{-1}(x)) &=& -\psi(\psi^{-1}(x))(\psi(\psi^{-1}(x)) + \psi^{-1}(x)) \IEEEeqnarraynumspace\\
&=& -x(x + \psi^{-1}(x)).\label{eq:eq0}
\end{IEEEeqnarray}
This implies that
\begin{IEEEeqnarray}{rCl}
 \IEEEeqnarraymulticol{3}{l}{\psi'(\psi^{-1}(\beta x)) - \beta \psi'(\psi^{-1}(x))}\nonumber\\
  &=& -\beta x(\beta x + \psi^{-1}(\beta x))+\beta x ( x+\psi^{-1}(x)))\label{eq:eq1}\\
&=& \beta x \left[x+\psi^{-1}(x)-\beta x - \psi^{-1}(\beta x)\right]\\
&>& 0.\label{eq:eq2}
\end{IEEEeqnarray}
Here, \eqref{eq:eq1} follows from \eqref{eq:eq0} and \eqref{eq:eq2} follows because $x + \psi^{-1}(x)$ is a decreasing function.
\end{IEEEproof}

\bibliographystyle{IEEEtranTCOM}
\bibliography{broadcast_trans}


\begin{IEEEbiographynophoto}{Kasper Fløe Trillingsgaard}
 (S'12) received his B.Sc. degree in electrical engineering, his M.Sc. degree in wireless communications, and his Ph.D. degree in electrical engineering from Aalborg University, Denmark, in 2011, 2013, and 2017, respectively. He is currently a postdoctoral researcher at the same institution. He was a visiting student at New Jersey Institute of Technology, NJ, USA, in 2012 and at Chalmers University of Technology, Sweden, in 2014. His research interests are in the areas of information and communication theory. 
\end{IEEEbiographynophoto}

\begin{IEEEbiographynophoto}
{Wei Yang}(S'09--M'15) received the B.E. degree in communication engineering and M.E. degree in communication and information systems from the Beijing University of Posts and Telecommunications, Beijing, China, in 2008 and 2011, and the Ph.D. degree in Electrical Engineering from Chalmers University of Technology, Gothenburg, Sweden, in 2015. In the summers of 2012 and 2014, he was a visiting student at the Laboratory for Information and Decision Systems, Massachusetts Institute of Technology, Cambridge, MA. From 2015 to 2017, he was a postdoctoral research associate at Princeton University, Princeton, NJ. In Sep. 2017, he joined Qualcomm Research, San Diego, CA, where he is now a senior engineer.
\end{IEEEbiographynophoto}

\begin{IEEEbiographynophoto}{Giuseppe Durisi}
(S'02--M'06--SM'12) received the Laurea degree summa cum laude and the Doctor degree both from Politecnico di Torino, Italy, in 2001 and 2006, respectively.
From 2006 to 2010 he was a postdoctoral researcher at ETH Zurich, Zurich, Switzerland.
In 2010, he joined Chalmers University of Technology, Gothenburg, Sweden, where he is now professor and co-director of Chalmers information and communication technology Area of Advance.

Dr. Durisi is a senior member of the IEEE. He is the recipient of the 2013 IEEE ComSoc Best Young Researcher Award for the Europe, Middle East, and Africa Region, and is co-author of a paper that won a ``student paper award" at the 2012 International Symposium on Information Theory, and of a paper that won the 2013 IEEE Sweden VT-COM-IT joint chapter best student conference paper award.  In 2015, he joined the editorial board of the \textsc{IEEE Transactions on Communications} as associate editor.
From 2011 to 2014, he served as publications editor for the \textsc{IEEE Transactions on Information Theory}. His research interests are in the areas of communication theory, information theory, and machine learning.
\end{IEEEbiographynophoto}

\begin{IEEEbiographynophoto}{Petar Popovski}
 (S’97–A’98–M’04–SM’10–F’16) is a Professor of Wireless Communications with Aalborg University. He received the Dipl. Ing. degree in electrical engineering and the Magister Ing. degree in communication engineering from the "Sts. Cyril and Methodius" University, Skopje, Republic of Macedonia, in 1997 and 2000, respectively, and the Ph.D. degree from Aalborg University, Denmark, in 2004. He has over 300 publications in journals, conference proceedings, and edited books. He holds over 30 patents and patent applications. He received an ERC Consolidator Grant (2015), the Danish Elite Researcher award (2016), the IEEE Fred W. Ellersick prize (2016), and the IEEE Stephen O. Rice prize (2018). He is currently a Steering Committee Member of IEEE SmartGridComm and previously served as a Steering Committee Member of the \textsc{IEEE Internet of Things Journal}. He is also an Area Editor of the \textsc{IEEE Transactions on Wireless Communications}. His research interests are in the area of wireless communication and networking, and communication/information theory.
\end{IEEEbiographynophoto}



%

\end{document}